\documentclass[12pt]{report}

\usepackage{epsfig,pbox}
\usepackage{amsmath,amsthm,amssymb}
\usepackage{enumitem}
\usepackage{tikz}
\usepackage[latin1]{inputenc}
\usetikzlibrary{trees}
\usepackage{float}
\usepackage{verbatim}
\usepackage{array}
\usepackage{graphics}
\usepackage{graphicx}
\usepackage{bm}

\usepackage[overload]{textcase}
\usepackage{graphics}
\usepackage{multirow}

\usepackage{color}

\newcommand{\R}{{\mathbb R}}

\newcommand{\Complex}{\mathbb{C}}
\newcommand{\Real}{\mathbb{R}}

\newcommand{\Var}{\textrm{Var}}

\newcommand {\sr}{\stackrel}

\newcommand {\eqv}{\equiv}

\newcommand {\td}{\tilde}

\newcommand {\ra}{\rightarrow}
\newcommand {\Ra}{\Rightarrow}

\newcommand {\La}{\Leftarrow}

\newcommand {\ral}{\longrightarrow}

\newcommand {\ub}{\underbrace}

\newcommand {\ep}{\epsilon}
\newcommand {\vep}{\varepsilon}

\newcommand {\del}{\partial}

\newcommand {\ld}{\lambda}

\newcommand {\al}{\alpha}

\newcommand {\bfr}{\begin{flushright}}
\newcommand {\efr}{\end{flushright}}
\newcommand {\bfl}{\begin{flushleft}}
\newcommand {\efl}{\end{flushleft}}

\newcommand {\nn} {\nonumber}

\newcommand {\txt}{\textrm}
\newcommand {\bd}{\begin{document}}
\newcommand {\ed}{\end{document}}

\newcommand {\be}{\begin{equation}}
\newcommand {\ee}{\end{equation}}
\newcommand {\bea}{\begin{eqnarray}}
\newcommand {\eea}{\end{eqnarray}}
\newcommand {\ba}{\begin{aligned}}
\newcommand {\ea}{\end{aligned}}

\newcommand{\bbib}{\begin{thebibliography}{99}}
\newcommand{\ebib}{\end{thebibliography}}
\newcommand {\bab}{\begin{abstract}}
\newcommand {\eab}{\end{abstract}}

\newcommand {\bc}{\begin{center}}
\newcommand {\ec}{\end{center}}
\newcommand {\bit}{\begin{itemize}}
\newcommand {\eit}{\end{itemize}}
\newcommand {\txtc}{\textcolor}

\newcommand{\argmin}{\mathop{\!\!~\textrm{argmin}\!\!~}}
\newcommand{\argmax}{\mathop{\!\!~\textrm{argmax}\!\!~}}

\def\A{{\cal A}}\def\B{{\cal B}}\def\C{{\cal C}}\def\D{{\cal D}}\def\E{{\cal E}}\def\F{{\cal F}}\def\G{{\cal G}}\def\H{{\cal H}}\def\I{{\cal I}}
\def\J{{\cal J}}\def\K{{\cal K}}\def\L{{\cal L}}\def\M{{\cal M}}\def\N{{\cal N}}\def\O{{\cal O}}\def\P{{\cal P}}\def\Q{{\cal Q}}\def\R{{\cal R}}
\def\S{{\cal S}}\def\T{{\cal T}}\def\U{{\cal U}}\def\V{{\cal V}}\def\W{{\cal W}}\def\X{{\cal X}}\def\Y{{\cal Y}}\def\Z{{\cal Z}}
\def\abar{\bar{a}}\def\bbar{\bar{b}}\def\cbar{\bar{c}}\def\dbar{\bar{d}}\def\ebar{\bar{e}}\def\fbar{\bar{f}}
\def\gbar{\bar{g}}\def\ibar{\bar{i}}\def\jbar{\bar{j}}\def\kbar{\bar{k}}\def\lbar{\bar{l}}
\def\mbar{\bar{m}}\def\nbar{\bar{n}}\def\obar{\bar{o}}\def\pbar{\bar{p}}\def\qbar{\bar{q}}\def\rbar{\bar{r}}
\def\sbar{\bar{s}}\def\tbar{\bar{t}}\def\ubar{\bar{u}}\def\vbar{\bar{v}}\def\wbar{\bar{w}}
\def\xbar{\bar{x}}\def\ybar{\bar{y}}\def\zbar{\bar{z}}
\def\Abar{\bar{A}}\def\Bbar{\bar{B}}\def\Cbar{\bar{C}}\def\Dbar{\bar{D}}\def\Ebar{\bar{E}}\def\Fbar{\bar{F}}
\def\Gbar{\bar{G}}\def\Hbar{\bar{H}}\def\Ibar{\bar{I}}\def\Jbar{\bar{J}}\def\Kbar{\bar{K}}\def\Lbar{\bar{L}}
\def\Mbar{\bar{M}}\def\Nbar{\bar{N}}\def\Obar{\bar{O}}\def\Pbar{\bar{P}}\def\Qbar{\bar{Q}}\def\Rbar{\bar{R}}
\def\Sbar{\bar{S}}\def\Tbar{\bar{T}}\def\Ubar{\bar{U}}\def\Vbar{\bar{V}}\def\Wbar{\bar{W}}
\def\Xbar{\bar{X}}\def\Ybar{\bar{Y}}\def\Zbar{\bar{Z}}
\def\0bar{\bar{0}}\def\1bar{\bar{1}}\def\2bar{\bar{2}}\def\3bar{\bar{3}}\def\4bar{\bar{4}}\def\5bar{\bar{5}}
\def\6bar{\bar{6}}\def\7bar{\bar{7}}\def\8bar{\bar{8}}\def\9bar{\bar{9}}
\def\cprod{\mathop{\prod\hspace{-0.457cm}\times}}
\def\tprod{\mathop{\prod\hspace{-0.457cm}\otimes}}
\def\dsum{\mathop{\sum\hspace{-0.458cm}\oplus}}
\def\Tr{\txt{Tr}}\def\tr{\txt{tr}}\def\str{\txt{Str}}
\def\bde{\begin{description}}\def\ede{\end{description}}

\renewcommand{\theequation}{\arabic{equation}}
\numberwithin{equation}{section}

\usepackage{chngcntr}
\counterwithout{equation}{section} 
\counterwithin{equation}{chapter}  

\newtheorem{thm}{Theorem}[chapter]
\newtheorem{lmm}[thm]{Lemma}
\newtheorem{dfn}[thm]{Definition}
\newtheorem{crl}[thm]{Corollary}
\newtheorem{prp}[thm]{Proposition}
\newtheorem{example}[thm]{Example}
\newtheorem{rmk}[thm]{Remark}
\newtheorem{prb}[thm]{Problem}
\newtheorem{alg}[thm]{Algorithm}
\newtheorem{qsn}[thm]{Question}
\newtheorem{recall}[thm]{Recall}
\newtheorem{recalls}[thm]{Recalls}
\newtheorem{exercise}[thm]{Exercise}
\newtheorem{exercises}[thm]{Exercises}
\newtheorem{note}[thm]{Note}
\newtheorem{notes}[thm]{Notes}
\newtheorem{caution}[thm]{Caution}
\newtheorem{claim}[thm]{Claim}
\newtheorem{fact}[thm]{Fact}

\theoremstyle{definition}\newtheorem*{thm*}{Theorem}
\newtheorem*{lmm*}{Lemma}
\newtheorem*{dfn*}{Definition}
\newtheorem*{crl*}{Corollary}
\newtheorem*{prp*}{Proposition}
\newtheorem*{example*}{Example}
\newtheorem*{examples*}{Examples}
\newtheorem*{rmk*}{Remark}
\newtheorem*{rmks*}{Remarks}
\newtheorem*{prb*}{Problem}
\newtheorem*{alg*}{Algorithm}
\newtheorem*{recall*}{Recall}
\newtheorem*{recalls*}{Recalls}
\newtheorem*{exercise*}{Exercise}
\newtheorem*{exercises*}{Exercises}
\newtheorem*{note*}{Note}
\newtheorem*{notes*}{Notes}
\newtheorem*{caution*}{Caution}
\newtheorem*{claim*}{Claim}
\newtheorem*{fact*}{Fact}  

\renewcommand{\baselinestretch}{1.5}

\begin{document}
%
\pagenumbering{roman}
\pagestyle{plain}
%

\thispagestyle{empty}
\begin{center}
{\Large Optimal Inference for Distributed Detection}\\[3em]
{Earnest Akofor}\\[1em]

Department of Electrical Engineering and Computer Science\\
Syracuse University, Syracuse, NY 13244, USA \\
Email: eakofor@syr.edu\\[8em]

(PhD Dissertation)\\
\let\thefootnote\relax\footnotetext{This work was supported by the National Science Foundation under Grant CCF1218289, the Army Research Office under Grant W911NF-12-1-0383, and the Air Force Office of Scientific Research, Arlington, VA, USA, under Grant FA9550-10-1-0458.}
\end{center}

\newpage
  \begin{center}
    {\large Abstract}\\[1em]
  \end{center}
\begin{quote}
In distributed detection, there does not exist an automatic way of generating optimal decision strategies for non-affine decision functions. Consequently, in a detection problem based on a non-affine decision function, establishing optimality of a given decision strategy, such as a generalized likelihood ratio test, is often difficult or even impossible.

In this thesis we develop a novel detection network optimization technique that can be used to determine necessary and sufficient conditions for optimality in distributed detection for which the underlying objective function is monotonic and convex in probabilistic decision strategies. Our developed approach leverages on basic concepts of optimization and statistical inference which are provided in sufficient detail. These basic concepts are combined to form the basis of an optimal inference technique for signal detection.

We prove a central theorem that characterizes optimality in a variety of distributed detection architectures. We discuss three applications of this result in distributed signal detection. These applications include interactive distributed detection, optimal tandem fusion architecture, and distributed detection by acyclic graph networks. In the conclusion we indicate several future research directions, which include possible generalizations of our optimization method and new research problems arising from each of the three applications considered.
\end{quote}

\let\thefootnote\relax

\newpage
\tableofcontents
\listoffigures

\pagenumbering{arabic} \pagestyle{headings}

\chapter{Introduction}\label{introduction}
\section{Problem description and relevance}
\subsection*{The problem}
In complex statistical decision problems such as in distributed, sequential, or dynamic settings, the decisions from earlier stages serve as part of the data for decisions in the later stages. Therefore, even if the decision function for the decision at the first stage is an affine function of the initial decision probabilities, the decision functions at later stages are in general nonlinear in the probabilities of earlier decisions.

For distributed detection in particular, various types of decision functions appear in the literature, along with a variety of numerical algorithms for optimizing seemingly different classes of decision functions. However, there does not seem to exist any attempt to provide an efficient optimization procedure capable of stating explicit model-independent decision rules applicable to all \emph{monotonic convex} decision functions (i.e., decision functions which are monotonic and convex in decision probabilities) without resorting to suboptimal techniques (e.g., numerical programming and simulation) even for the simplest types of problems.

We intend to provide such a decision optimization framework and, hopefully, generalize the discussion to include monotonic subharmonic decision functions. We will show, in particular, that given any convex decision function to be optimized, it is always possible to decrease the space of optimization variables (no matter how large) to a set whose cardinality is no larger than the product of the cardinalities of the sets of decisions, hypotheses, and network components such as sensors. This reduction is completely independent of any network model of distributed detection.

The key observation that makes the reduction noted above possible is the fact that every extremum, i.e., maximum or minimum, of a differentiable convex function is either a boundary point of its domain or a point where its derivative equals zero.

\subsection*{Importance}
It is not too difficult to observe that the optimization of two different decision functions $F_1$ and $F_2$ can yield two decision rules $R_1$ and $R_2$ that are identical or equivalent in the sense that they have decision regions of the same analytical form and there is a one-to-one correspondence between the set of threshold parameters $T_1$ that determines $R_1$ and the set of threshold parameters $T_2$ that determines $R_2$. Therefore it is clearly inefficient to directly compute $R_2$ when $R_1$ has already been computed.

We aim to show that there is only one type of decision rule or strategy (up to equivalence in analytical form as stated above) that optimizes every monotonic convex decision function, even in the distributed setting. This should significantly reduce the effort involved in computing decision rules for decision functions in the monotonic convex class. Moreover, this analysis reveals that if sensor observations are conditionally independent and follow certain simple distributions (e.g., exponential family), then the decision problem becomes analytically tractable even for certain complex situations, such as that of distributed detection over acyclic graphs, as long as the decision functions are monotonic and convex.

In distributed detection literature, apparently different algorithms exist for computing decision rules for objective functions in the monotonic convex class. However, with our analysis, only one such algorithm may be necessary.

\section{Related work and contributions}
Almost every research paper on distributed detection first specifies a decision function, and then proceeds to obtain decision rules serving as necessary (and sometimes sufficient) conditions for optimality. To provide these rules, the authors tend to rely on the following.
\begin{enumerate}
\item[(a)] Susceptibility of the optimization problem to person-by-person optimization (PBPO) methods, especially when the underlying objective function is affine in decision probabilities. Each local sensor rule is derived under the assumption that optimal rules of all other sensors are given. For example, PBPO methods have been employed in \cite{TPKa,TPKc,TPKbI,TPKbII,alhakeem-varshney-u,Gubner-et-al2,Zhang-et-al6,Varshney:book,veeravalli-basar-poor-93}.
\item[(b)] Suboptimal methods (e.g., generalized likelihood ratio tests) based on well known optimal solutions of simpler problems. At least one of the basic hypotheses is composite, and detection of \emph{ a given composite hypothesis} involves optimization over \emph{its components}. Generalized likelihood ratio tests have been used for example in \cite{zhu-et.al1,zhu-et.al2,song-et-al,song-et-al2}.
\item[(c)] Susceptibility of the optimization problem to dynamic programming techniques, especially in the context of sequential distributed detection. Optimization is performed repeatedly in several consecutive steps, where optimization at any given step utilizes suboptimal input from previous steps. For example, dynamic programming methods are found in \cite{teneketzis-ho-87,lavigna-makowski-baras-86,Veeravalli-92-thesis,veeravalli-basar-poor-93,veeravalli-99}.
\end{enumerate}
Any success with the first two (and possibly the third) methods above is mostly a consequence of the monotonic and convex nature of the underlying decision function. The third method, i.e., dynamic programming, attempts to avoid the problem of a large space of optimization variables by sequentially incrementing the number of active optimization variables until a desired level of accuracy is reached.

All of these methods fail to recognize, and to properly utilize, the automatic reduction in the space of optimization variables associated with convex decision functions in general, as well as automatic optimality conditions which hold for monotonic convex decision functions in particular. Consequently, much greater effort than necessary is often required in establishing sufficiency (and hence optimality) of necessary conditions given in the form of local sensor decision strategies. This is a problem we intend to address in some detail.

The main contributions of this thesis are the following.
\begin{enumerate}
\item \emph{Optimal hypothesis testing} (Chapter \ref{optimal-detection}):  We extend work on optimal detection initiated in \cite{akofor-chen-icassp,akofor-chen-it-paper}. Specifically, we prove that every monotonic convex decision function has a unique optimum. We derive the general structure of optimal decision rules that represent the necessary and sufficient conditions for this optimum.
\item \emph{Interactive distributed detection} (Chapter \ref{interactive-detection}): Based on the optimality criterion obtained in Chapter \ref{optimal-detection}, we present work done in \cite{akofor-chen-it-paper} on interactive distributed detection, which is related work done in \cite{akofor-chen-icassp,zhu-akofor-chen-wcnc}. We consider a decision fusion setup in which two sensors in tandem interact once in a memoryless fashion, by exchanging 1-bit decisions in a two-way communication process. It is shown that this interactive fusion can improve fixed sample performance of the Neyman-Pearson (NP) test but not large sample asymptotic performance of the test. This result is then extended to more realistic situations involving multiple rounds of memoryless interaction, multiple peripheral sensors, and the exchange of multibit decisions.
\item \emph{Optimal fusion architecture} (Chapter \ref{communication-direction}): Again, based on the optimality criterion in Chapter \ref{optimal-detection}, we present work done in \cite{akofor-chen-globalsip} on the problem of determining the preferred two-sensor tandem fusion architecture in distributed detection of a deterministic, or Gaussian-distributed random, signal in Gaussian noises. Using an optimal version of a suboptimal decision strategy employed in \cite{song-et-al,song-et-al2}, as well as some techniques used therein, we determine that for low signal-to-noise ratio (SNR), the better sensor, i.e., the one with a larger SNR, should serve as the fusion center.
\item \emph{Detection over acyclic graphs} (Chapter \ref{acyclic-graph-detection}): We present some preliminary work on Bayesian distributed detection with sensor networks in the form of acyclic directed graphs. Specifically, we prove that if the communicated messages among sensors are such that each sensor passes the same message to every sensor receiving from it, then the optimal local decision rules for such a network are not more complicated than those of the simple tandem and parallel networks. Similar work was done in \cite{TPKbI,TPKbII} under assumption of binary hypotheses, binary decisions, and at most a single connecting path between any two sensors. Our conclusions above do not require these assumptions.

\end{enumerate}

We would like to remark that the results of Chapter \ref{acyclic-graph-detection} in particular may, or may not, be known. However, what is important for us in that chapter is not novelty but the relative ease with which the results therein can be obtained with the help of Proposition \ref{region-proposition}. In other words, Chapter \ref{acyclic-graph-detection} is mainly illustrating applicability of optimal hypothesis testing as described in Section \ref{opt-hyp-testing}.

\section{Organization and prerequisite}
The material in this thesis can be subdivided into three parts as follows.

For completeness, we have provided a review of essential preliminary material as Part \ref{part-I}. This part contains a brief review of \emph{basic concepts} of optimal inference. These concepts include those of optimization of convex functions (Chapter \ref{optimization}) and of statistical information inference (Chapter \ref{info-theory}). The latter includes a discussion of probability, statistics, point estimation, and hypothesis testing. Part \ref{part-I} does not only make our work more self contained but also contains important results upon which the results of part \ref{part-II} are based.

Part \ref{part-II} considers statistical inference for \emph{signal detection}, and contains the formulation of an optimal inference procedure for signal detection based on the main results of Part \ref{part-I}. We begin with a brief nontechnical discussion of statistical decision theory in Chapter \ref{statistical-decision-theory}. This is then followed by a detailed discussion of optimal hypothesis testing in the context of signal detection in Chapter \ref{optimal-detection}. Here, we first formulate the optimization problem for convex decision functions and prove a central theorem that can be applied in a variety of distributed detection architectures. Then, for illustration of application of the results, we derive centralized and distributed sensor network decision rules for Bayesian detection.

Part \ref{part-III} deals with \emph{some applications} of the optimal inference procedure of Part \ref{part-II} in distributed detection. We summarize the main points of research work on distributed detection based on the methods we have developed in the previous chapters. In some cases, detailed proofs of theorems are not included since they can be found in the references. Each section is an overview of particular research papers. When possible, we indicate the papers that are being summarized, along with the references listed in those papers.

The applications considered in Part \ref{part-III} include interactive distributed detection (Chapter \ref{interactive-detection}), optimal two-sensor tandem fusion architecture (Chapter \ref{communication-direction}), and detection over acyclic graph networks (Chapter \ref{acyclic-graph-detection}). In the presentation of each application, we often begin with theoretical results which are essentially corollaries of the main results of Chapter \ref{optimal-detection}. This is then followed by performance analysis. In our case, performance analysis is done simply by plotting the optimal value of the decision function against different observational constraints (i.e., various possible types and qualities of data taken by the sensors), against different network patterns (i.e., the number and distribution of sensing nodes and links), or against different communication constraints (i.e., quality and capacity of the communication links).

We conclude the thesis in Chapter \ref{conclusion}, where we summarize our main results and applications, identify possible future directions of research, and briefly comment on why our central results from Chapter \ref{optimal-detection} can be applied in sequential detection in particular.

A reader who is familiar with techniques of convex optimization and statistical inference may proceed to Part \ref{part-II} after the introduction, and refer back to Part \ref{part-I} when necessary. Throughout the discussion, we take for granted that the reader is familiar with basic concepts of linear algebra such as spanning, independence, bases, dimension, and matrix representation of linear transformations.
We also assume acquaintance with basic notions of vector calculus in $\Real^n$, which include the volume integral, (total) derivative, partial derivative, gradient, and directional derivative of a function from $\Real^n$ to $\Real$. Some knowledge of basic probability and statistics would be helpful as well.

\section{Distributed detection}\label{distributed-detection}
Since this thesis is mainly concerned with distributed detection, we will now briefly introduce distributed detection before proceeding. As we will see in Section \ref{detection-problem}, detection is a means of data compression in which the resulting output directly infers the state of a physical phenomenon (such as the presence or absence of a signal).  Detection uses methods of optimization theory, statistical inference, and statistical decision theory. In the distributed detection setting, several detection devices called sensors perform detection separately to achieve a common goal. The main reason for studying distributed detection is contained in the following.

In practice, a distributed sensor network (i.e., a data processing system consisting of several sensors located far apart, in some precise sense) often has limited communication capabilities/resources. This makes distributed processing unavoidable. For example, two or more persons making a single decision together cannot function as a centralized system since they are only capable of exchanging summaries of their thoughts. Distributed detection provides a framework that can enhance data processing by such a system.

\begin{figure}[H]
\centering
\scalebox{0.8}{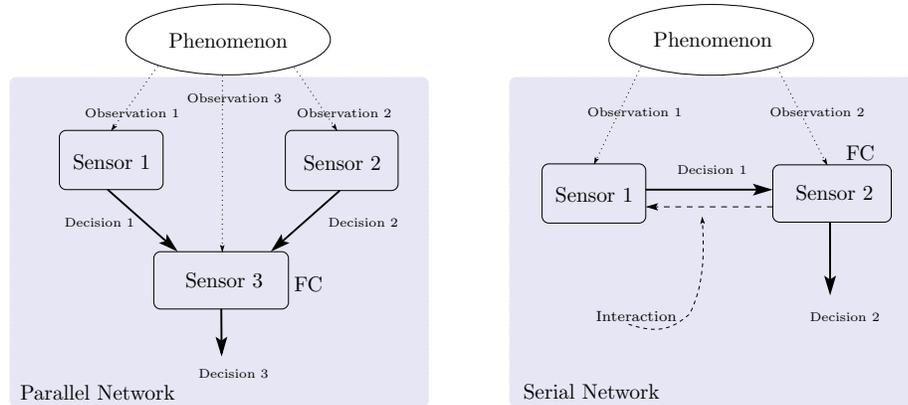}
\caption{Basic decision fusion networks}\label{network-diagrams}
\end{figure}

A distributed sensor network is often specified in the form of a graph consisting of a set of nodes and a set of arrows. Each node represents a sensor making an observation. Each arrow represents a communication link between two sensors and points in the direction in which information must flow. As shown in Fig \ref{network-diagrams}, which is based on diagrams found in \cite{tenney-sandell}, the simplest nontrivial distributed detection network contains about six basic elements - namely - at least two sensors, a phenomenon accessible to all sensors, sensor observations of the phenomenon as main input, communication links between sensors, sensor decision outputs, and sometimes a fusion center, i.e., a sensor whose output is considered to be ``the final decision''.

The following are some major benefits and advantages of distributed detection over centralized detection.
\begin{enumerate}
\item \emph{Amendable performance}: Detection performance can be improved by increasing inter-sensor connectivity or through interactive processing and feedback.
\item \emph{Robustness or fault tolerance}: If a few sensors fail, the distributed detection system can still function.
\item \emph{Reduced overload risk}: By distributing responsibility, the risk of over tasking (or overloading) one sensor is reduced.
\item \emph{Reduced communication cost}: Less communication resources/capabilities are required by a distributed detection network, since sensors exchange only summaries of their observations.
\end{enumerate}

The most significant disadvantage of distributed detection is delay in processing, i.e., a distributed system requires a longer processing time. Also, both the design and the performance analysis of a distributed detection system are more complex/challenging when compared with those of its centralized counterpart.

Other benefits and shortcomings of distributed detection can be deduced from the following discussion on distributed data compression for inference purposes.

\section*{Distributed quantization}
Quantization for inference is beneficial in a number of ways. Quantization can eliminate noise, as well as redundancies often contained in raw data collected for a specific purpose. Quantized data is easier to interpret, store, transfer, and the overall cost of processing is lower.

These benefits come at an expense. Raw data can be used for different purposes. However, quantized data can only be used for a specific purpose. That is, quantization eliminates some aspects of the data that could be relevant for other purposes. For example, it is more accurate to compare two data samples before compression than after compression.

Consequently, quantized data in general contains less information compared to the original raw data. Even when data compression is based on a sufficient statistic, there is always an underlying assumption that the data follow a specific class of distributions as determined by the underlying objective (See Sections \ref{detection-problem}, \ref{opt-hyp-testing}). These assumptions themselves can lead to a loss of information.

Nevertheless, the benefits of quantization for inference often outweigh its shortcomings due to limited capabilities of practical data processing systems. This point is strengthened by the related discussion in Section \ref{distributed-detection}.

In the literature on distributed quantization and inference, there are a number of network topologies, some of which have been studied extensively. Especially, linear and parallel networks, which are multi-sensor versions of the networks in Fig \ref{network-diagrams}, have received the greatest amount of attention because they are relatively easy to analyze. However, we will show in Chapter \ref{acyclic-graph-detection} that general networks can become equally easy to analyze under certain mild assumptions.


\part{Concepts} \label{part-I}  
\chapter{Optimization of Convex Functions}\label{optimization}

\section{The optimization problem}
We will briefly discuss \emph{optimization}\footnote[1]{A standard reference for the material in this chapter is \cite{boyed-vandenberghe}.} problems in general. Our main focus, however, will be on a class of problems called convex problems. For a fixed positive integer $d$, a real-valued function $f$ on the $d$-dimensional real vector space $\Real^d=\big\{x=(x_1,x_2,...,x_d):x_i\in\Real\big\}$ is a mapping expressed as
\bea
f:x\in D\subset\Real^d\mapsto f(x)\in\Real,\nn
\eea
where the domain $D$ is not necessarily all of $\Real^d$. For the purpose of optimization however, it is convenient to allow functions to take infinite values, in which case, we simply present every function in the form
\bea
S:\Real^d\ra\bar{\Real}=\Real\cup\{\pm\infty\},~x\mapsto S(x),\nn
\eea
where the \emph{natural domain} of $S$ is separately defined as
\bea
\txt{dom}~S=\{x\in\Real^d: S(x)\in\Real\}.\nn
\eea
The most basic optimization problem for $S$ can be presented in the form
\be\ba
\label{problem}&\txt{optimize}~~ S(x)\\
&\txt{subject to}~~ x\in C
\ea\ee
where $C$ is a subset of $\Real^d$ called the \emph{constraint set} of the problem, and $S$ is called the \emph{objective function} of the problem.

In the basic optimization problem (\ref{problem}), our \emph{objective} is either to \emph{minimize} (i.e., find the smallest value of) or to \emph{maximize} (i.e., find the largest value of) the function $S$. However, every maximization problem can be rewritten as a minimization problem, and likewise, every minimization problem is a maximization problem. Consequently, without loss of generality, we will \emph{temporarily} assume for convenience that every optimization problem is in the form
\be\ba
\label{min-problem}&\txt{minimize}~~ S(x)\\
&\txt{subject to}~~ x\in C
\ea\ee
The optimal value of $S$ will be denoted by $S_{\txt{opt}}$, and we will write
\bea
S_{\txt{opt}}=\min_{x\in C}S(x).\nn
\eea
We say a point $y\in\Real^d$ is an \emph{optimum} (or an \emph{optimal point}) of $S$ if $S(y)=S_{\txt{opt}}$, and we write
\bea
y\in\argmin_{x\in C}S(x)=\left\{z\in\Real^d:S(z)=S_\txt{opt}\right\},\nn
\eea
where the set~ $\argmin\limits_{x\in C}S(x)$~ is called the \emph{solution set} of the problem.

If the restriction $S|_C:C\ra \Real$ is a \emph{convex function} (Definition \ref{convex-function-dfn}), then the problem is called a \emph{convex problem}.

If the constraint set $C$ is not specified in the problem (\ref{problem}), then we assume $C=\Real^d$, and refer to the problem as \emph{unconstrained}. Otherwise it is a \emph{constrained optimization} problem. Most practical optimization problems are constrained in nature, and it is often possible to simplify the constraints by adjusting (or redefining) the objective function in some way. Some of these adjustment techniques are discussed next in Section \ref{constrained-optimization}.

\section{Constrained optimization}\label{constrained-optimization}
Recall that the basic problem (\ref{min-problem}) is constrained if $C\subsetneq \Real^d$, i.e., if $C$ is a proper subset of $\Real^d$. It is often possible to solve a complex optimization problem by solving a number of simpler optimization problems. However, such a possibility is difficult to uncover or identify when the geometric structure of the constraint set $C$ is sufficiently intricate. By trading the geometric complexity of $C$ for a relatively trivial algebraic refinement of the function $S$, the problem can become a lot easier to solve.

When the set $C$ is specified in terms of equality or inequality constraints, and the function $S$ satisfies some regularity conditions (e.g., differentiability), then the problem can be rewritten as an equivalent problem
\be\ba
\label{lagrangian-min-problem}&\txt{minimize}~~ L(x,\ld)\\
&\txt{subject to}~~ (x,\ld)\in \td{C}
\ea\ee
where the new objective function $L:\Real^d\times\Real^{\td{d}}\ra \Real,~(x,\ld)\mapsto L(x,\ld)$ depends on the original objective function $S$, and the new constraint set $\td{C}$ is geometrically simpler than the original constraint set $C$. The function $L$ is called a \emph{Lagrangian} function of the problem. The new optimization variables $\ld=(\ld_1,\ld_2,...,\ld_{\td{d}})$ are called \emph{Lagrange multipliers}.

We will now make the above discussion more explicit.

\subsection*{Equality constraints and the Lagrangian}
Consider an optimization problem with equality constraints:
\be\ba
\label{min-problem1}& \mathop{\txt{minimize}}_{x\in \Real^d}~~~~~S(x),\\
&\txt{subject to}~~~~h_i(x)=0,~~i=1,...,n.
\ea\ee
Let $C=\{x\in\Real^d:h_i(x)=0,~~i=1,...,n\}$ denote the constraint set as before, and let $\gamma:[0,1]\ra C,~t\mapsto\gamma(t)$ be any smooth curve in $C$. For simplicity, we will further make the following assumptions.
\begin{enumerate}
\item $S$ and $h_i$ are twice differentiable.
\item $S$ has local minima in $C$, which we wish to find.
\end{enumerate}
Then the constraints imply that
\bea
0={d\over dt}h_i(\gamma(t))=\gamma'(t)^T\cdot\nabla h_i(\gamma(t)),~~~~i=1,...,n,\nn
\eea
i.e., at the optimum, the hyperplane spanned by the gradients $\{\nabla h_i:i=1,...,n\}$ is orthogonal to $C$. Moreover, because this holds for all $\gamma$, the vectors $\{\nabla h_i:i=1,...,n\}$ span the \emph{orthogonal complement} of the \emph{tangent space} (i.e., the space of all vectors that are tangent or ``parallel'') to $C$ at the optimum.

Also, recall that at a local minimum, we have
\bea
&&0={d\over dt}S(\gamma(t))=\gamma'(t)^T\cdot\nabla S(\gamma(t)),\nn\\
\label{min-problem1-eq1}&&0\leq{d^2\over dt^2}S(\gamma(t))=\gamma'(t)^T\cdot\nabla^2 S\big(\gamma(t)\big)\cdot\gamma'(t).
\eea
The first of these relations says that at the optimum, $\nabla S$ is orthogonal to $C$. Since the orthogonal complement to $C$ at the optimum is spanned by the gradients $\{\nabla h_i:i=1,...,n\}$, it follows that at the optimum the vector $\nabla S$ must lie in the hyperplane spanned by the gradients $\{\nabla h_i:i=1,...,n\}$, so that
\bea
\nabla S(x) +\sum_{i=1}^n\ld_i\nabla h_i(x)=0,~~~~\txt{for some}~~\ld_i\in\Real.\nn
\eea

The optimization problem (\ref{min-problem1}) can now be restated as
\be\ba
\label{min-problem2}\mathop{\txt{minimize}}_{(x,\ld)\in \Real^{d+n}}~~~~~L(x,\ld)=S(x)+\sum_{i=1}^n\ld_ih_i(x).
\ea\ee
The optimality conditions (for a local minimum) are given by
\bea
&& \nabla_{x,\ld}L(x,\ld)=0,~~~~(\txt{d+n equations})\nn\\
&& \nabla_{x,\ld}^2L(x,\ld)\succeq 0,\nn
\eea
or equivalently, by
\bea
&& \nabla_xL(x,\ld)= \nabla S(x) +\sum_{i=1}^n\ld_i\nabla h_i(x)=0,\nn\\
&& \nabla_{\ld_i} L(x,\ld)= h_i(x)=0,~~i=1,...,n,\nn\\
&& \nabla^2_{x,\ld}L(x,\ld)=\left[
                              \begin{array}{cc}
                                \nabla_x^2S(x) & 0 \\
                                0 & 0 \\
                              \end{array}
                            \right]\succeq 0,\nn
\eea
where $\succeq$ denotes positive definiteness over the constraint set $C$ as implied by the relation (\ref{min-problem1-eq1}) which holds for every smooth curve $\gamma$ in $C$ that passes through the optimum.

These optimality conditions show that the problems (\ref{min-problem1}) and (\ref{min-problem2}) are equivalent for the objective of finding local minima of $S$.

\subsection*{Inequality constraints and the KKT Lagrangian}
Consider a problem with inequality constraints:
\be\ba
\label{min-problem3}& \mathop{\txt{minimize}}_x~~~~~S(x)\\
&\txt{subject to}~~~~\substack{ f_i(x)\leq 0,~~i=1,...,m\\
                       h_j(x)=0,~~j=1,...,n}
\ea\ee
We again assume for simplicity that $S,f_i,h_j$ are twice differentiable, and that $S$ has local minima in the constraint set, $C=\{x\in\Real^d:f_i(x)\leq 0,~h_j(x)=0,~i=1,...,m,~j=1,...,n\}$, that we wish to find.

The inequality constraints $f_i(x)\leq 0$ hold if and only if
\bea
f_i(x)+s_i^2=0,~~\txt{for some}~~s_i\in \Real,
\eea
where $s_i$ are known as \emph{slack variables} and their actual values need to be optimal. Therefore the problem becomes
\bea
&& \mathop{\txt{minimize}}_{x,\{s_i\}}~~~~~S(x)\nn\\
&&\txt{subject to}~~~~\substack{ f_i(x)+s_i^2= 0,~~i=1,...,m\\
                       h_j(x)=0,~~j=1,...,n}
\eea
As before, we can write down a Lagrangian
\bea
&& L(x,s,\ld,\nu)=S(x)+\sum_{i=1}^m\ld_i\big(f_i(x)+s_i^2\big)+\sum_{j=1}^n\nu_jh_j(x),\nn
\eea
in terms of which the optimization problem (\ref{min-problem3}) becomes
\bea
&& \mathop{\txt{minimize}}_{x,s,\ld,\nu}~~L(x,s,\ld,\nu).\nn
\eea
The optimality conditions (for a local minimum) are given by
\bea
&& \nabla_{x,s,\ld,\nu}L(x,s,\ld,\nu)=0,\nn\\
&& \nabla^2_{x,s,\ld,\nu}L(x,s,\ld,\nu)\succeq 0,\nn
\eea
which are equivalent (after $\{s_i\}$ has been completely eliminated) to
\bea
&& \nabla S+\sum_{i=1}^m\ld_i\nabla f_i(x)+\sum_{j=1}^n\nu_j\nabla h_j(x)=0,\nn\\
&& f_i(x)\leq 0,~~~~i=1,...,m,\nn\\
&& \ld_i\geq 0~~~~i=1,...,m,\nn\\
&& \ld_if_i(x)=0,~~~i=1,...,m,\nn\\
&& h_j(x)=0,~~~~j=1,...,n,\nn\\
&& \nu_j\in \Real~~~~i=1,...,n.\nn
\eea
The above relations, called \emph{KKT conditions}, show that the original problem (\ref{min-problem3}) is equivalent to the problem
\be\ba
\label{min-problem4}& \mathop{\txt{minimize}}_{x,\{\ld_i\},\{\nu_j\}}~~~~~L(x,\ld,\nu)=S(x)+\sum_{i=1}^m\ld_if_i(x)+\sum_{j=1}^n\nu_jh_j(x)\\
&\txt{subject to}~~~~~x\in\Real^d,~~\ld\in[0,\infty)^m,~~\nu\in\Real^n
\ea\ee

\section{Convex functions}\label{convex-functions}
Many problems that arise in practice are convex. Convex functions possess nice properties which make their optimization relatively easy to handle computationally. We will present some basic properties of convex functions in this section. The optimization of convex functions is considered in Section \ref{convex-function-opt}.

The discussion in this section pays special attention to the following points:
\begin{enumerate}
\item The description of a convex set in terms of line segments through the set, and basic operations that preserve set convexity.
\item The behavior of a convex function along line segments through its domain, and basic operations that preserve function convexity.
\end{enumerate}
These points provide a way of understanding maxima and minima of convex functions in terms of two-dimensional geometry. They are also useful for identifying those optimization problems that are convex, as well as constructing convex functions.

To simplify our discussion, we will denote the \emph{oriented line segment} between two points $x,y\in\Real^n$ by $[x,y]$. It is convenient to view $[x,y]$ as the image of the parametrization
\bea
\label{oriented-line-segment}l_{x,y}:[0,1]\ra\Real^n,~t\mapsto l_{x,y}(t)=(1-t)x+ty.
\eea

\subsection*{Set convexity}
\begin{dfn}[Convex set]\label{convex-set-dfn}
A set $D\subset\Real^n$ is convex if $[x,y]\subset D$ for any two points $x,y\in D$.
\end{dfn}
The following are some operations that preserve set convexity, and they are not difficult to check using Definition \ref{convex-set-dfn}.
\begin{enumerate}
\item \emph{Composition of operations that each preserve set convexity}: It is clear that if $f:U\subset\Real^m\ra V\subset\Real^n$, $g:V\subset\Real^n\ra W\subset\Real^k$ are mappings that each preserve set convexity, then their composition $g\circ f:U\subset\Real^m\ra W\subset\Real^k$ also preserves set convexity.
\item \emph{Set intersection}: If $A,B\subset\Real^n$ are two convex sets, let $D=A\cap B$. Then for any $x,y\in D$, $[x,y]\subset A$ and $[x,y]\subset B$, and so $[x,y]\subset D$, i.e., the intersection of convex sets is a convex set.
\item \emph{Affine transformation}: If $D\subset \Real^n$ is convex and $f:\Real^n\ra\Real^m,~~x\mapsto Ax+b$ is an affine function, then $f(D)\subset \Real^m$ is convex. More precisely, we have the following.

Let $f:U\subset\Real^m\ra V\subset\Real^n,~x\mapsto Ax+b$, be a surjective affine function, where $A$ is an $n\times m$ matrix with real entries and $b\in\Real^n$. Observe that for $x,y\in U$, we have
\bea
f\big((1-t)(Ax+b)+t(Ay+b)\big)=Af\big((1-t)x+ty\big)+b,~~~~\txt{for all}~~t\in[0,1],\nn
\eea
and so $f([x,y])=[f(x),f(y)]$. Thus, if $[x,y]\subset U$, then $[f(x),f(y)]\subset V$. This shows that affine mappings preserve set convexity.


\item \emph{Perspective transformation}: A map of the form $P:\Real^{n+1}\ra\Real^n,~~(x,t)\mapsto x/t$ is called a perspective function. This function simply uses the last component of $(x,t)$ to scale the rest and drops that last component, and thus preserves set convexity.
\item \emph{Fractional linear transformation}: This is the composition of an affine transformation and a perspective transformation $P$. Let $g:\Real^n\ra \Real^{m+1},~x\mapsto (Ax+b,c^Tx+d)$, where $A\in\Real^{m\times n},~~c^T\in \Real^{1\times n},~b,d\in\Real$. Since $P:\Real^{m+1}\ra \Real^m$, we have $\Real^n\sr{g}{\ral}\Real^{m+1}\sr{P}{\ral}\Real^m$, i.e., we have the composition $P\circ g:~\Real^n\ra\Real^m$, which is given by
    \bea
    P\circ g(x)=P(Ax+b,c^Tx+d)=(Ax+b)/(c^Tx+d),~~~~\txt{for}~~~~x\in\Real^n.\nn
    \eea
\end{enumerate}

\subsection*{Function convexity}
\begin{dfn}[Convex function]\label{convex-function-dfn}
A function $f:D\subset\Real^n\ra\Real$ is convex if $D$ is a convex set, and for all $x,y\in D$, we have
\bea
f\big((1-t)x+ty\big)\leq (1-t)f(x)+tf(y),~~~~\txt{for all}~~t\in[0,1].\nn
\eea
\end{dfn}
\begin{rmk*}
It follows immediately from Definition \ref{convex-function-dfn} that a function is convex if and only if it is convex along every line segment through its domain. For this reason, any characterization of convexity in one dimension may be readily extended to higher dimensions simply by considering it along every line segment through the function's domain.
\end{rmk*}

The following are some operations which preserve function convexity. They are not difficult to check using Definition \ref{convex-function-dfn}, but some of them can be more conveniently visualized with the help of simple geometric pictures.
\begin{enumerate}
\item \emph{Nonnegative weighted sum:} If $\{f_\al(x)\}_\al$ is a collection of convex functions, and $w_\al\geq 0$ for each $\al$, then the function $\sum_\al w_\al f_\al(x)$ is convex.
\item \emph{Composition with an affine mapping:} If $f:D\subset\Real^n\ra\Real$ is convex, and $L:\Real^m\ra\Real^n,~x\mapsto Ax+b$, where $A\in\Real^{n\times m}$, $b\in \Real^n$, then the function
    \bea
    g=f\circ L:~L^{-1}(D)\subset\Real^m\ra\Real,~x\mapsto f(Ax+b)\nn
    \eea
    is convex.
\item \emph{Pointwise supremum:} If $f_\al(x)$ is convex for each $\al$ then $\sup_\al f_\al(x)$ is convex over ~$\txt{dom}~\sup_\al f_\al=\bigcap_\al \txt{dom}f_\al.$  In particular, if $f(x,y)$ is convex in $x$ for each $y$, then $\sup_{y\in \D}f(x,y)$ is convex for any set $\D$.
\item \emph{Pointwise infimum:} If $f(x,y)$ is convex in $(x,y)$ and $C$ is a nonempty convex set, then $g(x)=\inf_{y\in C}f(x,y)$ is convex if $-\infty<g(x)$ for all $x$.

\item \emph{Perspective of a function:} If $f:\Real^n\ra \Real$ is convex then the function
\bea
g:\Real^{n+1}\ra \Real,~(x,t)\ra tf({x/t}),~~\txt{dom}g=\{(x,t):~{x/t}\in\txt{dom}f,~t>0\},\nn
\eea
is convex.

\item \emph{Composition of convex functions:} Let $h:\Real^m\ra \Real$ and $g:\Real^n\ra\Real^m$ be twice differentiable. Then  $f=h\circ g:~\Real^n\ra\Real$ satisfies
\bea
&& \nabla_{x_i}\nabla_{x_j}f=\sum_{k=1}^m\nabla_{x_i}\nabla_{x_j} g_k(x)~\nabla_{g_k}h(g)+\sum_{k,k'}\nabla_{x_i} g_k(x)~~\nabla_{g_k}\nabla_{g_{k'}}h(g)~~\nabla_{x_j}g_{k'}.\nn
\eea
Therefore, if $g$ is convex, and $h$ is both convex and increasing in each of its arguments (or if $g$ is concave, and $h$ is both convex and decreasing in each of its arguments), then $f$ is convex.
\end{enumerate}

Based on the remark following Definition \ref{convex-function-dfn}, convexity of a differentiable function of several variables can be described in terms of the following result for a function of a single variable.
\begin{thm}\label{diff-convexity-thm}~
\bit
\item[(a)] If $f:(a,b)\ra\Real$ is differentiable, then $f$ is convex if and only if $f'$ is monotonically increasing.
\item[(b)] If $f:(a,b)\ra\Real$ is twice differentiable, then $f$ is convex if and only if $f''(x)\geq 0$ for all $x\in (a,b)$.
\eit
\end{thm}
\begin{proof}~
\bit
\item[(a)] Assume $f$ is differentiable on $(a,b)$.
\bit
\item[$\ast$]($\Ra$)~ Let $f$ be convex. Then for all $\ld\in(0,1)$ and $x,y\in (a,b)$,
\bea
&& f(\ld x-(1-\ld)y)\leq \ld f(x)+(1-\ld)f(y),\nn\\
&&~~\Ra~~{f(\ld[x-y]+y)-f(y)\over\ld}\leq f(x)-f(y).\nn
\eea
By taking the limit $\ld\ra 0$, and by interchanging $x$ and $y$, we obtain
\bea
\label{DCeq1}f'(y)(x-y)\leq f(x)-f(y),~~~~f'(x)(y-x)\leq f(y)-f(x).
\eea
If $x<y$, then (\ref{DCeq1}) implies
\bea
f'(x)\leq{f(x)-f(y)\over x-y}\leq f'(y),\nn
\eea
and thus $f'$ is monotonically increasing.
\item[$\ast$]($\La$)~ Conversely, let $f'$ be monotonically increasing on $(a,b)$. Let $x,y\in (a,b)$ such that $x<y$. For $\ld\in (0,1)$, let $z=\ld x+(1-\ld)y$. Then $x<z<y$, and
\be\ba
&f(\ld x+(1-\ld)y)-[\ld f(x)+(1-\ld)f(y)]\nn\\
&~~=f(z)-[\ld f(x)+(1-\ld)f(y)]\nn\\
&~~=\ld [f(z)-f(x)]+(1-\ld)[f(z)-f(y)]\nn\\
&~\sr{\txt{mvt}}{=}\ld (z-x)f'(c)+(1-\ld)(z-y)f'(d),~~~~(~x<c<z<d<y~)\nn\\
&~~=\ld (1-\ld)(y-x)[f'(c)-f'(d)]\nn\\
&~~\leq 0,\nn
\ea\ee
where $mvt$ denotes the mean value theorem. Hence $f$ is convex.
\eit

\item[(b)] Assume $f''$ exists on $(a,b)$.
\bit
\item[$\ast$]($\Ra$)~ Let $f$ be convex. Then for $x,y\in (a,b)$, $x<y$,
\be\ba
&f'(x)\leq f'(y)~~\Ra~~f'(x)-f'(y)=(x-y)f''(c)\leq 0,~~~~(x<c<y),\nn\\
&~~\Ra~~f''(c)\geq 0.\nn
\ea\ee
Since $x,y$ were arbitrary, we have that $f''(t)\geq 0$ for all $t\in (a,b)$.
\item[$\ast$]($\La$)~ Conversely, let $f''(x)\geq 0$ for all $x\in (a,b)$. Then $f'$ is monotonically increasing, and thus $f$ is convex by part (a).
\eit
\eit
\end{proof}

By applying Theorem \ref{diff-convexity-thm} along every line segment in the function's domain, the following results are immediate.
\begin{crl}\label{diff-convexity-crl}~
\bit
\item[(a)] If $f:D\subset\Real^n\ra\Real$ is differentiable, then $f$ is convex if and only if
\bea
&&(x-y)^T\cdot\nabla f(y)\leq f(x)-f(y),~~~~\txt{for all}~~x,y\in D.\nn
\eea
\item[(b)] If $f:D\subset\Real^n\ra\Real$ is twice differentiable, then $f$ is convex if and only if
\bea
\nabla^2f(x)\succeq 0,~~~~\txt{for all}~~x\in D,\nn
\eea
i.e., if and only if the Hessian matrix $\nabla^2f(x)$ is positive semi-definite for all $x\in D$.
\eit
\end{crl}

In the optimization of convex functions, the following inequality is frequently useful.
\begin{thm}[Jensen's Inequality]\label{jensen-inequality}
Let $f:D\subset\Real^n\ra\Real$ be integrable, and let $p:D\subset\Real^n\ra[0,\infty)$ be a probability mass/density function, i.e., $\sum_{x\in D}p(x)=1$. Let $E_p[f]=\sum_{x\in D}f(x)p(x)$ denote the average of $f$ with respect to $p$.

If $S:(a,b)\supset f(D)\ra\Real$ is a differentiable convex function, then
\bea
S\big(E_p[f]\big)\leq E_p\big[S(f)\big].\nn
\eea
\end{thm}
\begin{proof}
By Corollary \ref{diff-convexity-crl},
\bea
\label{jensen-eq1}(u-v)S'(v)\leq S(u)-S(v),~~~~\txt{for all}~~u,v\in (a,b).
\eea
To obtain the result, we set $u=f(x)$, $v=E_p[f]$, and take the expectation $E_p[\cdot]$ of both sides of the inequality (\ref{jensen-eq1}).
\end{proof}
Note that the conclusion of the theorem does not require $S$ to be differentiable, i.e., differentiability was included for simplicity only. This is because the definition of convexity of a function implies \emph{a convex function is differentiable almost everywhere in its domain}.

\section{Maximization and minimization of a convex function}\label{convex-function-opt}
The goal in this section is to determine necessary and sufficient conditions for the maximum, and for the minimum, of a convex function. With respect to optimization, (differentiable) convex functions are nice because they fall in a class of functions whose maxima and minima on any domain (i.e., a connected open set) occur either on the boundary of the domain or at points where the derivative equals 0. Such functions are called \emph{subharmonic} functions.

The important thing about subharmonic behavior is the following.  Optimization problems involving a large (often infinite) number of optimization variables arise in detection theory. However, mere knowledge of the fact that the maxima and minima of the underlying objective function lie on the boundary of the function's domain (although a necessary condition only) greatly reduces the number of optimization variables. Frequently, the reduction in cardinality of the space of optimization variables is from \emph{infinite} to \emph{at most countable}. (See Theorem \ref{region-proposition}). Moreover, this necessary condition can sometimes directly yield the optimal solution if the objective function and constraints are sufficiently simple. A remark on this last point is given after Theorem \ref{convex-max-theorem}, where monotonicity is essential for necessary conditions to become sufficient.

If $f:D\subset\Real^n\ra R$ is a differentiable convex function, then it is easy to see that the necessary and sufficient condition for $y\in\Real^n$ to be the minimum of $f$ is given by
\bea
\label{minimum-condition1}{d\over dt}f\big(l_{x,y}(t)\big)\bigg|_{t=1}\leq 0~~~~\txt{for all}~~x\in D,
\eea
where $l_{x,y}(t)=l_{x,y}(t)=(1-t)x+ty$ was defined in (\ref{oriented-line-segment}). This condition simply says that while we approach $y$ along any line segment, the derivative of the function at $y$ is either 0 or negative. The condition (\ref{minimum-condition1}) is of course equivalent to
\bea
\label{minimum-condition2}(x-y)^T\cdot\nabla f(y)\geq 0~~~~\txt{for all}~~x\in D.
\eea

The necessary and sufficient conditions for the maximum of a convex function are a bit more involved because an additional property, which is \emph{monotonicity} of the objective function over the constraint set, is required to establish sufficiency of basic necessary conditions. The main results are presented in Theorem \ref{convex-max-theorem}. Note that in this theorem, the derivative $\nabla f$ is denoted by $f'$ for convenience.

\begin{thm}[Convex maximization theorem]\label{convex-max-theorem}
Let~ $f:\Real^n\ra\bar{\Real}=\Real\cup\{\pm\infty\}$~ be differentiable and convex in its natural domain, $\txt{dom}f=\{x\in\Real^n:f(x)\in\Real\}$. Let $D\subset\txt{dom}f$ be any convex set. Then for any point $z\in \overline{D}=D\cup\del D$,
\bea
\label{maxeq0}f(z)=\max_{x\in D}f(x)
\eea
if and only if the following conditions hold.
\begin{enumerate}
\item[(a)] $(x-z)\cdot f'(z)\leq 0$ for all $x\in D$.
\item[(b)] $f(z)\geq f(y)$ for every point $y\in\Real^n$ satisfying $(x-y)\cdot f'(y)\leq 0$ for all $x\in D$.
\end{enumerate}
\end{thm}
\begin{proof}
Let $z\in\overline{D}$.
\bit
\item($\Ra$): Assume $z$ satisfies (\ref{maxeq0}). Let $y\in \overline{D}$. By the convexity of $f$ in $D$, Corollary \ref{diff-convexity-crl} implies
\bea
\label{maxeq1}(x-y)\cdot f'(y)\leq f(x)-f(y)~~~~\txt{for all}~~x\in D.
\eea
Setting $y=z$ in (\ref{maxeq1}), we see that ``$f(x)-f(z)\leq 0$ for all $x\in D$'' implies ``$(x-z)\cdot f'(z)\leq 0$ for all $x\in D$'', which verifies condition (a). Condition (b) is also trivially satisfied.

\item($\La$): Assume $z$ satisfies (a) and (b). Let $y\in\overline{D}$ be any point satisfying ``$(x-y)\cdot f'(y)\leq 0$ for all $x\in D$''. Then for each $x\in D$, the function
    \bea
    g_x(t)=f\left(l_{x,y}(t)\right)=f(x+(y-x)t),~~~~0\leq t\leq 1,\nn
    \eea
    is nondecreasing at $y$, i.e., $g_x'(1)=(y-x)\cdot f'(y)\geq 0$ for all $x\in D$. This says that as we approach the point $y$ along any line segment, the function cannot decrease. Thus $y$ is a \emph{relative local maximum}, since $f(y)\geq f(x)$ for all $x\in B_\vep(y)\cap D$, where $B_\vep(y)$ is a ball of some radius $\vep>0$ centered at $y$.

    Now, by (a), $z$ is a relative local maximum of $f$ on $D$ and, by (b), $f(z)\geq f(y)$ for every relative local maximum, $y$, of $f$. This means $z$ is a global maximum of $f$ on $D$, and so $z$ satisfies (\ref{maxeq0}).
\eit
\end{proof}
\begin{rmks*}~
\begin{enumerate}
\item The condition given in (a) of the theorem is necessary but not sufficient for a global maximum as one can easily verify using a simple quadratic function on the real line. However, if for all $x\in D$ the derivative $g_x'(t)=(y-x)\cdot f'(x+(y-x)t)$ of
    \bea
    g_x(t)=f(x+(y-x)t),~~~~0\leq t\leq 1,\nn
    \eea
has the same sign for all $0\leq t\leq 1$, then condition (a) is necessary and sufficient for a global maximum. In other words, if $f$ is \emph{monotonic} in $D$, then (a) is a complete characterization for a maximum of $f$. In particular, if $f$ is an affine function, then (a) is necessary and sufficient for a global maximum of $f$.

Monotonicity as described above is too strong. In the following remarks, we will see that the maximum is always a boundary point, and so monotonicity is only required with respect to one point of the boundary $\del D$ in order for (a) to be necessary and sufficient for a global maximum. I.e., if there is a point $w\in \del D$ such that $f'$ is monotonic along every line segment through $w$ in $\overline{D}$, then (a) is both necessary and sufficient for a global maximum.

\item Note that the derivative $f'$ does not have to be zero at a relative local maximum. Also, every global maximum is a relative local maximum.
\item Let $L:\Real^n\ra \Real,~y\mapsto \mathop{\max}\limits_{x\in D\backslash\{y\}}(x-y)\cdot f'(y)$. Then the theorem says $z\in\overline{D}$ is a global maximum of $f$ on $D$ if and only if
\bea
f(z)=\max_{y\in\overline{D}~\!\!:~\!\!L(y)\leq 0}f(y).\nn
\eea
\item Note that because $f$ is convex, if $y\in\overline{D}$ satisfies $(x-y)\cdot f'(y)\leq 0$ for all $x\in D$, then $y$ must be a boundary point of $D$. That is, every local maximum lies on the boundary $\del D$ of $D$. This can be seen geometrically by recalling that a function is convex if and only if it is convex along each line segment in its domain.

    Therefore, $z\in\overline{D}$ is a global maximum of $f$ on $D$ if and only if $z\in \del D$, and
\bea
f(z)=\max_{y\in\del D~\!\!:~\!\!L(y)\leq 0}f(y).\nn
\eea

\item Algorithms exist for solving the optimization problem in Theorem \ref{convex-max-theorem}. See \cite{enhbat} for example.
\item In typical problems that arise in detection theory with a huge number of optimization variables, the role of condition (a) is to cut down the space of optimization variables to an at most countable number of threshold variables. Condition (b) then guarantees that (direct) optimization over these threshold parameters will yield an optimal solution, provided the function is monotonic. The above two-step optimization procedure is explicitly carried out in Chapters \ref{optimal-detection} to \ref{acyclic-graph-detection}.
\end{enumerate}
\end{rmks*}

\begin{crl}[Convex minimization theorem]\label{convex-min-theorem}
A point $z\in\Real^n$ is the \emph{global minimum} of the function $f$ given in Theorem \ref{convex-max-theorem} if and only if it satisfies $(x-z)\cdot f'(z)\geq0$ for all $x\in D$, which is the reverse inequality version of condition (a) of the theorem.
\end{crl}
\begin{proof}
This follows the same arguments as in the proof of Theorem \ref{convex-max-theorem}. Also see the discussion leading to the conditions (\ref{minimum-condition1}) and (\ref{minimum-condition2}).
\end{proof}

\chapter{Statistical Information Inference}\label{info-theory}
The term \emph{statistics}\footnote[2]{A more complete discussion of the concepts in this chapter can be found in \cite{casella-berger}.} refers to a collection of conceptual methods for quantifying and processing experimental observation. Some of these methods include probability in Section \ref{probability}, random variables in Section \ref{random-variables}, point estimation in Section \ref{estimation-I}, and hypothesis testing in Section \ref{estimation-II}.

Given a relatively new physical system, one would like to be able to predict its behavior under certain desired operating conditions. Accordingly, one performs an \emph{experiment} on the system by first subjecting it under specific (external or environmental) conditions, and then monitoring and recording the system's \emph{basic behavioral responses} to the conditions. In a typical experiment, the above process may be repeated as many times as necessary. From a practical standpoint, it is observed that accuracy in predicting the system's behavior using experimental results increases with the number of repetitions.

The basic behavioral responses of the system noted above are called \emph{outcomes} of the experiment. The set of all possible outcomes of the experiment is called the \emph{sample space} of the experiment. The sample space will be denoted by $S$. Subsets of $S$ are called \emph{events} of the experiment.

\section{Probability}\label{probability}
For computational convenience, the experiment is often specified in the form $(S,\Sigma,P)$, and called a  \emph{probability measure space}, where the entries are defined as follows.
\bit
\item $S$ is the sample space of the experiment as defined above.
\item $\Sigma$ is a nonempty collection of events (i.e., subsets of $S$) which is closed under complement and countable union, in the sense that $\Sigma$ contains the complements and countable unions of its elements. $\Sigma$ is called a \emph{$\sigma$-algebra} (sigma algebra) over $S$.
\item $P$ is a real function of the form $P:\Sigma\ra [0,1]$, with the following defining properties.
\bit
\item[(i)] $P(S)=1$.
\item[(ii)] $P(U)\leq P(V)$, whenever $U\subset V$.
\item[(iii)] $P(U\cup V)=P(U)+P(V)$, whenever $U\cap V=\emptyset$.
\eit
$P$ is called a \emph{probability measure} over $S$. Note that property (iii) can be extended to any countable collection of sets in $\Sigma$.
\eit

The probability $P(U)$ of an event $U\subset S$ is a measure of its \emph{likelihood of occurrence} in the experiment. Since events do intersect (so that the ``previous'' occurrence of one affects the likelihood of ``subsequent'' occurrence of another), a useful concept is that of \emph{conditional probability}, where the probability of an event $U$ given that another event $V$ has already occurred is defined as
\bea
P(U|V)~\triangleq~{P(U\cap V)\over P(V)},~~~~\txt{or by}~~~~P(U\cap V)=P(U|V)P(V).\nn
\eea

If $\{U_i,~i=1,...,n\}\subset\Sigma$ is a partition of the sample space $S$, then
\bea
&& P(V)=P\left(V\cap\bigcup_{i=1}^nU_i\right)=\sum_{i=1}^nP(V\cap U_i)=\sum_{i=1}^nP(V|U_i)P(U_i)\nn\\
\label{bayes-rule}&&~~\Ra~~P(U_i|V)={P(U_i\cap V)\over P(V)}={P(V|U_i)P(U_i)\over \sum_{j=1}^nP(V|U_j)P(U_j)},
\eea
where the relation (\ref{bayes-rule}) is known as \emph{Bayes rule}.

\section{Random variables}\label{random-variables}
Random variables are functions on sample spaces. More precisely, let $(S,\Sigma,P)$ be the probability measure space representing an experiment. Then any function $X:S\ra\X$ is called a \emph{random variable}, where $\X$ is a \emph{vector space}. Note that the probability measure $P$ is seen as summarizing all possible results of the experiment, meanwhile a random variable $X$ is seen as isolating a particular aspect or realization of the experiment.

It is not difficult to observe that every random variable $X$ gives rise to a measure space $(\X,\Sigma_X,P_X)$, where $\Sigma_X$ is a $\sigma$-algebra over $\X$ such that $X^{-1}(A)\in \Sigma$ for all $A\in \Sigma_X$, and the function
\bea
P_X:\Sigma_X\ra[0,1],~A\mapsto P(X^{-1}(A))\nn
\eea
is called the \emph{probability distribution} of $X$. Therefore, apart from isolating a certain aspect of the experiment, a random variable also summarizes the results of the experiment through its probability distribution. Note that $P_X(A)=P(X^{-1}(A))$ is often simply written as $P(X\in A)$, or as $P(X=x)$ if $A=\{x\}$ consists of a single point $x\in \X$.

Given $s\in S$, let $x=X(s)\in\X$, and let $U_x=X^{-1}(x)=X^{-1}(X(s))$. We say $X=x$ (in a random manner) with probability
\bea
P(X=x)=P_X\big(\{x\}\big)=P\big(X^{-1}(x)\big)=P(U_x).\nn
\eea
In other words, $X$ can take on any value $x\in X(S)$ but with a certain degree of uncertainty determined by the probability function $P$. The \emph{expected value} of a random variable $X$ is defined as
\bea
\label{expectation}E_P[X]=\sum_{x\in\X}x~p_X(x),\nn
\eea
where the function $p_X:\X\ra[0,\infty)$, defined such that
\bea
 P(X\in A)=\sum_{x\in A}p_X(x),\nn
\eea
is called the \emph{probability mass function} (pmf) of $X$ if $X$ is discrete, or \emph{probability density function} (pdf) of $X$ if $X$ is continuous. Note that $\sum_{x\in A}$ denotes integration over $A\subset \X$ if $X$ is continuous. The existence of the function ${dP_X\over dx}\triangleq p_X$ is determined by the \emph{Radon-Nikodym theorem}.

A function (or \emph{transformation}) of a random variable is again a random variable, in the following sense. If $X:S\ra\X$ is a random variable and $f:\X\ra\Y$ is any function (or transformation), then the composition $Y=f\circ X:S\sr{X}{\ra}\X\sr{f}{\ra}\Y$, written simply as $Y=f(X)$, is a random variable. A collection of random variables $X^n=(X_1,...,X_n)$, $X_i:S\ra\X_i$, is again a random variable given by
\bea
X^n:S\ra \X^n=\X_1\times\cdots\times\X_n,~s\mapsto X^n(s)=\big(X_1(s),...,X_n(s)\big).\nn
\eea
Verification of the above claims, based on the definitions, is straightforward.

Note that we can have a possibly continuous collection of random variables, an example of which is the following.
\begin{dfn}[Random process]
A random process $\{X(t):t\in\Real\}$ is a collection of random variables indexed by time. That is, for each value of $t$, $X(t)$ is a random variable.
\end{dfn}

\subsection*{Basics of computation with random variables}
In this section, for simplicity, we set $\X=\Real$. Thus, a (univariate) random variable $X$ is a mapping from the sample space to the reals:
\bea
X:S\ra\Real.\nn
\eea
The measure space associated with $X$ is $(\Real,\Sigma_X,P_X)$, where the \emph{probability distribution} $P_X:\Sigma_X\ra [0,1],~A\mapsto P_X(A)$ is given by
\bea
P_X(A)\eqv P(X\in A)\triangleq P\left(X^{-1}(A)\right)=P\big(\{s\in S:X(s)\in A\}\big).\nn
\eea
A random variable is said to be \emph{discrete} if its image is a discrete set in $\Real$, or \emph{continuous} if its image is a continuous set in $\Real$. We note however that the description of a continuous random variable is similar to that of a discrete random variable, except that summation $\sum$ is replaced by integration $\int$.

For a discrete random variable $X$, the evaluation of $P_X$ is often for convenience specified in terms of a \emph{probability mass function} (pmf)~ $f_X$ for $X$. Likewise, if $X$ is continuous, the evaluation of $P_X$ is specified in terms of a \emph{probability density function} (pdf)~ $f_X$  for $X$. The pmf or pdf is given by
\bea
P(X\in A)=\sum_{x\in A}f_X(x)~~~~\txt{or}~~~~P(X\in A)=\int_A f_X(x)dx.\nn
\eea
The \emph{cumulative distribution function} (cdf) of a random variable $X$ is defined as
\bea
&&F_X(x)~\triangleq~P(X\leq x)~=~\left\{
                               \begin{array}{ll}
                                 \sum_{x'=-\infty}^xf_X(x'), & \hbox{if $X$ is discrete} \\
                                 \int_{-\infty}^x f_X(x')dx', & \hbox{if $X$ is continuous}
                               \end{array}
                             \right.\nn\\
&&~~\Ra~~0\leq F_X(x)\leq 1.\nn
\eea
Therefore,
\bea
f_X(x)~=~ \left\{
            \begin{array}{ll}
              F_X(x)-F_X(x-1), & \hbox{if $X$ is discrete}, \\
              {dF_X(x)\over dx}, & \hbox{if $X$ is continuous.}
            \end{array}
          \right.\nn
\eea
The \emph{expected value} (or \emph{mean}) and \emph{variance} of a function $g(X)$ of the random variable $X$ are respectively defined by
\bea
&&\mu_{g(X)}~=~E[g(X)]~=~\left\{
                      \begin{array}{ll}
                        \sum_{x=-\infty}^\infty g(x)f_X(x), & \hbox{if $X$ is discrete,} \\
                        \int_{-\infty}^\infty g(x)f_X(x)dx, & \hbox{if $X$ is continuous,}
                      \end{array}
                    \right.\nn\\
&&\sigma_{g(X)}^2~=~\txt{Var}[g(X)]~=~E\left[\big(g(X)-E[g(X)]\big)^2\right].\nn
\eea
Note that~ $Z=g(X):S\ra \Real,~s\mapsto Z(s)=(g\circ X)(s)=g(X(s))$~ is itself a random variable with distribution function given by
\bea
&&P_Z(A)=P(Z\in A)=P(Z^{-1}(A))=P(X^{-1}(g^{-1}(A)))=P_X\left(g^{-1}(A)\right),\nn\\
&& F_Z(z)=P(Z\leq z)=P\big(g(X)\leq z\big)=P\left(X\in g^{-1}\big((-\infty,z]\big)\right)\nn\\
&&~~~~~=~\left\{
                      \begin{array}{ll}
                        \sum_{x\in A_g(z)}f_X(x), & \hbox{if $X$ is discrete,} \\
                        \int_{A_g(z)}f_X(x)dx, & \hbox{if $X$ is continuous,}
                      \end{array}
                    \right.\nn
\eea
where
\bea
&&A_g(z)=g^{-1}\big((-\infty,z]\big)=\big\{x\in\Real:~g(x)~\in~(-\infty,z]\big\}\nn\\
&&~~~~=\big\{x\in\Real:~g(x)\leq z\big\}.\nn
\eea

For simplicity, we assume that the random variables are \emph{continuous} in what follows, while noting that the case of \emph{discrete} as well as \emph{mixed} random variables is much the same. Note that a \emph{mixed random variable} is one whose range in $\Real$ has both discrete and continuous subsets that are disjoint. Also, we will sometimes denote a pmf, or pdf, $f_X$ by $p_X$.

Analogously to the univariate random variable, we define a \emph{bivariate random variable} $(X,Y)$, its \emph{inherited probability distribution} $P_{X,Y}$, its pdf $f_{X,Y}$, and its cdf $F_{X,Y}$ as follows:
{\small
\bea
(X,Y):S\ra \Real^2,~s\mapsto (X(s),Y(s)),~~~~~~~~P_{X,Y}:\Sigma_{X,Y}\ra [0,1],~~A\mapsto P_{X,Y}(A),\nn
\eea
}
where
{\small
\be\ba
&P_{X,Y}(A)\eqv P((X,Y)\in A)\triangleq P\left((X,Y)^{-1}(A)\right)=P\big(\{s\in S:(X(s),Y(s))\in A\}\big),\nn\\
& P((X,Y)\in A)=\int_A f_{X,Y}(x,y)dxdy,\nn\\
&F_{X,Y}(x,y) \triangleq P(X\leq x,Y\leq y)=\int_{-\infty}^y\int_{-\infty}^x f(x',y')dx'dy',~~\Ra~~0\leq F_{X,Y}(x,y) \leq 1,\nn\\
& f_{X,Y}(x,y)={\del^2\over\del x\del y}F_{X,Y}(x,y).\nn
\ea\ee
}$P_{X,Y}$  is said to be the \emph{joint probability distribution} for the pair of random variables $(X,Y)$, while the \emph{component random variables} $X$ and $Y$ are said to have \emph{marginal distribution functions}
\bea
&&f_X(x)=\int_{-\infty}^\infty f_{X,Y}(x,y)dy={F_{X,Y}(x,\infty)\over dx}-{F_{X,Y}(x,-\infty)\over dx},\nn\\
&&f_Y(y)=\int_{-\infty}^\infty f_{X,Y}(x,y)dx={dF_{X,Y}(\infty,y)\over dy}-{dF_{X,Y}(-\infty,y)\over dy}\nn
\eea
associated with (i.e., due to) the joint distribution. The expected value and distribution of a new random variable $Z=g(X,Y)$ are given by
\bea
&& E[g(X,Y)]=\int_{-\infty}^\infty \int_{-\infty}^\infty g(x,y)f_{X,Y}(x,y)dxdy,\nn\\
&& F_Z(z)=P(Z\leq z)=P\big(g(X,Y)\leq z\big)=\int_{A_g(z)}f_{X,Y}(x,y)dxdy,\nn
\eea
where~ $A_g(z)=\left\{(x,y)\in\Real^2:~g(x,y)\leq z\right\}$.

We can similarly proceed to describe \emph{multivariate random variables}~ $X=(X_1,...,X_n)$, where the \emph{mean vector} $M_X$ and \emph{covariance matrix} $\Sigma_X$ of $X$ are defined as
{\small
\bea
&&M_X=E[X]=(E[X_1],...,E[X_n]),\nn\\
&&\Sigma_X=E\left[ (X-E[X])(X-E[X])^T\right]=\bigg[E\big[ (X_i-E[X_i])(X_j-E[X_j])\big]\bigg]_{ij}.\nn
\eea
}If $X$ is a Gaussian-distributed real multivariate random variable, then a basic fact is that its pdf is completely determined by the pair $(M_X,\Sigma_X)$ and given by
\bea
f_X(x)={e^{-{(x-M_X)^T\Sigma_X^{-1}(x-M_X)\over 2}}\over \sqrt{(2\pi)^n\det\Sigma_X}}.\nn
\eea
It is also useful to note that if $\{X_i\}$ are jointly Gaussian-distributed, then so is any collection of variables $\{Y_k=\sum_i a_{ki}X_i\}$ that each depend linearly on the variables $\{X_i\}$.

Given any two random variables $X$ and $Y$ (each of which may be multivariate), it is often convenient to write
\bea
f_{X,Y}(x,y)={f_{X,Y}(x,y)\over f_Y(y)}f_Y(y)=f_{X|Y}(x|y)f_Y(y),~~~~f_{X|Y}(x|y)={f_{X,Y}(x,y)\over f_Y(y)},\nn
\eea
where $f_{X|Y}(x|y)$ is referred to as the \emph{conditional pdf} of $X$ given $Y$. Equivalently, in a shorthand notation where $X~\sim~f_X(x)$ means $X$ is distributed according to the density function $f_X(x)$, we can also write
\bea
X|Y~\sim~f_{X|Y}(x|y).\nn
\eea

\begin{dfn}[Characteristic function, Moment generating function]
The characteristic function of a random variable $X$ is defined to be the function
\bea
M_X(t)=E[e^{tX}]=\sum_{x\in\X}f_X(x)e^{tx},\nn
\eea
for every complex number $t\in\Complex$ for which the expectation exists. The restriction of $M_X(t)$ to $t\in\Real$ is called the moment generating function (mgf) of $X$.
\end{dfn}
Note that the characteristic function of a random variable can be used to determine its distribution.

\begin{dfn}[Independence, Conditional independence, Identical distribution, iid sequence]~\\
Let $X^n=(X_1,X_2,...,X_n)$, $Y$ be random variables. We say the random variables $X_1,X_2,...,X_n$ are independent, or that the sequence of random variables $X^n=(X_1,X_2,...,X_n)$ is independent, if
$$p(x^n)=p(x_1)p(x_2)\cdots p(x_n).$$
Similarly, we say $X_1,X_2,...,X_n$ are conditionally independent (or that $X^n$ is conditionally independent) with respect to $Y$ if
$$p(x^n|y)=p(x_1|y)p(x_2|y)\cdots p(x_n|y).$$

A sequence of random variables $X_1,X_2,\cdots$ is identically distributed if for all $i,j$, we have $p_{X_i}=p_{X_j}$, i.e., $p_{X_i}(x)=p_{X_j}(x)$ for all $x\in \X$. We say the sequence $X_1,X_2,\cdots$ is iid if it is both independent and identically distributed.
\end{dfn}

\begin{dfn}[Convergence in distribution, Convergence almost surely, Convergence in probability]~\\
A sequence of random variables $X_1,X_2,\cdots$ converges in distribution to a random variable $X$ if
\bea
\lim_{n\ra\infty}F_{X_n}(x)=F_X(x)\nn
\eea
whenever $F_X$ is continuous at $x\in\X$.

The sequence $X_1,X_2,\cdots$ converges almost surely to $X$ if for any $\vep>0$, we have
\bea
P\left(\lim|X_n-X|<\vep\right)\triangleq P\left(\limsup E_n^\vep\right)=1,\nn
\eea
where~ $E_n^\vep=\{s\in S:|X_n(s)-X(s)|<\vep\}$,~ and~ $\limsup E_n^\vep=\mathop{\cap}_{m=1}^\infty\mathop{\cup}_{n=m}^\infty E_n^\vep$.

The sequence $X_1,X_2,\cdots$ converges in probability to $X$ if for any $\vep>0$, we have
\bea
\lim P(|X_n-X|<\vep)\triangleq\lim P(E_n^\vep)=1.\nn
\eea
\end{dfn}

\begin{thm}[Central limit theorem]\label{CLT}
Let $X_1,X_2,...$ be a sequence of iid random variables. Then the random variable $Y_n={\sqrt{n}(\bar{X}_n-\mu)\over \sigma}$, where $\bar{X}_n={1\over n}\sum_{i=1}^nX_i$, converges in distribution to the standard normal random variable, i.e.,
\bea
\lim_{n\ra\infty}f_{Y_n}(x)=N(0,1)(x)={1\over\sqrt{2\pi}}e^{-{x^2\over2}},~~~~x\in \Real.\nn
\eea

Alternatively, for sufficiently large $n$ we have
\bea
\label{asymptotic-mean-distribution}f_{\bar{X}_n}(x)\approx N(\mu,\sigma^2/n)(x)={1\over\sqrt{2\pi \sigma^2/n}}e^{-{(x-\mu)^2\over 2\sigma^2/n}}.
\eea
\end{thm}
\begin{proof}
For simplicity, assume the mgf of the $X_i$'s exists near $t=0$. Then the mgf of $Y_n={\sqrt{n}(\bar{X}_n-\mu)\over \sigma}={1\over\sqrt{n}}\sum_{i=1}^n{X_i-\mu\over\sigma}$ is given by
\bea
&&M_{Y_n}(t)=M_{{1\over\sqrt{n}}\sum_{i=1}^n{X_i-\mu\over\sigma}}(t)=M_{\sum_{i=1}^n{X_i-\mu\over\sigma}}\left({t\over\sqrt{n}}\right)=\left[M_{X-\mu\over\sigma}\left({t\over\sqrt{n}}\right)\right]^n\nn\\
&&~~=\left[\sum_{k=0}^\infty {1\over k!}M^{(k)}_{X-\mu\over\sigma}(0){t^k\over n^{k\over2}}\right]^n=\left[1+0+{t^2\over 2n}+\sum_{k=3}^\infty{1\over k!} M^{(k)}_{X-\mu\over\sigma}(0){t^k\over n^{k\over2}}\right]^n\nn\\
&&~~\sr{n\ra\infty}{\ral}e^{{t^2\over 2}+\lim_{n\ra\infty}\sum_{k=3}^\infty{1\over k!} M^{(k)}_{X-\mu\over\sigma}(0){t^k\over n^{k-2\over2}}}=e^{t^2\over 2},\nn
\eea
where ~$\lim_n(1+a_n)^n=e^{\lim_n\!\!~na_n}$, and the limiting mgf is that of the standard normal distribution $N(0,1)$.
\end{proof}

\begin{lmm}[Chebychev-Markov inequality]~
Let $X$ be a random variable, and let $g:\X\ra (0,\infty)$ be an integrable function. Then
\bea
P(g(X)\geq c)\leq {1\over c}E[g(X)],~~~~\txt{for any}~~c>0.\nn\nn
\eea

\end{lmm}
\begin{proof}
\bea
P(g(X)\geq c)=\sum_{x\in\X:g(x)\geq c}f_X(x)\leq{1\over c}\sum_{x\in\X:g(x)\geq c}g(x)f_X(x)\leq {1\over c}\sum_{x\in\X}g(x)f_X(x).\nn
\eea
\end{proof}

\begin{thm}[Laws of large numbers]~
Let $X_1,X_2,...$ be iid random variables with $EX_i=\mu,~\Var X_i=\sigma^2<\infty$, and $\bar{X}_n={1\over n}\sum_{i=1}^nX_i$. Then we have the following.
\begin{enumerate}
\item Strong law: $\bar{X}_n$ converges almost surely to $\mu$.
\item Weak law: $\bar{X}_n$ converges in probability to $\mu$.
\end{enumerate}
\end{thm}
\begin{proof}
Observe that
\bea
P\left(\lim|\bar{X}_n-\mu|<\vep\right)=1~~\iff~~P\left(\lim|\bar{X}_n-\mu|\geq\vep\right)=0.\nn
\eea
\begin{enumerate}
\item Let
{\footnotesize
\be\ba
&A_n=\{s\in S:|\bar{X}_n(s)-\mu|\geq\vep\}=\{s\in S:|[X_1(s)+\cdots+X_n(s)]/n-\mu|\geq\vep\}\nn\\
&~~~~\simeq\{(x_1,...,x_n,0,0,\cdots)\in\Real^{\infty}:~|(x_1+\cdots+x_n)/n-\mu|\geq\vep\},\nn
\ea\ee
} where $\simeq$ denotes equivalence of sets with respect to cardinality, i.e., $A\simeq B$ if $A$ and $B$ have the same cardinality. Then {\footnotesize
    \be\ba
    &P\left(\lim|\bar{X}_n-\mu|\geq\vep\right)=P\left(\limsup A_n\right)=P\left(\mathop{\cap}\limits_{m=1}^\infty\mathop{\cup}\limits_{n=m}^\infty A_n\right)\leq \inf_{m\geq 1}P\left(\mathop{\cup}\limits_{n=m}^\infty A_n\right)\nn\\
    &~~~~\leq \inf_{m\geq 1}\sum_{n=m}^\infty P(A_n)=\inf_{m\geq 1}\sum_{n=m}^\infty P(|\bar{X}_n-\mu|\geq\vep).\nn
    \ea\ee
    }
    Thus, it is sufficient to show that $\sum_{n=1}^\infty P(|\bar{X}_n-\mu|\geq \vep)<\infty$. This finiteness easily follows from the central limit theorem, i.e., from the fact that $\bar{X}_n$ is distributed according to (\ref{asymptotic-mean-distribution}) for large $n$. Hence,
    \bea
    0\leq P\left(\lim|\bar{X}_n-\mu|\geq\vep\right)\leq \inf_{m\geq 1}\sum_{n=m}^\infty P(|\bar{X}_n-\mu|\geq\vep)=0.\nn
    \eea

\item Almost sure convergence implies convergence in probability for the following reason: Let $B_n=\bigcup_{k=n}^\infty A_k$, where $A_k$ is as defined in part 1 above. Then $P(|\bar{X}_n-\mu|\geq\vep)=P(A_n)\leq P(B_n)$, which implies
\be\ba
&\lim P(|\bar{X}_n-\mu|\geq\vep)=\lim P(A_n)\leq \lim P(B_n)=P(\limsup A_n)\nn\\
&~~~~=P(\lim|\bar{X}_n-\mu|\geq\vep),\nn\\
&~~\Ra~~\lim P(|\bar{X}_n-\mu|\geq\vep)\leq P(\lim|\bar{X}_n-\mu|\geq\vep).\nn
\ea\ee

Alternatively, by the Chebychev-Markov inequality, we get
{\small
\bea
P(|\bar{X}_n-\mu|\geq\ep)=P((\bar{X}_n-\mu)^2\geq\ep^2)\leq {E(\bar{X}_n-\mu)^2\over\ep^2}={\sigma^2\over n\ep^2}\ra0~~\txt{as}~~n\ra\infty.\nn
\eea
}

\end{enumerate}
\end{proof}

\section{Statistical information }\label{information-statistics}
A function of several random variables is called a \emph{statistic}. If $X^n=(X_1,...,X_n)$ is a collection of random variables and $f:\X^n\ra\Y$ is any function, then the random variable $Y=f\circ X^n:S\sr{X^n}{\ra}\X^n\sr{f}{\ra}\Y$, written $Y=f(X^n)=f(X_1,...,X_n)$, is said to be a \emph{statistic} based on $X^n$. In particular, $X^n$ is a statistic based on $X^n$.

A basic property of every random variable is \emph{uncertainty} or \emph{entropy}, and is defined as a measure of the amount of \emph{randomness} in the variable. Therefore, we can view the space $\R=\{X:S\ra \X\}$ consisting of all random variables as a ``\emph{field of uncertainty}'', with the random variables being the points of the space. \emph{Information} is a measure of how much two variables in $\R$ are separated in randomness. That is, information is randomness distance, or distance with respect to randomness, between random variables in $\R$. Therefore, information is \emph{relative uncertainty}, and one may of course loosely refer to the randomness of a random variable as the ``\emph{information content}'' of the variable.

A basic example of an information measure $d_P:\R\times\R\ra\Real$ is given by
\bea
\label{error-distance}d_P(X,Y)=P(X\neq Y),
\eea
which is a familiar quantity known as \emph{error probability} in a context where one of the variables is viewed as an \emph{estimate} of the other. Other examples, called \emph{distortions}, are given by

\bea
\label{distortion-distance}d_P(X,Y)=E_P[d(X,Y)],
\eea
where $d:\X\times\Y\ra\Real$ is a \emph{deterministic} ``\emph{distance}'' function.

In general, \emph{information metrics} are real-valued functions of statistics. Examples include asymptotic detection and estimation performance metrics such as Shannon, Kullback, Chernoff, and Fisher information. Note that all of these metrics are special instances of the quantities in (\ref{error-distance}) and (\ref{distortion-distance}), which have certain convenient properties, including \emph{additivity} over independent random variables in particular.

\section{Estimation I: Point estimators and sufficiency}\label{estimation-I}
Consider a sequence of random variables~ $X^n=(X_1,...,X_n):S\ra\X^n,~s\mapsto (X_1(s),...,X_n(s))$. Any value $x^n=(x_1,...,x_n)\in\X^n$ of $X^n$ is called a \emph{data sample}, where $n$ is the \emph{sample size}. If the sequence $x^n$ is generated according to the distribution of $X^n$, it is called a \emph{random data sample}. Consequently, we may loosely refer to $X^n$ itself as a ``\emph{random sample}''. Because $X^n$ summarizes the results of a \emph{composite experiment} in the form of a \emph{series of experiments}, each variable $X_i$ in the random sample $X^n$ is called an (experimental) \emph{observation}.

Assume we have a system with a property $\theta$ that can take values in a set $\Theta$, but we do not know its true (current) value. Then in order to determine the true value of $\theta$, we further assume that we have conducted an experiment on the system and made observations $X^n=X_1,...,X_n$. The observations are presumed to have been randomly (and independently) generated from the system according to a distribution $p_\theta(x)=p(x|\theta)$, written $X_i\sim p(x_i|\theta)$, so that $p(x^n|\theta)=\prod_{i=1}^np(x_i|\theta)$, which is just another way of saying that the sample $X^n$ summarizes the results of an experiment on the system (by means of its distribution $p(x^n|\theta)$). Consequently, $X^n$ contains information about the true value of $\theta$. Accordingly, we have the following definitions.

\begin{dfn}[Point estimator]
Any statistic $T(X^n)$ for the purpose of inferring the true value of $\theta$ is called a \emph{point estimator} of $\theta$.
\end{dfn}
By the reasoning in (\ref{distortion-distance}), estimation performance of an estimator $T(X^n)$ can be investigated using a metric of the form
\bea
\label{estimation-metric}D=E_{p(x,\theta)}\big[d\big(T(X^n),\vartheta\big)\big],
\eea
where $\vartheta$ is $\theta$ viewed as a random variable.

\begin{dfn}[Sufficient statistic]\label{sufficient-statistic-dfn}
$T(X^n)$ is a sufficient statistic for $\theta$ if $X^n$, as an estimator of $\theta$, is no better than $T(X^n)$, i.e., if
\bea
\label{sufficient-estimate}E_{p(x,\theta)}d\big(T(X^n),\vartheta\big)\leq\inf_{T'}~E_{p(x,\theta)}d\big(T'(X^n),\vartheta\big),
\eea
where the infimum is taken over all possible estimators of $\theta$ based on $X^n$.
\end{dfn}
\begin{prp}
The following are equivalent.
\begin{enumerate}
\item $T(X^n)$ is a sufficient statistic for $\theta$.
\item We have a Markov chain~ $\vartheta\ra T(X^n)\ra X^n$, i.e.,
    \bea
    p(\theta,x^n|T(x^n))=p(\theta|T(x^n))p(x^n|T(x^n)).\nn
    \eea
    Equivalently,
    \bea
    p(\theta|x^n,T(x^n))=p(\theta|T(x^n)).\nn
    \eea

\item The conditional distribution
\bea
h(x^n)=p_\theta(x^n|T(x^n))\triangleq p(x^n|T(x^n),\theta)\nn
\eea
is independent of $\theta$.

\item For any points $x^n,y^n,z^n\in\X^n$ satisfying the redundancy condition $T(x^n)=T(y^n)=T(z^n)$, the function
\bea
h(x^n,y^n,z^n)={p_\theta(y^n|T(x^n))\over p_\theta(z^n|T(x^n))}\nn
\eea
is independent of $\theta$, where
{\small
\be\ba
&p_\theta(y^n|T(x^n))=p_\theta(X^n=y^n|T(X^n)=x^n)=p(X^n=y^n|T(X^n)=x^n,\theta)\nn\\
&~~~~={p_\theta(y^n,T(x^n))\over p_\theta(T(x^n))}.\nn
\ea\ee
}
\item For all $x^n,y^n\in \X^n$,
\bea
T(x^n)=T(y^n)~~\Ra~~{\del\over\del\theta}{p(x^n|\theta)\over p(y^n|\theta)}=0.\nn
\eea
(Note that some continuity and differentiability are assumed in this case)
\end{enumerate}
\end{prp}
\begin{proof} The equivalences~ $2\iff 3\iff 4\iff 5$~ are straightforward.

The main challenge is with $1\iff 2$. For this case, we must choose the function $d$ in (\ref{sufficient-estimate}) in such a way that the following conditions hold.
\bit
\item[(a)] An estimator $T_1(X^n)$ is closer in randomness to $\vartheta$ than another estimator $T_2(X^n)$, i.e.,
\bea
E_{p(x,\theta)}d\big(T_1(X^n),\vartheta\big)\leq E_{p(x,\theta)}d\big(T_2(X^n),\vartheta\big),\nn
\eea
if and only if we have a Markov chain~ $\vartheta\ra T_1(X^n)\ra T_2(X^n)$.
\item[(b)] For every estimator $T'(X^n)$, we have a Markov chain $\vartheta\ra X^n\ra T'(X^n)$, i.e., $X^n$ is the best possible estimator.
\eit
It then follows that a statistic $T(X^n)$ satisfies a Markov chain $\vartheta\ra T(X^n)\ra X^n$ ~[~in addition to a Markov chain $\vartheta\ra X^n\ra T(X^n)$~]~ if and only if it satisfies (\ref{sufficient-estimate}).
\end{proof}

\begin{thm}[Factorization theorem] A statistic $T(X^n)$ is sufficient for $\theta$ if and only if there are functions $\al$, $\beta_\theta$ (with $\al$ independent of $\theta$) such that
\bea
\label{sufficient-statistic-formula}p(x^n|\theta)=\al(x^n)~\!\beta_\theta\big(T(x^n)\big).
\eea
\end{thm}
\begin{proof}~
\bit
\item($\La$): If $p(x^n|\theta)$ satisfies (\ref{sufficient-statistic-formula}), then it follow immediately from the definitions that $T(X^n)$ is a sufficient statistic for $\theta$.
\item($\Ra$): Assume $T(X^n)$ is a sufficient statistic for $\theta$. Then for all $x^n,y^n\in\X^n$,
\bea
T(x^n)=T(y^n)~~\Ra~~{\del\over\del\theta}{f(x^n|\theta)\over f(y^n|\theta)}=0.\nn
\eea
Since
\bea
&&{\del\over\del\theta}{p(x^n|\theta)\over p(y^n|\theta)}=0~~\iff~~{\del_\theta p(x^n|\theta)\over p(x^n|\theta)}={\del_\theta p(y^n|\theta)\over p(y^n|\theta)},\nn
\eea
we have
\bea
T(x^n)=T(y^n)~~\Ra~~\del_\theta\ln p(x^n|\theta)=\del_\theta\ln p(y^n|\theta),~~~~\txt{for all}~~x^n,y^n\in\X^n.\nn
\eea
This means $\del_\theta\ln p(x^n|\theta)=g(\theta,T(x^n))\eqv g_\theta(T(x^n))$ for some function $g_\theta$. Integration of this relation with respect to $\theta$ yields a formal solution of the form
\bea
p(x^n|\theta)=e^{\int^\theta d\theta'~g_{\theta'}(T(x^n))+K(x^n)}\eqv \al(x^n)~\beta_\theta(T(x^n)),\nn
\eea
which is (\ref{sufficient-statistic-formula}).
\eit
\end{proof}

\begin{dfn}[Necessary statistic]
$T(X^n)$ is a necessary statistic for $\theta$ if for all $ x^n,y^n\in\X^n$,
\bea
{\del\over\del\theta}{p(x^n|\theta)\over p(y^n|\theta)}=0~~\Ra~~T(x^n)=T(y^n).\nn
\eea
\end{dfn}

\begin{dfn}[Efficient statistic]
A statistic $T(X^n)$ is an efficient statistic for $\theta$ if it is both a necessary and a sufficient statistic for $\theta$, i.e., if for all $x^n,y^n\in\X^n$,
\bea
T(x^n)=T(y^n)~~\Longleftrightarrow~~{\del\over\del\theta}{p(x^n|\theta)\over p(y^n|\theta)}=0.\nn
\eea

Note that an efficient statistic is also called a minimal sufficient statistic.
\end{dfn}

\begin{thm}(Efficient statistic formula)
An efficient statistic $T(X^n)$ has the form
\bea
\label{efficient-statistic-formula}T(X^n)=h_\theta\big(\del_\theta \log p(X^n|\theta)\big),
\eea
where $h_\theta$ is any invertible function which is at least capable of removing all of the $\theta$ dependence from $\del_\theta \log p(X^n|\theta)$ as its argument.
\end{thm}
\begin{proof}
Recall that $T(X^n)$ is an efficient statistic iff for all $x^n,y^n\in\X^n$,
\bea
&&T(x^n)=T(y^n)~~\iff~~{\del\over\del\theta}{f(x^n|\theta)\over f(y^n|\theta)}=0.\nn
\eea
Since
\bea
&&{\del\over\del\theta}{p(x^n|\theta)\over p(y^n|\theta)}=0~~\iff~~{\del_\theta p(x^n|\theta)\over p(x^n|\theta)}={\del_\theta p(y^n|\theta)\over p(y^n|\theta)},\nn
\eea
we have
\bea
T(x^n)=T(y^n)~~\iff~~\del_\theta\ln p(x^n|\theta)=\del_\theta\ln p(y^n|\theta),~~~~\txt{for all}~~x^n,y^n\in\X^n.\nn
\eea
This means $\del_\theta\ln p(x^n|\theta)=g(\theta,T(x^n))\eqv g_\theta(T(x^n))$, where $g_\theta$ is an invertible function such that the quantity $g_\theta^{-1}(\del_\theta \log p(x^n|\theta))$ is independent of $\theta$. Setting $h_\theta=g_\theta^{-1}$, we obtain the formula (\ref{efficient-statistic-formula}).
\end{proof}

\section{Estimation II: Set estimators and hypothesis testing}\label{estimation-II}
Recall that a point estimator is a statistic of the form $T(x^n)\in\Theta$, $x^n\in\X^n$. In general, estimators of the form $T(x^n)\subset\Theta$, $x^n\in\X^n$, are more practical. These are called \emph{set estimators} (or \emph{confidence sets}).
\begin{dfn}[Set estimator]
Any statistic $T(X^n)$ for the purpose of inferring a reasonably small subset of $\Theta$ containing the true value of $\theta$ is called a \emph{set estimator} of $\theta$.
\end{dfn}

With a point estimator, one reports the result of estimation based on $x^n\in\Real^n$ by saying ``given $\theta\in\Theta$, we have $\theta=T(x^n)$ with probability $P(T(X^n)=\theta)$''. With a set estimator, one similarly reports the result of estimation based on $x^n\in\Real^n$ by saying ``given $\theta\in\Theta$, we have $\theta\in T(x^n)$ with probability $P\big(\theta\in T(X^n)\big)$''.

Note that a point estimator is a special case of a set estimator. This implies, in particular, that the notion of sufficiency discussed earlier for point estimators can, at least formally, be extended to set estimators. Moreover, concepts we will introduce for interval estimators apply to point estimators as well.

\begin{dfn}[Degree of confidence, Percentage of confidence]
The degree of confidence (or confidence coefficient) of a set estimator $T(X^n)$ is~ $c=\min\limits_{\theta\in\Theta}P\big(\theta\in T(X^n)\big)$. The percentage of confidence of $T(X^n)$ is $100c~\!\%$, and we say $T(X^n)$ is a $100c~\!\%$ confidence set for $\theta$.
\end{dfn}

A method of statistical estimation that involves set estimators in a natural way is called \emph{hypothesis testing}. Hypothesis testing, like point estimation, is a method of inference (of a parameter $\theta$) based on observations. In the discussion that follows, it is assumed we have observations $\{x^n\in\X^n\}$ based on at least one \emph{known} family of probability distributions $\{p(x^n|\theta):~\theta\in\Theta\}$.

\begin{dfn}[Hypothesis, Simple hypothesis]
A hypothesis, denoted by $H$, is a statement about the inference parameter $\theta$ (i.e., a parameter whose value we wish to infer), which is in the form of a constraint or restriction $R_H$ on the value of $\theta$. By convention, we write
\bea
H:~R_H,\nn
\eea
which reads ``$H$ stands for, or represents, the value restriction $R_H$ on $\theta$''.

A simple hypothesis, $H$, is a statement of the form
\bea
H:\theta=\theta_0,\nn
\eea
for some fixed value $\theta_0\in\Theta$.
\end{dfn}
We will say that two statements are \emph{mutually exclusive} if they cannot be both valid simultaneously.

\begin{dfn}[Hypothesis test]
Given a set of (mutually exclusive) hypotheses on $\theta$, one and only one of which is valid, a hypothesis test is a method for deciding (based on observations $x^n\in\X^n$) the hypothesis that is most likely to be the valid one.
\end{dfn}

\begin{rmks*}~
\bit
\item[I.] Observe that by definition, the hypothesis test is determined by statistics which are real valued functions of the observations. Consequently, every hypothesis test can be specified as a solution of some optimization problem, as discussed in Chapter \ref{optimization}. Moreover, in practice, the decision involved in the test is of course made so as to meet a given objective, which is often the optimization (minimization or maximization) of some information measure.
\item[II.] Note that the above discussion indicates that the notion of an optimal hypothesis test (Definition \ref{optimal-objective-test}) is a natural generalization of the notion of a sufficient statistic (Definition \ref{sufficient-statistic-dfn}).
\eit
\end{rmks*}

\begin{dfn}[Objective hypothesis test, Optimal hypothesis test]\label{optimal-objective-test}
A hypothesis test is objective if it is specified as a solution of some optimization problem. An optimal hypothesis test is an objective hypothesis test for which the decision on the valid hypothesis is optimal with respect to the underlying objective of the test.
\end{dfn}

The following is a preview of some basic points which are relevant in \emph{statistical decision theory} (the subject of Chapter \ref{statistical-decision-theory}) and \emph{optimal hypothesis testing} (the main subject of Chapter \ref{optimal-detection}).

\begin{rmk*}
Although the eventual or end objective in a hypothesis test is to \emph{decide} the valid hypothesis among a set of say $M$ hypotheses, it is often more useful to consider intermediate decision operations that can take values in a set whose cardinality is different from $M$. This is important in distributed detection where some local sensors may only need to forward quantized versions of their observations to a fusion center which actually decides the true hypothesis based on the quantized observations. In general therefore, the intermediate decision output from a local sensor may not have the same alphabet as the hypothesis.
\end{rmk*}
The above remark is also emphasized in Chapters \ref{statistical-decision-theory} and \ref{optimal-detection}.

\begin{dfn}[Decision rule, Decision function]\label{decision-rule-dfn}
A decision rule is a point estimator of the form
\bea
\label{decision-rule}\gamma:x^n\in\X^n\mapsto u=\gamma(x^n)\in\{0,1,\cdots,N-1\},
\eea
i.e., an estimator that takes on a discrete set of values.

In objective hypothesis testing, a decision function is an information measure that depends on both the decision rule and the hypothesis.
\end{dfn}
We can, for example, consider a decision function of the form
\bea
\label{decision-function}S=E_{p(u,h)}d(U,H),
\eea
where $H$ denotes the hypothesis and $u$ the decision. The primary objective of a hypothesis test is often to select a decision rule that optimizes an underlying decision function such as the function $S$ in (\ref{decision-function}).

\begin{dfn}[Binary hypothesis test, Null hypothesis, Alternative hypothesis]
A hypothesis test involving two complementary hypotheses is the simplest type of hypothesis test, and is called a binary hypothesis test. One of the hypotheses is denoted by ~$H_0:~\theta\in\Theta_0$ and called the \emph{null hypothesis}, while the other is denoted by ~$H_1:~\theta\in\Theta_1$ and called the \emph{alternative hypothesis}, where~ $\Theta_1\cup\Theta_2=\Theta$.
\end{dfn}

\begin{rmks*}[Computational Setup and Results]~
\begin{enumerate}
\item \emph{\textbf{Indicator variables}}: If we let $s=s(\theta)=I_{\Theta_1}(\theta)=\left\{
                                  \begin{array}{ll}
                                    0, & \theta\in\Theta_0 \\
                                    1, & \theta\in\Theta_1
                                  \end{array}
                                \right\}
$, then the binary hypothesis test becomes a problem of estimating the value of the binary variable $s$. The hypotheses become $H_0:s=0$, $H_1:s=1$. The family of distributions $\{p(x^n|s):s=0,1\}$ associated with $s$, i.e., the distribution of the observation conditioned on $s$, is given by
{\small
\bea
p(x^n|s=i)={p(x^n,s=i)\over p(s=i)}={p(x^n,\theta\in\Theta_i)\over p(\theta\in\Theta_i)}={\sum_{\theta\in\Theta_i}p(x^n|\theta)p(\theta)\over \sum_{\theta\in\Theta_i}p(\theta)},~~~~~i=0,1,\nn
\eea
}
where $p(\theta)$ is a \emph{prior probability} distribution on $\Theta$. Note that if $p(\theta)$ is unknown, then it must be treated as an optimization variable in the objective function of the test.

\item For notational convenience, we will often write $p_i(x^n)=p(x^n|s=i)$. Also, the hypotheses
\bea
H_0:s=0,~~~~H_1:s=1\nn
\eea
are equivalently expressed in terms of the conditional distribution of the observations as
\bea
H_0:x^n\sim p(x^n|s=0),~~~~H_1:x^n\sim p(x^n|s=1),\nn
\eea
where $x^n\sim p(x^n|s=i)$ means ``\emph{$x^n$ is distributed according to $p(x^n|s=i)$}''.

\item \emph{\textbf{Neyman-Pearson lemma}}: Now suppose the binary hypothesis test satisfies the following two conditions.
\bit
\item[(a)] The decision rule (\ref{decision-rule}) is binary, i.e., $M=2$, where the decision $u=0$ is interpreted as acceptance of $H_0$ (or rejection of $H_1$) while $u=1$ is acceptance of $H_1$ (or rejection of $H_0$).
\item[(b)] The test's underlying objective function, such as (\ref{decision-function}), to be maximized is a convex function of the conditional probabilities $p(u|x^n)$, viewed as the main optimization variables.
\item[(c)] The observations $X^n$ are continuous variables and the distribution $p_i(x^n)$ is continuous.
 \eit
Then it can be shown (see Proposition \ref{region-proposition}) that under these conditions, the \emph{optimal decision rule} for the binary hypothesis test takes the form
\bea
p_{\txt{opt}}(\theta\in\Theta_i|x^n)=p_{\txt{opt}}(u=i|x^n)=I_{R_{u=i}}(x^n),~~~~i=0,1,
\eea
where the \emph{decision regions} $R_{u=i}$ are given by
\bea
R_{u=1}=\left\{x^n\in\X^n:~p_1(x^n)/p_0(x^n)>\ld\right\},~~~~R_{u=0}\cup R_{u=1}=\X^n.\nn
\eea
Here, $\ld\in\Real$ is a threshold parameter. A formal statement of this particular result is well known as the \emph{Neyman-Pearson lemma}.

\item \emph{\textbf{A set estimator of $\theta$}}: Let us fix $\vep>0$, and let $\theta_0\in\Theta$. Denote by the pair $(\vep,\theta_0)$ the binary hypothesis test with
\bea
H_0:\theta\in B(\theta_0,\vep),~~~~H_1:\theta\not\in B(\theta_0,\vep),\nn
 \eea
where $B(\theta_0,\vep)\subset\Theta$ is the open ball of radius $\vep$ centered at $\theta_0$. Let the acceptance region for $H_0$ be
\bea
R_{(\vep,\theta_0)}=R_{u=0}^{(\vep,\theta_0)}=\left\{x^n\in\X^n:~{p_1^{(\vep,\theta_0)}(x^n)/p_0^{(\vep,\theta_0)}(x^n)}<\ld_{(\vep,\theta_0)}\right\},\nn
\eea
where
{\small
\bea
p_0^{(\vep,\theta_0)}(x^n)={\sum_{\theta\in B(\theta_0,\vep)}p(x^n|\theta)p(\theta)\over \sum_{\theta\in B(\theta_0,\vep)}p(\theta)},~~~~p_1^{(\vep,\theta_0)}(x^n)={\sum_{\theta\not\in B(\theta_0,\vep)}p(x^n|\theta)p(\theta)\over \sum_{\theta\not\in B(\theta_0,\vep)}p(\theta)}.\nn
\eea
}
Then a natural set estimator for $\theta$ is given by
\bea
T^{(\vep)}(x^n)=\left\{\theta_0\in\Theta:x^n\in R_{(\vep,\theta_0)}\right\}.\nn
\eea
Given $\theta\in\Theta$, we have
\bea
P\left(\theta\in T^{(\vep)}(X^n)\right)=P\left(X^n\in R_{(\vep,\theta)}\right)=P\left(R_{(\vep,\theta)}\right).\nn
\eea
Thus, the confidence coefficient of $T^{(\vep)}(x^n)$ satisfies
\bea
c^{(\vep)}~=~\min_{\theta\in\Theta}~P\left(\theta\in T^{(\vep)}(X^n)\right)~=~\min_{\theta\in\Theta}~P\left(R_{(\vep,\theta)}\right).\nn
\eea

\item \emph{\textbf{Hypothesis test sequences}}: Although the result of a single binary hypothesis test does not necessarily yield a direct estimate for the true value of $\theta$ (except when $\Theta$ is a binary set), it does reduce the \emph{search space} for the true value of $\theta$ from $\Theta$ to $\Theta_0\subsetneq\Theta$ or $\Theta_1\subsetneq\Theta$. Thus, if we consider a sequence of consecutive binary hypothesis tests $\tau_1,\tau_2,\tau_3,\cdots$ and let $\Theta^{(\tau_k)}$ denote the search space in the $k$th test $\tau_k$, then we have
\bea
\Theta=\Theta^{(\tau_1)}\subsetneq\Theta^{(\tau_2)}\subsetneq\Theta^{(\tau_3)}\subsetneq\cdots
\eea
That is, the result of a sufficiently long sequence of binary hypothesis tests will yield a reasonable estimate for the true value of $\theta$. Of course if we consider a sequence of tests with more than two hypotheses, then the length $N'$ of a sequence of such tests required to reach a certain desired level of accuracy will be smaller than the length $N$ of a sequence of binary hypothesis tests required to reach the same level of accuracy.

\item \emph{\textbf{Multiple hypothesis tests}}: The description of binary hypothesis tests given above extends in a straightforward way to \emph{multiple hypothesis tests}. The basic idea remains the same: To split up $\Theta$ into $N$ disjoint subsets $\Theta_0,\Theta_1,...,\Theta_{N-1}$, consider hypotheses $H_i:\theta\in\Theta_i$, and then find the conditional probabilities $p(\theta\in \Theta_i|x^n)$ that best suite a given objective.

Once again, the multiple hypothesis test for $\theta$ with hypotheses $H_i:\theta\in\Theta_i$ is equivalent to a multiple hypothesis test for a discrete indicator variable $s$ with hypotheses $H_i:s_i=\mu_i$, where ~$s=s(\theta)=\sum_{i=0}^{N-1}\mu_iI_{\Theta_i}(\theta)\in\{\mu_0,\mu_1,...,\mu_{N-1}\}$, and the distribution of $s$ is computed as
\bea
p(x^n|s=\mu_i)={p(x^n,s=\mu_i)\over p(s=\mu_i)}={p(x^n,\theta\in\Theta_i)\over p(\theta\in\Theta_i)}={\sum_{\theta\in\Theta_i}p(x^n|\theta)p(\theta)\over \sum_{\theta\in\Theta_i}p(\theta)},\nn
\eea
for each $i\in\{0,1,...,N-1\}$.

\item Note that a multiple hypothesis test can be approximated by a number of binary hypothesis tests. Also, a binary hypothesis test can be approximated by a \emph{union-intersection} (or \emph{intersection-union}) test, which is a combination of a number of binary hypothesis tests.
\end{enumerate}
\end{rmks*}

\subsection*{Generalized likelihood ratio tests}
Consider a general binary hypothesis test
\bea
\label{GLRT-eq1}H_0:\theta\in\Theta_0,~~~~H_1:\theta\in\Theta_1,~~~~\Theta_0\cup\Theta_1=\Theta.
\eea
For a fixed $\theta_0\in \Theta_0$ and a fixed $\theta_1\in\Theta_1$, we have the simple test
\bea
H^{(\theta_0,\theta_1)}_0:\theta=\theta_0,~~~~H^{(\theta_0,\theta_1)}_1:\theta=\theta_1.\nn
\eea
Thus, we have a family of simple tests $\left\{\left(H^{(\theta_0,\theta_1)}_0,H^{(\theta_0,\theta_1)}_1\right):\theta_0\in\Theta_0,\theta_1\in\Theta_1\right\}$.
For each pair $(\theta_0,\theta_1)\in\Theta_0\times\Theta_1$, let the decision region for the test $\left(H^{(\theta_0,\theta_1)}_0,H^{(\theta_0,\theta_1)}_1\right)$ be
\bea
R^{(\theta_0,\theta_1)}_{u=1}=\{x^n\in\X^n:p(x^n|\theta_1)/p(x^n|\theta_0)>\ld_{\theta_0,\theta_1}\}.\nn
\eea

If we choose to accept a data point $x^n$ under $H_1$ in (\ref{GLRT-eq1}) whenever it falls in any one of the regions $R^{(\theta_0,\theta_1)}_{u=1}$, then the test (\ref{GLRT-eq1}) has a (suboptimal) decision rule given by the decision region
{\small
\bea
&&R_{u=1}=\bigcup_{(\theta_0,\theta_1)\in\Theta_0\times\Theta_1}R^{(\theta_0,\theta_1)}_{u=1}=\left\{x^n\in\X^n:\sup_{\theta_0,\theta_1}{p(x^n|\theta_1)/\ld_{\theta_0,\theta_1}\over p(x^n|\theta_0)}>1\right\}\nn\\
\label{GLRT-eq2}&&~~~~\subset \left\{x^n\in\X^n:{\sup_{\theta_1}p(x^n|\theta_1)\over\sup_{\theta_0}p(x^n|\theta_0)}>\sup_{\theta_0,\theta_1}\ld_{\theta_0,\theta_1}\right\}=\td{R}_{u=1}.
\eea
}

Tests with a decision region of the form $\td{R}_{u=1}$ in (\ref{GLRT-eq2}) are called generalized likelihood ratio tests (GLRT's). Although such tests are clearly suboptimal in general, a remark we made earlier says that if such a test is repeated a sufficiently large number of times, it can yield very good results that may even be judged to be \emph{asymptotically optimal} depending on the underlying objective of the test.

\subsection*{Union-Intersection (or Intersection-Union) tests}
There are situations where a binary test $(H_0,H_1)$ is seen to be composed of elements of a family of binary hypothesis tests ~$\{(H_{0\al},H_{1\al}):~\al\in\A\},$ where $\A$ is an index set.

Consider a test with hypotheses~ $H_0:\theta\in \Theta_0,~~H_1:\theta\in \Theta_0^c$. Suppose that $\Theta_0=\bigcap_{\al\in \A}\Theta_{0\al}$. Then
\bea
&&H_0:~\theta\in \Theta_0=\bigcap_{\al\in \A}\Theta_{0\al},~~~~H_1:~\theta\in \Theta_0^c= \bigcup_{\al\in \A}\Theta_{0\al}^c.\nn
\eea
Notice that the test involves separate tests of the form
\bea
\label{UI-hypotheses}H_{0\al}:~\theta\in \Theta_{0\al},~~~~H_{1\al}:~\theta\in \Theta_{0\al}^c,~~~~\al\in\A.
\eea
Thus, if $R^{(\al)}_{u=1}=\left\{x^n:L_\al(x^n)>\ld_\al\right\}$ is the decision region of the test $(H_{0\al},H_{1\al})$ for each $\al\in\A$, then a (suboptimal) decision region for the test $(H_0,H_1)$ is given by
{\footnotesize
\bea
R_{u=1}=\bigcup_{\al\in\A}R^{(\al)}_{u=1}=\bigcup_{\al\in\A}\left\{x^n\in\X^n:L_\al(x^n)>\ld_\al\right\}=\left\{x^n\in\X^n:\sup_{\al\in\A}L_\al(x^n)/\ld_\al>1\right\}.\nn
\eea
}

Similarly, if we consider a test with hypotheses~ $H_0:\theta\in \Theta_0,~~H_1:\theta\in \Theta_0^c$, and suppose that $\Theta_0=\bigcup_{\al\in \A}\Theta_{0\al}$, then
\bea
&&H_0:~\theta\in \Theta_0=\bigcup_{\al\in \A}\Theta_{0\al},~~H_1:~\theta\in \Theta_0^c= \bigcap_{\al\in \A}\Theta_{0\al}^c.\nn
\eea
We again observe that the test involves separate tests of the form (\ref{UI-hypotheses}). Thus, if
\bea
R^{(\al)}_{u=1}=\left\{x^n\in\X^n:L_\al(x^n)>\ld_\al\right\}\nn
\eea
is the decision region of the test $(H_{0\al},H_{1\al})$, then a (suboptimal) decision region for the test $(H_0,H_1)$ is
\bea
R_{u=1}=\bigcap_{\al\in\A}R^{(\al)}_{u=1}=\left\{x^n\in\X^n:\inf_{\al\in\A}L_\al(x^n)/\ld_\al>1\right\}.\nn
\eea

\part{Detection}\label{part-II} 
\chapter{Statistical Decision Theory}\label{statistical-decision-theory}
\section{Introduction}
This chapter is intended to provide motivation for, as well as improve our understanding of the practical significance of, the analysis to be presented in the next chapter. Since it is a special introduction to Chapter \ref{optimal-detection}, we will be brief and concerned mainly with nontechnical aspects of the basic structure of a simple decision process. The question of how we can actually make certain types of decisions in practice is the subject of Chapter \ref{optimal-detection}. The discussion here will illustrate the usefulness of statistical hypothesis testing in general.

A concise introduction to statistical decision theory is found in \cite{parmigiani-inoue}, and nontechnical introductions to the same are found in \cite{north,hansson}. Other useful references include \cite{mazumder,berger,wald}.

The main motivation for, as well as the general definition of, decision theory are contained in the following. Making decision under uncertainty is a task that increases in difficulty as society grows and increases in complexity. Decision theory provides a general structure or framework aimed at simplifying the decision making task.

Statistical decision theory is a method that uses observational data to enhance the decision making process  when uncertainty is involved. It is a very interdisciplinary subject with varying perspectives, approaches, and applications. Nevertheless, the basic decision structure is the same in all cases.

The items we will discuss include ``basic elements of a simple decision process'', ``classification of simple decision processes'', ``extensions to more complex decision processes'', and ``some applications of statistical decision theory''.


\section{Basic elements of a simple decision process}
In a simple decision process, there are three basic elements, the \emph{decision}, the (often unknown) \emph{circumstance}, and the \emph{consequence}.

For concreteness, we will work directly with an example. An example that easily illustrates the statistical aspect of a simple decision process is that of deciding whether an accused person is \emph{guilty}, \emph{partly guilty}, or \emph{not guilty} of a crime.

\begin{figure}
\centering
\scalebox{0.7}{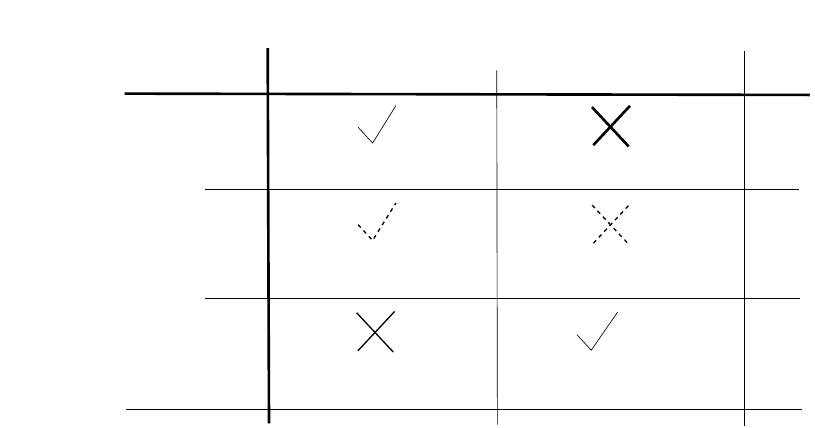}
\caption{Table of consequences and associated costs}\label{dt}
\end{figure}

When given this decision task, we have the following three basic components (See Figure \ref{dt}).
\begin{enumerate}
\item \emph{The decision}: This is one of a number of alternative actions to choose from. For our example, these actions or decision values are ``The accused is guilty'', ``The accused is partly guilty'', ``The accused is not guilty''.
\item \emph{The circumstance}: This is the existing one among a number of possible natural conditions/states on which the objective or appropriate decision value depends. The circumstance is often uncertain, i.e., not completely known to the decision maker. For our example, these conditions are ``The accused committed the crime'', ``The accused did not commit the crime''.
\item \emph{The consequence}: This is one of a number of (anticipated) decision-circumstance outcomes, to each of which a cost of some sort is assigned. For our example, these outcomes are ``Punish a guilty person'', ``Punish an innocent person'', ''Warn a guilty person'', ``Warn an innocent person'', ``Free a guilty person'', ``Free an innocent person''.
\end{enumerate}

We will refer to the above three elements as \emph{internal elements} of the simple decision process. In order to classify simple decision types, a few more conceptual elements of the simple decision process are necessary.

\section{Classification of simple decision processes}
Our classification of simple decision processes is based on a number of \emph{external elements} -- namely -- preference, prior information, and data.
\begin{enumerate}
\item \emph{Preference}: The decision process is objective if it is based on a function of the internal elements (i.e., the decision, circumstance, and consequence) that the decision seeks to optimize, otherwise, the decision process is subjective.
\item \emph{Prior information}: The decision process is Bayesian if it uses prior knowledge (i.e., past experience with the circumstances), and it is frequentist or classical otherwise.
\item \emph{Data}:
    The decision process is statistical if it uses data, which consists of observations from experiment on any systems that are directly or indirectly affected by the existing circumstance. Otherwise, the decision process is non-statistical.
\end{enumerate}

In a statistical decision process, data can reduce uncertainty of the circumstance, i.e., it can partly reveal the circumstance. Consequently, data can improve decision quality. This is the main motivation for considering a statistical decision process.

We will mainly be concerned with objective statistical decision processes, which involve the following. In order to reduce uncertainty in the circumstance, we carry out a statistical investigation or experiment by collecting data from systems whose behaviors depend on the circumstance. The data is then used to improve decision making. The decision as a function of data is called a \emph{decision strategy}. Our decision preference is represented by a function that depends on the decision strategy, the circumstances, and any costs associated with the consequences. Such a function is called a \emph{decision function}, \cite{wald}. The \emph{decision objective} is to select an \emph{optimal decision strategy}, which is any decision strategy that minimizes the decision function.

Based on the above discussion, a convenient mathematical tool for handling statistical decision problems is hypothesis testing, which is introduced in Section \ref{estimation-II}. Based on the discussion above and that in Section \ref{estimation-II}, a statistical decision process is, equivalently, a hypothesis test. We will see in Chapter \ref{optimal-detection} that the problem of detecting a signal embedded in corrupted measurements is a statistical decision problem, which can therefore be equivalently expressed as a hypothesis testing problem.

\section{Extensions to decision processes in practice}
In practice a typical decision process can contain several simple decisions, and may also involve several decision makers. For our purpose, these more complex decision processes can take one of the following labels.
\begin{enumerate}
\item \emph{Sequential}:  A decision process is sequential if it consists of several consecutive simple decisions.
\item \emph{Distributed}:  A decision process is distributed (or decentralized) if it involves several decision makers.
\item \emph{Hybrid}: A decision process is hybrid if it is both sequential and distributed.
\end{enumerate}

In Chapters \ref{optimal-detection}, \ref{interactive-detection}, \ref{communication-direction}, \ref{acyclic-graph-detection}, we will encounter applications involving decision processes of the above types. This will be in the context of signal detection.

\section{Some applications of statistical decision theory}
In each of the cases below, the role of statistical decision theory becomes apparent when one attempts to answer the posed questions.
\begin{enumerate}
\item \emph{Signal detection}:  Is there a signal or no signal? How do we statistically extract it from noisy observations?
\item \emph{Marketing}:  Is there demand for a given product? How do we statistically determine it?
\item \emph{Management}:  Which task or who needs a resource? How can we be statistically sure?
\item \emph{Forecasting/Prediction}:  What is going to happen? How can we find out statistically?
\end{enumerate}
Discussions on various applications of statistical decision theory can be found with the help of \cite{parmigiani-inoue,north,hansson,mazumder,berger}.

\chapter{Optimal Signal Detection}\label{optimal-detection}
\section{Introduction}\label{optimal-detection-intro}
In this chapter, which is a synthesis of preliminary results discussed in some detail in Chapters \ref{optimization}, \ref{info-theory}, we study optimization of convex functions of \emph{decision rules} or of \emph{decision probability functions} for the purpose of distributed signal detection. Here, we should be mindful of the fact that the objective functions we shall deal with in real applications are not merely functions on $\Real^n$ as discussed in Chapter \ref{optimization}, but functions on function spaces, i.e., functions whose arguments are themselves functions. Such functions are also called functionals.

Routine problems considered in convex optimization mostly involve either the minimization of a convex function or the maximization of a concave function. However, problems that require maximization of a convex function, or minimization of a concave function, also arise. For example, in certain distributed detection problems the Kullback-Leibler distance is a performance metric that is convex in the variables Pr(decision$|$data), which are the pmf's of local sensor decisions conditioned on data. The optimal decision rules are those that maximize this function.

The goal is to first provide differential relations that serve as necessary and sufficient conditions for the maximum of any detection performance metric that is a differentiable monotonic convex function of Pr(decision$|$data). Next, we then express optimal local sensor decision rules in terms of these differential relations.

Our approach is based on the following. Consider a real-valued differentiable convex function, defined on the $n$-dimensional real space, which we wish to maximize over a convex subset of the space (See Chapter \ref{optimization}). By carefully studying the geometry of the graph of the function, we can derive \emph{optimality conditions} (i.e., necessary and sufficient conditions for optimality) in the form of differential inequalities involving the derivative of the function at an optimal point (See Theorem \ref{convex-max-theorem}). Once this has been done, the problem can then be solved with the help of standard algorithms for solving differential inequalities.

In order to present local sensor decision rules in terms of the optimality conditions, we will first restate the detection problem as a general optimization problem in which the optimization variables in the objective function are Pr(decision$|$data), i.e., the pdf's of local decision rules conditioned on data. Optimal decision regions will then consist of those data points that satisfy the optimality conditions. The above procedure is presented in Section \ref{opt-hyp-testing}.


\subsection*{Notation}
For convenience, we will adopt the following conventions from now on. Lower case letters such as $x,y,...$ denote both random variables and their values. The symbol $\sum\limits_x$ denotes summation when $x$ is a discrete variable, or integration when $x$ is a continuous variable. The expression $\delta_{xx'}$ denotes the Kronecker delta when $x$ and $x'$ are discrete, or the Dirac delta when $x$ and $x'$ are continuous. Therefore, wherever the identity
\bea
{\del f(x)/\del f(x')}=\delta_{xx'}\nn
\eea
appears, the variables $x$, $x'$ can be tuples $(x_1,...,x_N)$, $(x'_1,...,x'_N)$ of several discrete or continuous variables, for which we naturally define $\delta_{xx'}=\delta_{(x_1,...,x_N)(x'_1,...,x'_N)}$ by
\bea
 \delta_{(x_1,x_2,\cdots
 ,x_N)(x'_1,x'_2,\cdots,x'_N)} = \delta_{x_1x'_1} \delta_{x_2x'_2}\cdots \delta_{x_Nx'_N}.\nn
\eea
In addition to the above conventions, we will denote sensors by upper case letters and their observations by the corresponding lower case letters. For example, $x$ will denote the observation of sensor $X$, $y$ the observation of sensor $Y$, and so on.

\section{The detection problem}\label{detection-problem}
According to the preliminary discussion in Chapter \ref{info-theory}, the estimation problem refers to any situation in which one is faced with the task of determining the state of a system based on a given data sample, i.e., a set of experimental observations on the system.

\begin{figure}[H] 
\centering
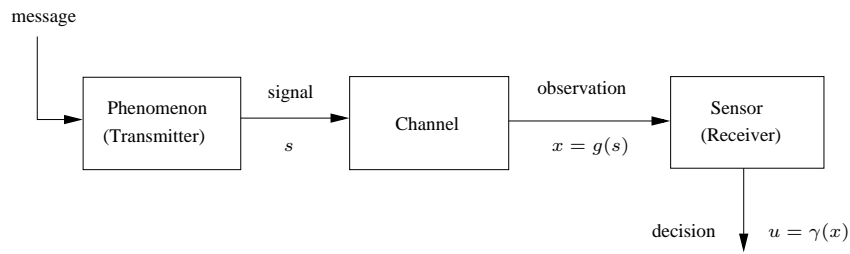
  \caption{Communication system}\label{communication-system}
\end{figure}

In the detection problem in particular, we are faced with the task of extracting the value of a discrete signal
\bea
s\in\S=\{\mu_0,\mu_1,\cdots,\mu_{M-1}\},\nn
\eea
(representing the unknown state of some system) from a given data sample $x^n=\{x_i\in\X:i=1,2,\cdots,n\}$. The system here refers to one or more components, such as the encoder, the channel, or the decoder, of a typical transmitting system. Each observation $x$ in the sample $x^n$ is typically viewed as a (known) \emph{random function} of the signal. That is, the observation can be expressed as
\bea
\label{transmit-filter}x=g(s),
\eea
where the random function $g:s\in\S\mapsto x\in\X$ represents possible effects of known system properties on the signal. Such a function $g$ is often called a \emph{filter}, owing to its role in the signal extraction process.

We will not be dealing with the details of the encoding, channeling, and decoding rules whose composition determines $g$ in general. Instead, for the most part in applications, we will consider the simplest case in which we assume that
\bea
\label{additive-filter}x=s+b,
\eea
where $b$ is white noise, i.e., a Gaussian-distributed random variable, and the signal $s$ may be random as well. This will be sufficient for our main application interests.

Given the observation $x$ as in (\ref{additive-filter}), we wish to know how much noise $b$ there is in $x$ so that we can remove it and be left with $s$, i.e., we wish to know the value of $b=x-s$. However, we do not know the actual value of $s$. Since the value of $b$ is known whenever that of $s$ is given, we are therefore faced mainly with the problem of statistically deciding the true value of $s$ from the given observation $x$. This is precisely a hypothesis testing problem involving simple hypotheses
\bea
H_i:s=\mu_i,~~~~i=0,1,\cdots,M-1,\nn
\eea
or equivalently,
\bea
\label{multi-hypothesis-test}H_i: x\sim p_i(x)=p(x|s=\mu_i),~~~~i=0,1,\cdots,M-1.
\eea
Note that the distribution $p_i(x)$ is known since we have already assumed that the distribution of $b$ in (\ref{additive-filter}), or of $g(\cdot)$ in (\ref{transmit-filter}), is known. With a slight abuse of notation, $p(x|s=\mu_i)$ will also be written as $p(x|H_i)$. Therefore the expressions $p_i(x)$, $p(x|H_i)$, $p(x|s=\mu_i)$ will all mean the same thing. Also, for notational convenience, we will not distinguish between a single observation $x\in x^n$ and the whole sample $x^n$, i.e., $x$ will stand for a single sample $x\in x^n$ as well as for the whole sample $x^n$ sometimes.

In the detection problem as described above, it is sufficient to consider decision rules $\gamma(x)$ that take the same number of values as the number of hypotheses, so that $\gamma(x)=i\in\{0,1,\cdots,M-1\}$ stands for acceptance of the $j$th hypothesis. However, in the more general context of statistical decision theory, discussed in Chapter \ref{statistical-decision-theory}, the number of decision values can be different from the number of circumstance (or hypothesis) values. Moreover, in applications that require data quantization or compression in general, we may sometimes wish to first convert a relatively large set of observations into a smaller data set that represents the original set of observations as best as possible for the purpose at hand.

The above comment is especially true in distributed detection with fusion, where peripheral sensors generally compress their observations and pass them onto a fusing sensor that makes a decision based on the compressed data. In such a setting, even when all sensors are using the same set of hypotheses, the output of a peripheral sensor may, or may not, be of the same alphabet type as the output of the fusing sensor.

We will therefore take into account the above situations in our analysis of optimal hypothesis testing. In particular, the set of decision values will have a cardinality which is different from that of the set of hypothesis values.


\section{Optimal hypothesis testing}\label{opt-hyp-testing} 
This section contains an extension of preliminary work found in \cite{akofor-chen-icassp,akofor-chen-it-paper}.

Recall that hypothesis testing was introduced in Section \ref{estimation-II}, where optimal hypothesis testing was identified as a generalization of the notion of a sufficient statistic. We will now discuss optimal hypothesis testing and derive optimality conditions that are valid for all convex decision functions.

Consider a test of the $M$ simple hypotheses in (\ref{multi-hypothesis-test}). Under $H_i$, we will denote the probability of a data set $A\subset \X$ by
\bea
P_i(A)=\sum_{x\in A}p_i(x).\nn
\eea
The observation space $\mathcal{X}$ may be of arbitrary dimension. As in Definition \ref{decision-rule-dfn}, a \emph{decision rule} is a mapping defined as
\bea
\gamma:~x~\mapsto ~j~\in~\{0,1,...,N-1\},\nn
\eea
where $\gamma$ is a deterministic function. We refer to the assignment ~$\gamma(x)=j$~ as a \emph{decision} based on the observation $x$. In problems where the decision output has the same alphabet as the underlying hypothesis, i.e., $N=M$, the decision $\gamma(x)=j$ may be interpreted as acceptance of the $j$th hypothesis $H_j$.

The desired decision rule $\gamma$ so defined is deterministic in the sense that $p(\gamma(x)=j|x)=\delta_{j,\gamma(x)}$, where $\delta_{a,b}\triangleq 1$ if $a=b$ and $0$ otherwise. Therefore, once $x$ is given $\gamma(x)$ is precisely known. As the optimum decision rule is not necessarily deterministic, we consider the larger set containing all deterministic and nondeterministic decision rules. Let us write the generic decision rule as
\bea
\Gamma:~x~ \mapsto ~j~\in~\{0,1,...,N-1\},
\eea
and let $u=\Gamma(x)$.
Then $\Gamma=\gamma$ denotes a deterministic choice of the decision rule. Recall as in \cite{tsitsiklis} that the set of $\Gamma$ is the convex hull of the set of $\gamma$. Therefore
\bea
p(\Gamma(x)=j)=\sum_{g}p(g)~p\big(\gamma_g(x)=j\big)
 \eea
where $g$ is a random variable with probability mass, or density, function $p(g)$ and is independent of $x$. The decision optimization process simply picks the appropriate $p(g)$, and hence the desired $p(\Gamma(x)=j)$.

As $u$ is a random variable, making an optimal guess $u=j$ is equivalent to choosing $p(u=j|x)$ such that some \emph{objective function}, which we denote by $S$, is optimized. Here $S$ is a function of $p(u=j|x)$ for all $j=0,1,...,N-1$ and for all data points $x\in\X$. Note that we also refer to $S$ as the \emph{decision function} (See Definition \ref{decision-rule-dfn}).

In general, $0\leq p(u=j|x)\leq 1$, for each $j\in\{0,1,...,N-1\}$. A \emph{deterministic decision rule} is one for which $p(u=j|x)$ takes on only the boundary values $0$ and $1$. For such cases, we will see in Proposition \ref{region-proposition} that the decision rule can be expressed as a partition of the data space into disjoint decision regions, i.e.,
\be\ba
\label{optimal1} p(u=j|x)~\triangleq~p(\gamma(x)=j|x)= I_{R_{u=j}}(x),
\ea\ee
where $I_{R_{u=j}}(x)=\delta_{j,\gamma(x)}$ is the indicator function of the set $R_{u=j}=\{x:~\gamma(x)=j\}$, which we call the \emph{decision region} for the $j$th decision, and
\be
R_{u=j}\cap R_{u=j'}=\emptyset,~~~~\txt{if}~~~~j\neq j'.\nn
\ee
When the number of decision values $N$ equals the number of hypothesis values $M$, we may also refer to $R_{u=j}$ as the \emph{acceptance region} for the $j$th hypothesis.

In the following proposition, we establish the general structure of the optimal decision rule for an important class of decision problems namely, those with monotonic convex objective functions. In other words, this proposition solves all monotonic convex versions of the optimization problem
\be\ba
\label{functional-max-problem}&\txt{maximize}~~ S\big(p(u|x)\big)\\
&\txt{subject to}~~p(u|x)\in Cs
\ea\ee

\begin{prp}\label{region-proposition}
Let $x$ be a random variable or vector, and suppose the objective function $S$ and constraint set $Cs$ in the problem (\ref{functional-max-problem}) satisfy the following conditions.

\begin{enumerate}
\item $Cs$ is a convex set.
\item $S$ is nonconstant, differentiable, and \underline{convex} on $Cs$ in the multi-variable $\{p(u|x):u=0,1,...,N-1,~x\in\X\}$.
\item $S$ is \underline{monotonic} with respect to a point of the boundary $\del Cs$, i.e., there is a point $p\in\del Cs$ such that $S$ is monotonic along every line segment through $p$ in the closure $\overline{Cs}=Cs\cup\del Cs$. (In other words, $S$ has the property given in Remark 1. immediately after Theorem \ref{convex-max-theorem}).
\item For each $j$, the set of data points
{\small
\be
\label{null-set}C_{u=j}=\bigcup_{j'\neq j}\left\{x:~{\del S/\del p(u=j|x)}={\del S/\del p(u=j'|x)}\neq 0\right\}
\ee
}has zero probability.
\end{enumerate}

Then every optimal decision rule is deterministic, and is uniquely (i.e., necessarily and sufficiently) given by
\bea
\label{solution}p_{\txt{opt}}(u=j|x)=I_{R_{u=j}}(x),~~~~j=0,1,...,N-1,
\eea
where the $j$th decision region $R_{u=j}$ is specified as
{\small
\be
\label{decision-regions}R_{u=j}=\bigcap_{j'\neq j}\left\{x:~{\del S/\del p_{\txt{opt}}(u=j|x)}>{\del S/\del p_{\txt{opt}}(u=j'|x)}\right\}.
\ee
}
\end{prp}

Note that in (\ref{decision-regions}), the expression $\del S/\del p_{\txt{opt}}(u=j|x)$ denotes the derivative of $S$ evaluated at the optimal point, i.e.,
$${\del S/\del p_{\txt{opt}}(u=j|x)}={\del S/\del p(u=j|x)}|_{p(u=j|x)=p_{\txt{opt}}(u=j|x)}.$$

\begin{proof}
Since $Cs$ can be simplified to a polygonal (or simplicial) set by replacing $S$ with a convex Lagrangian, we will assume without loss of generality that $Cs$ is simplicial, i.e., we will choose the constraint set to be the \emph{free} (or extended Cartesian) product $Cs=\prod_{x\in\X}\Delta_N(x)$, where $\Delta_N(x)$ is the $N$-dimensional probability simplex given by

{\footnotesize
\be
\Delta_N(x)\!=\!\left\{\vec{r}(x)\!=\!\big(p(u\!=\!0|x)\!,...,\!p(u\!=\!N-1|x)\big)\!:p(u=i|x)\geq 0,\!\sum_{i=0}^{N-1}p(u\!=\!i|x)=1\!\right\}\subset[0,1]^N.\nn
\ee
}

Moreover, because $S$ is convex in each variable, it is clear that we can proceed by optimizing $S$ over one variable at a time while the others are held constant. Thus, it suffices to optimize $S$ over $\Delta_N(x)$ for an arbitrary $x\in \X$. In other words, we want to maximize $S$ with respect to $\vec{r}(x)=\big(p(u=0|x),...,p(u=N-1|x)\big)\in [0,1]^N$, with $\sum_{i=0}^{N-1} p(u=j|x) =1$ and $p(u=j|x)\geq 0$ for $j=0,\cdots, N-1$.

If $S$ is convex in $\vec{r}(x)$, then by Remark 4. after Theorem \ref{convex-max-theorem}, its maximum value occurs at one or more corner points of $\Delta_N(x)$, i.e.,
\bea
\vec{r}_{\txt{opt}}(x)=\vec{e}_j=(0,...,0,\ub{1}_{j\txt{th spot}},0,...,0),~~~~\txt{for some}~~j\in \{0,...,N-1\}.\nn
 \eea
For each $x\in X$ and each $j\in\{0,1,...,N-1\}$, condition (a) of Theorem \ref{convex-max-theorem} implies
{\footnotesize
  \be\ba
  \label{rg} &\vec{r}_{\txt{opt}}(x)=\vec{e}_j~\iff~p_{\txt{opt}}(u=j|x)=1\\
  &~~\iff~\txt{for all}~\vec{r}(x)\in \Delta_N(x)\backslash\{\vec{e}_j\},~~\big(\vec{e}_j-\vec{r}(x)\big)\cdot{\del S/\del \vec{r}_{\txt{opt}}(x)}>0,
  \ea\ee
}and,
{\footnotesize
  \be\ba
  &\label{rg2}\vec{r}_{\txt{opt}}(x)\neq\vec{e}_j~\iff~p_{\txt{opt}}(u=j|x)=0\\
  &~~\iff~\txt{for some}~\vec{r}(x)\in \Delta_N(x)\backslash\{\vec{e}_j\},~~\big(\vec{e}_j-\vec{r}(x)\big)\cdot{\del S/\del \vec{r}_{\txt{opt}}(x)}<0,
  \ea\ee
}where~ {\small$\del S/\del \vec{r}_{\txt{opt}}={\del S/\del \vec{r}}\big|_{\vec{r}=\vec{r}_{\txt{opt}}}$},~ $\big(\vec{e}_j-\vec{r}(x)\big)\cdot{\del S/\del \vec{r}_{\txt{opt}}(x)}$ is the dot-product of the vectors $\vec{e}_j-\vec{r}(x)$ and ${\del S/\del \vec{r}_{\txt{opt}}(x)}$, and $A\backslash B$ denotes the set difference, i.e., $A\backslash B=\{a: a\in A \mbox{ and } a \notin B\}$.

The region defined by $x$ satisfying~ $p_{\txt{opt}}(u=j|x)=1$~ is
{\footnotesize
\be\ba
\label{dregion-proof}& R_{u=j}=\left\{x:~\big(\vec{e}_j-\vec{r}(x)\big)\cdot{\del S/\del \vec{r}_{\txt{opt}}(x)}>0~~\txt{for all}~\vec{r}(x)\in\Delta_N(x)\backslash\{\vec{e}_j\}\right\}\\
&~~~~\sr{(a)}{=} \left\{x:~{\del S/\del p_{\txt{opt}}(u=j|x)}>{\del S/\del p_{\txt{opt}}(u=j'|x)}~\txt{for all}~j'\neq j \right\}\\
&~~~~=\bigcap_{j'\neq j}\left\{x:~{\del S/\del p_{\txt{opt}}(u=j|x)}>{\del S/\del p_{\txt{opt}}(u=j'|x)}\right\},
\ea\ee
}which by (\ref{null-set}) is probability-wise complementary to the region defined by $x$ satisfying $p_{\txt{opt}}(u=j|x)=0$. Therefore, (\ref{rg}) covers both cases, and is equivalent to the deterministic rule (\ref{solution}). Note that step (a) in (\ref{dregion-proof}) is due to the following.

For fixed $j$, let $\Delta_{N-1}^{(j)}(x)$ be the convex hull of $\{\vec{e}_{j'}:~\txt{for all}~j'\neq j\}$, which is the face of $\Delta_N(x)$ opposite to $\vec{e}_j$. Then every point $\vec{s}(x)\in \Delta_{N-1}^{(j)}(x)$ can be expressed as
\bea
\label{variable-reduction1}\vec{s}(x)=\sum_{j'\neq j}\al_{j'}\vec{e}_{j'},
\eea
for some nonnegative numbers $\al_{j'}=\al_{j'}(x)\geq 0$ such that $\sum_{j'\neq j}\al_{j'}=1$. Now, observe that for any point $\vec{r}(x)\in\Delta_N(x)$, we can write
{\small
\bea
\label{variable-reduction2}\vec{e}_j-\vec{r}(x)=\vec{e}_j-\big(\ld\vec{s}(x)+(1-\ld)\vec{e}_j\big)=\ld\big(\vec{e}_j-\vec{s}(x)\big)\sr{(\ref{variable-reduction1})}{=}\sum_{j'\neq j}\ld\al_{j'}\big(\vec{e}_j-\vec{e}_{j'}\big),
\eea
}
for some $\ld=\ld(x)\in[0,1]$ and some $\vec{s}(x)\in \Delta_{N-1}^{(j)}(x)$.

\begin{figure}
\centering
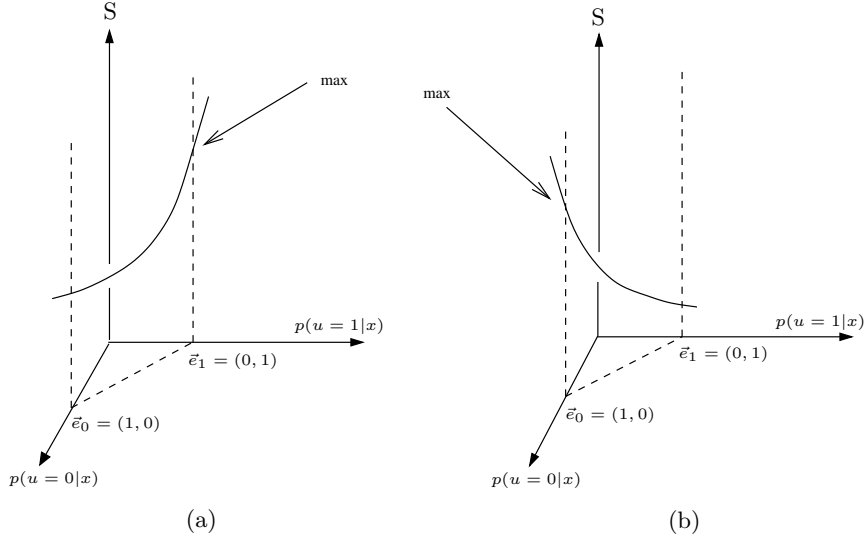
  \caption{Visualization of the decision function $S$ for $N=2$:~ In case (a), $x\in R_{u=1}$, and in case (b), $x\in R_{u=0}$.}\label{region-proof-diag}
\end{figure}
\end{proof}

The following is a series of important remarks regarding applicability and possible extensions of Proposition \ref{region-proposition}.

\begin{rmks*}~
\begin{enumerate}[leftmargin=0.5cm]
\item~ The binary decision rule can be simplified further. In this case, the probability simplex $\Delta_2(x)$ is the single line with equation $p(u=0|x)+p(u=1|x)=1$. Thus, by the chain rule of differentiation, the differential operator $\big(\vec{e}_j-\vec{r}(x)\big)\cdot\del /\del \vec{r}_{\txt{opt}}(x)$ along $\Delta_2(x)$ is equivalent to a derivative, which we denote by $\del^B /\del p_{\txt{opt}}(u=j|x)$, with the property
\bea
\label{binary-derivative}{\del^B p(u|x)/\del p(u'|x')}=(-1)^{u-u'}\delta_{xx'}
\eea
in addition to linearity and the Leibnitz rule. Hence, the binary decision regions take the compact form
\bea
\label{binary-decision-regions}R_{u=j}=\left\{x:~{\del^B S/\del p_{\txt{opt}}(u=j|x)}>0 \right\},
\eea
where the superscript $B$ in $\del^B$ serves as a reminder to the reader of the property (\ref{binary-derivative}) which ensures that the derivative is restricted to the probability simplex $\Delta_2(x)$. The compact form of the binary decision rule as implemented by (\ref{binary-derivative}) and (\ref{binary-decision-regions}) will greatly simplify calculations later on.

\item~ Notice that (\ref{solution}) is an implicit equation in $p_{\txt{opt}}(u=j|x)$ since the region $R_{u=j}$ also depends on $p_{\txt{opt}}(u=j|x)$. Therefore we must proceed to substitute the equations
    \be
    \left\{p_{\txt{opt}}(u=j|x)=I_{R_{u=j}}(x): j=0,1,...,N-1\right\}\nn
    \ee
    into the objective function $S$, and then compute the optimal threshold values that explicitly determine the decision regions. In the case of distributed networks of sensors where more than one set of local decision rules are involved, the resulting system of equations is often analytically intractable and one has to resort to numerical computation. This is especially the case when sensor observations are conditionally dependent, i.e., they remain dependent or correlated under at least one of the hypotheses.

    In sufficiently simple problems for which the decision regions in (\ref{decision-regions}) can be specified in terms of a number of threshold parameters, the optimization problem can be \emph{explicitly} solved in a natural way simply by optimizing $S$ over the threshold variables. This, we will refer to as \emph{the second stage} of the optimization problem, \emph{the first stage} being the \emph{implicit} solution as stated in the proposition.

\item~ Recall that for the optimal decision rule to be deterministic, as given in (\ref{solution}), the data sets (\ref{null-set}) must be null with respect to the probability measure. Otherwise, the deterministic rule (\ref{solution}) is replaced by a randomized version
\bea
\label{solution-randomized}p_{\txt{opt}}(u=j|x)=I_{R_{u=j}}(x)+\sum_k\rho_{jk}I_{C_k}(x),
\eea
where $\{C_k\}$ is a partition of the set
{\small
\be
\bigcup_jC_{u=j},~~~~C_{u=j}=\bigcup_{j'\neq j}\left\{x:~{\del S/\del p_{\txt{opt}}(u=j|x)}={\del S/\del p_{\txt{opt}}(u=j'|x)}\neq0\right\},\nn
\ee
}and $\rho_{jk}\in [0,1],~\sum_j\rho_{jk}=1,$ are arbitrary (i.e., free) coefficients but which must be consistent with every constraint of the optimization problem. It is worthwhile to remark that the deterministic rule (\ref{solution}) is more easily realized when $x$ is continuous than when $x$ is discrete. Thus, the randomized rule (\ref{solution-randomized}) is often required when $x$ is a discrete random variable.

The fact that randomization depends on nullness of the sets $C_{u=j}$ generalizes a similar observation that was made in \cite{willett-warren-1990,willett-warren-1992,tsitsiklis-0} under the Neymann-Pearson framework.

\item~ Since the Bayes risk is an affine (hence a convex) function of $p(u|x)$, Proposition \ref{region-proposition} is a direct generalization of the familiar procedure whereby the unconditional Bayes risk
{\footnotesize
\bea
R(\gamma)=\sum_{i,j}C_{ij}p(\gamma(x)=i,H_j)=\sum_{i,x}p(\gamma(x)=i|x)R_i(x),
\eea
}is minimized over $\gamma$ simply by separately minimizing the associated conditional Bayes risks $R_i(x)=\sum_jC_{ij}p(x|H_j)p(H_j)$ over $i$ by means of the choice
 \bea
 p(\gamma(x)=i|x)=\left\{
                    \begin{array}{ll}
                      1, & R_i(x)<R_j(x)~\txt{for all}~j\neq i, \\
                      0, & \txt{otherwise},
                    \end{array}
                  \right.\nn
 \eea
where we have assumed for simplicity that $P\big(R_i(x)=R_j(x)\big)=0$ for all $i$ and $j\neq i$.

\item If condition 3 of the proposition (i.e., the monotonicity condition) fails, then exactly the same conclusions hold, except for the sufficiency of the decision rule for optimality. That is, if conditions 1, 2, and 4 hold, then the optimal decision rule is deterministic, and satisfies (\ref{solution}) and (\ref{decision-regions}) as necessary conditions. In that case, we must proceed, according to condition (b) of Theorem \ref{convex-max-theorem}, to select a decision rule that maximizes $S$ among all possible decision rules that satisfy (\ref{solution}) and (\ref{decision-regions}).

    In sufficiently simple problems for which monotonicity fails, the second stage of the optimization (as described in Remark 2. above) will in general suffer from the \emph{local optimum problem} in the sense that $S$, as a function of the thresholds, possesses two or more local optima from which a global optimum must then be picked by some other method.

\item If both conditions 3 and 4 of the proposition fail, then the randomized rule (\ref{solution-randomized}) is necessary but not sufficient for optimality. Once again, we must proceed, according to condition (b) of Theorem \ref{convex-max-theorem}, to select a decision rule that maximizes $S$ among all possible decision rules that satisfy (\ref{solution-randomized}).

\item Using standard methods of convex optimization, Proposition \ref{region-proposition} can be further extended to include non-differentiable convex objective functions by replacing the derivative $\del S/\del p(u=j|x)$ with a subdifferential. The proposition can also be refined to include the optimization of subharmonic objective functions.
\end{enumerate}
\end{rmks*}

We will now state a corollary of Proposition \ref{region-proposition} for the minimization of a convex function. Consider the problem
\be\ba
\label{functional-min-problem}&\txt{minimize}~~ S\big(p(u|x)\big)\\
&\txt{subject to}~~p(u|x)\in Cs
\ea\ee

\begin{crl}\label{region-corollary}
Let $x$ be a random variable or vector, and suppose the objective function $S$ and constraint set $Cs$ in the problem (\ref{functional-min-problem}) satisfy the following conditions.

\begin{enumerate}
\item $Cs$ is a convex set.
\item $S$ is nonconstant, differentiable, and \underline{convex} on $Cs$ in the multi-variable $\{p(u|x):u=0,1,...,N-1,~x\in\X\}$.
\item For each $j$, the set of data points
{\small
\be
\label{null-set-min}C_{u=j}=\bigcup_{j'\neq j}\left\{x:~{\del S/\del p(u=j|x)}={\del S/\del p(u=j'|x)}\neq 0\right\}
\ee
}has zero probability.
\end{enumerate}

Then every optimal decision rule is uniquely (i.e., necessarily and sufficiently) given by
\bea
\label{solution-min}p_{\txt{opt}}(u=j|x)=I_{R_{u=j}}(x)+p^0(u=j|x)I_{R^0}(x),~~~~j=0,1,...,N-1,
\eea
where
{\small
\be
\label{decision-regions-min}R_{u=j}=\bigcap_{j'\neq j}\left\{x:~{\del S/\del p_{\txt{opt}}(u=j|x)}<{\del S/\del p_{\txt{opt}}(u=j'|x)}\right\},
\ee
and
\bea
R^0=\{x\in\X:~\del S/\del p(u=j|x)=0~~\txt{for all}~~j=0,1,\cdots,N-1\},\nn
\eea
and for each $j,x$ pair, $p^0(u=j|x)$ is the $j$th component of the solution of the system of equations
\bea
\{\del S/\del p(u=j|x)=0,~~j=0,1,\cdots,N-1\}.\nn
\eea
}
\end{crl}
\begin{proof}
With the help of Corollary \ref{convex-min-theorem}, the proof follows the same arguments as in the proof of Theorem \ref{region-proposition}. We simply need to (1) reverse the inequalities that determine the decision regions, (2) account for the possibility of the minimum occurring at the point where the derivative vanishes, and (3) recognize that monotonicity is not necessary for uniqueness of the solution.
\end{proof}

\begin{rmks*}~
\begin{enumerate}
\item If the derivative of $S$ with respect to $p(u|x)$ is nonzero for all $x\in \X$, then the second term in (\ref{solution-min}) disappears.

\item If the derivative of $S$ with respect to $p(u|x)$ equals $0$ for all $x\in \X$, then the first term in (\ref{solution-min}) disappears.

\item If condition 3 of the corollary fails, then we must include a randomization term in the rule (\ref{solution-min}), as in (\ref{solution-randomized}), to obtain the randomized version
\bea
\label{solution-randomized-min}p_{\txt{opt}}(u=j|x)=I_{R_{u=j}}(x)+p^0(u=j|x)I_{R^0}(x)+\sum_k\rho_{jk}I_{C_k}(x),
\eea
where $\{C_k\}$ is a partition of the set $\bigcup_jC_{u=j}$, and $\rho_{jk}\in [0,1],~\sum_j\rho_{jk}=1$, are arbitrary coefficients but which must be consistent with every constraint of the optimization problem.
\end{enumerate}
\end{rmks*}

Proposition \ref{region-proposition} will be used in Chapters \ref{interactive-detection}, \ref{communication-direction}, \ref{acyclic-graph-detection} to determine optimal decision regions with various types of objective functions, including the probability of detection, KL distance, and Bayesian probability of error. First, however, we would like to illustrate how this result can be applied. We will use the Bayesian objective function in this illustration, while noting that the results for more general objective functions are similar. The main purpose is to demonstrate how Proposition \ref{region-proposition} can be used in practical distributed detection problems. For concreteness, we begin with a discussion of centralized sensor rules in Section \ref{single-sensor-rules}. The centralized detection process is then upgraded to a discussion of distributed sensor rules in Section \ref{sensor-network-rules}.

\section{Single sensor rules}\label{single-sensor-rules}
From Section \ref{detection-problem}, suppose a \emph{discrete} random signal $s\in \S$ is observed by an isolated sensor $X$ as
\bea
\label{signal-transform1}x=s+b~\in~\X,
\eea
where $b\in\B$ is a continuous random parameter whose distribution is continuous and known, and $s,b$ are \emph{statistically independent} of one another. Let the signal alphabet be given by $\S=\{\mu_0,\mu_1,...,\mu_{M-1}\}$. Then we have a set of $M$ \emph{hypotheses}
\bea
H_j:~s=\mu_j,~~\txt{i.e.,}~~x=\mu_j+b,~~~~~~~~j=0,1,...,M-1,\nn
\eea
each of which represents a possible value of the signal $s$.
For the special case where $M=2$ and $\S=\{0,1\}$, the hypothesis $H_0$ denoting ``target absent'' is called the \emph{null} hypothesis, while the hypothesis $H_1$ denoting ``target present'' is called the \emph{alternative}  hypothesis. As usual, we denote the conditional pdf, $p(x|H_j)=p(x|s=\mu_j)$ of $x$ under $H_j$ by $p_j(x)$. We emphasize here that $p_j(x)$ is known for each $j$ since the distribution of $b$ in (\ref{signal-transform1}) is known.

A \emph{decision rule} of the sensor $X$ is a mapping
\bea
\gamma:x\in\X~\longmapsto~ u=\gamma(x)\in \U=\{0,1,...,N-1\},\nn
\eea
where we often denote the rule $\gamma$ by its value $u$ as a \emph{variable}.

Our objective is to choose the rule $\gamma$ such that some function $S=S(\gamma(x))$ of $u=\gamma(x)$ is optimized. Because $u$ is a random variable, it is sufficient to treat $S$ as a function of the conditional distributions $p(u|x)$, for all $x\in \X$. For our illustration, we consider the Bayesian objective
\bea
\label{bayes-objective1}S = \sum_{i,j}C_{ij}p(u=i,H_j)=\sum_{i,x,j}C_{ij}p(u=i|x)p_j(x)\pi_j,
\eea
where $p_j(x)=p(x|H_j)$, $\pi_j=p(H_j)=p(s=\mu_j)$ is the probability that $H_j$ is true, and $C_{ij}$ is a nonnegative number denoting the cost of the decision $u=i$ when $H_j$ is true. Note that if $N=M$ (i.e., if the number decision values equals the number of hypothesis values), the decision $u=i$ may be viewed as acceptance of $H_i$, in which case $C_{ij}$ is the cost of accepting $H_i$ when $H_j$ is true.

Since $S$ is an affine function of the conditional probabilities $p(u|x)$ and the observation $x$ is continuously distributed, by Proposition \ref{region-proposition} the optimal decision rule is deterministic and, for each $i=0,1,...,N-1$, is given by
\bea
\label{optimal-rule1}p_{\txt{opt}}(u=i|x)=I_{R_{u=i}}(x)=\left\{
                                                            \begin{array}{ll}
                                                              1, & \hbox{if}~~x\in R_{u=i}, \\
                                                              0, & \hbox{if}~~x\not\in R_{u=i},
                                                            \end{array}
                                                          \right.
\eea
where the \emph{decision region} $R_{u=i}$ is given by

{\small
\be\ba
&\label{optimal-region1}R_{u=i}=\bigcap_{j\neq i}\left\{x:{\del S\over\del p_{\txt{opt}}(u=i|x)}-{\del S\over\del p_{\txt{opt}}(u=j|x)}<0\right\}\\
&~~~~=\bigcap_{j\neq i}\left\{x:\sum_lC_{il}p_l(x)\pi_l-\sum_lC_{jl}p_l(x)\pi_l<0\right\}.
\ea\ee
}

In terms of probabilities of the decision regions, the optimal value of $S$ is
\bea
\label{optimal-bayes-objective1}S_{\txt{opt}}=\sum_{i,j}C_{ij}~\pi_j~p_j(R_{u=i}).
\eea

It is clear in this case that the decision regions, and hence $S_{\txt{opt}}$, are determined by a fixed set of known thresholds. These thresholds are directly determined by the costs $\{C_{ij}\}$ and the prior probabilities $\{\pi_i\}$. As we will soon see, the situation is no longer so simple in distributed sensor settings.

\section{Sensor network rules}\label{sensor-network-rules}
Now suppose we have a distributed network of $n$ sensors $\{X_k,~k=1,...,n\}$, each sensor observing the same signal $s\in \{\mu_0,\mu_1,...,\mu_{M-1}\}$. Here, the main difference with the preceding section is that each $X_k$ must now make its decision $u_k$ based not only on its own observation $x_k$, but as well on the set of decisions $\td{u}_k$ of all sensors forwarding their decisions to $X_k$, i.e., $u_k=\gamma_k(x_k,\td{u}_k)$ for some integer-valued function $\gamma_k$. Thus, $X_k$ makes an observation
\bea
\label{signal-transform2}x_k=s+b_k~\in~\X_k,~~~~b_k\in\B_k,
\eea
considers a set of hypotheses
\bea
H_j:~s=\mu_j,~~\txt{i.e.,}~~x_k=\mu_j+b_k,~~~~j=0,1,...,M-1,\nn
\eea
and applies a decision rule
\bea
\gamma_k:(x_k,\td{u}_k)\in \X_k\times\td{\U}_k~\longmapsto~u_k=\gamma_k(x_k,\td{u}_k)\in \U_k,\nn
\eea
where $\U_k=\{0,1,...,N_k-1\}$, and $\td{u}_k\in \td{\U}_k$ denotes the \emph{set} of decision variables of all sensors transmitting their decisions to $X_k$. Here, for each $k$, we again assume the parameters $s$, $b_k$ in (\ref{signal-transform2}) have the same properties as $s,b$ in (\ref{signal-transform1}). Note that if we fix $N_k=N$ for all $k=0,1,...,n$, then $\td{\U}_k=\{0,1,...,N-1\}^{I_k}$, where $I_k$ is the number of sensors transmitting their decisions to $X_k$.

Without loss of generality, we will let sensor $X_1$ serve as the fusion center for the network. Once again, our objective is to choose the \emph{decision strategy} $\{\gamma_1,\gamma_2,...,\gamma_n\}$ such that some function
$$S=S\big(\gamma_1(x_1,\td{u}_1),\gamma_1(x_2,\td{u}_2),...,\gamma_k(x_n,\td{u}_n)\big)$$
of $u_1=\gamma_1(x_1,\td{u}_1)$, $u_2=\gamma_2(x_2,\td{u}_2)$, $\cdots$, $u_n=\gamma_n(x_n,\td{u}_n)$ is optimized. For our illustration, we consider $S$ to be the Bayesian function at the fusion center $X_1$, i.e.,
\be\ba
&\label{bayes-objective2}S = \sum_{i,j}C_{ij}p(u_1=i,H_j)=\sum_{u_1,i}C_{u_1i}p(u_1,H_i)\\
&~~~~=\sum_{u^n,x^n,i}C_{u_1i}~\prod_{k=1}^np(u_k|x_k,\td{u}_k)~p_i(x^n)\pi_i,
\ea\ee
where $u^n=(u_1,...,u_n)$, $x^n=(x_1,...,x_n)$.


Since $S$ is an affine function of the conditional probabilities $p(u_k|x_k,\td{u}_k)$ and the observations $x_k$ are continuously distributed, by Proposition \ref{region-proposition} the optimal decision rule for each sensor $X_k$ is deterministic, and is given by
\bea
\label{optimal-rule2}p_{\txt{opt}}(u_k|x_k,\td{u}_k)=I_{R_{u_k|\td{u}_k}}(x_k)=\left\{
                                                            \begin{array}{ll}
                                                              1, & \hbox{if}~~x_k\in R_{u_k|\td{u}_k}, \\
                                                              0, & \hbox{if}~~x_k\not\in R_{u_k|\td{u}_k},
                                                            \end{array}
                                                          \right.
\eea
where $u_k\in\{0,1,...,N_k-1\}$, and the \emph{decision region} $R_{u_k|\td{u}_k}$ is given by

{\footnotesize
\bea
\label{optimal-region2}R_{u_k|\td{u}_k}=\bigcap_{u\neq u_k}\left\{x_k:{\del S\over\del p_{\txt{opt}}(u_k|x_k,\td{u}_k)}-{\del S\over\del p_{\txt{opt}}(u|x_k,\td{u}_k)}<0\right\}.
\eea
}The derivative of $S$ in (\ref{optimal-region2}) can be express as

{\footnotesize
\be\ba
\label{optimal-derivative2}&{\del S\over\del p_{\txt{opt}}(u_k|x_k,\td{u}_k)}=\sum_{\{u^n,x^n\}_k,~i}C_{u_1i}~\prod_{k'\neq k}p_{\txt{opt}}(u_{k'}|x_{k'},\td{u}_{k'})~p_i(x^n)\pi_i\\
&~~~~=\sum_{\{u^n,x^n\}_k,~i}C_{u_1i}~\prod_{k'\neq k}I_{R_{u_{k'}|\td{u}_{k'}}}\left(x_{k'}\right)~p_i(x^n)\pi_i,
\ea\ee
}where $\{u^n,x^n\}_k=\{u^n,x^n\}\backslash\{u_k,x_k,\td{u}_k\}$. With \emph{conditionally independent observations}, we have $p_i(x^n)=\prod_{k=1}^np_i(x_k)$. If we further assume there are \emph{no closed processing paths} in the sensor network that can lead to \emph{overlaps} among the decision regions, then (\ref{optimal-derivative2}) can be written as

{\footnotesize
\be\ba
\label{optimal-derivative2.2}{\del S\over\del p_{\txt{opt}}(u_k|x_k,\td{u}_k)}=\sum_{\{u^n,x^n\}_k,~i}C_{u_1i}~\prod_{k'\neq k}p_i\left(R_{u_{k'}|\td{u}_{k'}}\right)~p_i(x_k)\pi_i,
\ea\ee
}in which case, the optimal value of $S$ in terms of probabilities of the decision regions is
\bea
\label{optimal-bayes-objective2}S_{\txt{opt}}=\sum_{u^n,i}C_{u_1i}~\pi_i~\prod_{k=1}^np_i\left(R_{u_k|\td{u}_k}\right).
\eea
We have thus proved the following result.
\begin{thm}\label{acyclic-prop}
Suppose we are given a network of $n$ sensors $X_1,X_2,...,X_n$ with conditionally independent and continuously distributed observations $x_1,x_2,...,x_n$. Suppose further that there are no cyclic communication paths in the sensor network. Then the decision rules for detection, based on the Bayes function (\ref{bayes-objective2}), by the sensor network are given by (\ref{optimal-rule2}), (\ref{optimal-region2}), and (\ref{optimal-derivative2.2}). Moreover, the optimal value of the Bayes function is given by (\ref{optimal-bayes-objective2}).
\end{thm}

In further applications in Chapters \ref{interactive-detection}, \ref{communication-direction}, \ref{acyclic-graph-detection}, we will mostly consider \emph{binary} hypothesis and \emph{binary} decisions, i.e., we set $M=N_k=2$. As already noted in previous remarks, when the problem is sufficiently simple, the decision regions (\ref{optimal-region2}) can be completely specified in terms of a number of threshold parameters that do not depend on the observations. In that case, we only need to optimize $S$ as a function of the thresholds.

\part{Applications} \label{part-III}  
\chapter{Interactive Distributed Detection}\label{interactive-detection}
\section{Introduction}
This chapter is based mainly on \cite{akofor-chen-icassp,akofor-chen-it-paper}, where detection is done in the Neyman-Pearson (NP) framework, but \cite{zhu-akofor-chen-wcnc} also contains similar results under the Bayesian framework. We are going to study the impact of interactive fusion on detection performance in tandem fusion networks with conditionally independent observations. Both the fixed sample size NP test and the large sample NP test will be analyzed. There exist related work on fusion architecture in \cite{song-et-al,song-et-al2,Papastavrow&Athans:92AC}, on parallel and noninteractive feedback settings in \cite{alhakeem-varshney,Tay&Tsitsiklis,Shalaby&Papamarcou}, on various forms of asymptotic in \cite{Tay&Tsitsiklis,Shalaby&Papamarcou,Papastavrou-Athans-2,Tay-Tsitsiklis-Win}, and on parley in \cite{swaszek-willett-1995}.

For the fixed sample test in Section \ref{section_NP}, we will find that interactive distributed detection may strictly outperform the one-way tandem fusion structure. For the large sample test in Section \ref{section_KL}, however, we will see that interactive fusion and one-way tandem fusion achieve the same asymptotic detection performance. (Note that this conclusion may no longer hold if certain communication constraints are imposed, \cite{Xiang-Kim:Allerton12}). Also, we will find in Section \ref{generalize} that these results remain valid in the following more general settings:
\bit
\item The two sensors undergo multiple steps of memoryless interaction.
\item The peripheral sensor is replaced by multiple peripheral sensors.
\item Sensor outputs (before the final output) are multibit.
\eit

A simple tandem sensor network is a sequence of two or more sensors in which each sensor makes a single decision using its own observation and the output of its predecessor, and then passes its decision to the next sensor, i.e., its successor. The last sensor serves as a fusion center, and its decision is considered the final decision.

If the sensor outputs are single bit decisions, and sensor observations are independent conditioned on any given hypothesis, the optimal decision rule is determined by a likelihood ratio test \cite{Varshney:book,tsitsiklis-0}. Note that this result assumes that every sensor makes only one decision.

\begin{figure}[H]
\centering
\scalebox{1.5}{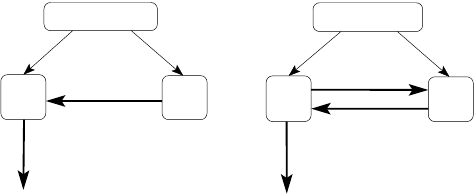}
\caption{(a) One-way tandem fusion (YX process),~~ (b) Interactive fusion (XYX process).}\label{one-and-two-way-diagram}
\end{figure}

For a two-sensor tandem network, we will replace the above static message passing with an interactive one: the fusion center (FC) sends an initial bit to the peripheral sensor (PS) based on its observation. The PS then makes a decision based on its own observation as well as the input from the FC and passes it back to the FC as partial input for the final decision. Fig.~\ref{one-and-two-way-diagram} illustrates the difference between the one-way tandem and interactive fusion networks.

In Fig \ref{one-and-two-way-diagram}(b),  $x$ is the observation of sensor X, $y$ the observation of sensor Y, $u$ the initial decision of $X$, $v$ the decision of Y based on $x,u$, and $w$ the final decision of X based on $y,v$. The random variables $x$ and $y$ are real-valued and assumed to be conditionally independent with respect to the hypothesis, i.e., $p_i(x,y)=p_i(x)p_i(y)$, meanwhile $u=\gamma(x),~v=\delta(y,u),~w=\rho(x,v)$ are binary, where $\gamma$, $\delta$, $\rho$ are integer-valued mappings. For simplicity, we refer to the fusion architecture in Fig.~\ref{one-and-two-way-diagram}(a) as the YX process whereas to that in Fig.~\ref{one-and-two-way-diagram}(b) as the XYX process.

Later in Section \ref{generalize}, we will also consider more general versions of the above situation, which involve multiple rounds of interaction, multiple peripheral sensors, and exchange of multi-bit decisions.

\section{The fixed sample size Neyman-Pearson test}\label{section_NP}
Our objective in the NP test is to maximize the probability of detection in such a way that the probability of false alarm does not exceed a given value $\al$, i.e., we have the constrained optimization problem
\be\ba
\label{NP-problem}& \txt{maximize}~~~~ P_d = p_1(w=1)\\
&\txt{subject to}~~~~ P_f =
p_0(w=1)\leq \al
\ea\ee
The Lagrangian for the problem is
\bea
\label{NP-Lagrangian-0} L=p_1(w=1)+\ld~\big(\al-p_0(w=1)\big), ~~\ld\geq 0.
\eea
For the YX process, $L$ is a function of $\ld$, $p(v|y)$, and $p(w|x,v)$, with
\bea
\label{NP-YX-probabilty}p_i(w)=\sum_{x,y,v}p(w|x,v)p(v|y)~p_i(x)p_i(y),
\eea
while for the XYX process, $L$ is a function of $\ld$, $p(u|x)$, $p(v|y,u)$, and $p(w|x,v)$, with
\bea
\label{NP-XYX-probabilty}p_i(w)=\sum_{x,v,y,u}p(w|x,v)p(v|y,u)p(u|x)~p_i(x)p_i(y).
\eea

By applying Proposition \ref{region-proposition}, we obtain the following result, the proof of which is given in \cite{akofor-chen-it-paper}.
\begin{thm}\label{np-theorem}
The optimal decision rules for the NP test with objective (\ref{NP-Lagrangian-0}) are as follows. For the YX process, we have
\bea
p_{\txt{opt}}(v|y)=I_{R_v}(y),~~~~p_{\txt{opt}}(w|x,v)=I_{R_{w|v}}(x),\nn
\eea
with the decision regions given by
\be\ba
\label{NP-YX-regions}R_{v=1}=\left\{y:~{p_1(y)\over p_0(y)}>\ld^{(2)}\right\},~~~~R_{w=1|v}=\left\{x:~{p_1(x)\over p_0(x)}>\ld_{v}^{(3)}\right\},
\ea\ee
where
\be
\ld^{(2)}=\ld~{P_0(R_{w=1|v=1})-P_0(R_{w=1|v=0})\over P_1(R_{w=1|v=1})-P_1(R_{w=1|v=0})}~~~~\txt{and}~~~~\ld_{v}^{(3)}=\ld~{P_0(R_{v})\over P_1(R_{v})}.\nn
\ee
For the XYX process, we have
\bea
p_{\txt{opt}}(u|x)=I_{R_u}(x),~~~~p_{\txt{opt}}(v|y,u)=I_{R_{v|u}}(y),~~~~p_{\txt{opt}}(w|x,v)=I_{R_{w|v}}(x),\nn
\eea
with the decision regions given by
\be\ba
\label{NP-XYX-regions}& R_{u=1}=\left\{x:~{p_1(x)\over p_0(x)}Q(x)>\ld^{(1)}Q(x)\right\},~~~~R_{v=1|u}=\left\{y:~{p_1(y)\over p_0(y)}>\ld^{(2)}_u\right\},\\
&R_{w=1|v}=\left\{x:~{p_1(x)\over p_0(x)}>\sum_u\ld_{vu}^{(3)}~I_{R_u}(x)\right\},
\ea\ee
where
\be\ba
&\ld^{(1)}=\ld~{P_0(R_{v=1|u=1})-P_0(R_{v=1|u=0})\over P_1(R_{v=1|u=1})-P_1(R_{v=1|u=0})},~~~~ Q(x)=I_{R_{w=1|v=1}}(x)-I_{R_{w=1|v=0}}(x),\nn\\
&\ld^{(2)}_u=\ld~{P_0(R_{w=1|v=1}\cap R_u)-P_0(R_{w=1|v=0}\cap R_u)\over P_1(R_{w=1|v=1}\cap R_u)-P_1(R_{w=1|v=0}\cap R_u)},~~~~\txt{and}~~~~ \ld_{vu}^{(3)}=\ld~{P_0(R_{v|u})\over P_1(R_{v|u})}.\nn
\ea\ee
\end{thm}

Observe that in the interactive process, even though sensor observations are conditionally independent, the decision regions at the FC are not determined by simple likelihood ratio tests. A similar phenomenon was noted in \cite{MMM1,MMM2}.

In terms of the obtained decision regions in Theorem \ref{np-theorem}, the Lagrangian in (\ref{NP-Lagrangian-0}) can be written as
{\small
\be\ba
L_{YX}=\sum_{v}P_1(R_{v})P_1(R_{w=1|v}) +\ld~\left[\al-\sum_{v}P_0(R_{v})P_0(R_{w=1|v})\right]\nn
\ea\ee
}
for the YX process, and as
{\small
\be\ba
L_{XYX}=\sum_{u,v}P_1(R_{v|u})P_1(R_{w=1|v}\cap R_u)+\ld\left[\al-\sum_{u,v}P_0(R_{v|u})P_0(R_{w=1|v}\cap R_u)\right]\nn
\ea\ee
}
for the XYX process.

\subsection*{Example: Constant Signal in White Gaussian Noise}
\begin{figure}
\centering
\includegraphics[width=10cm,height=7cm]{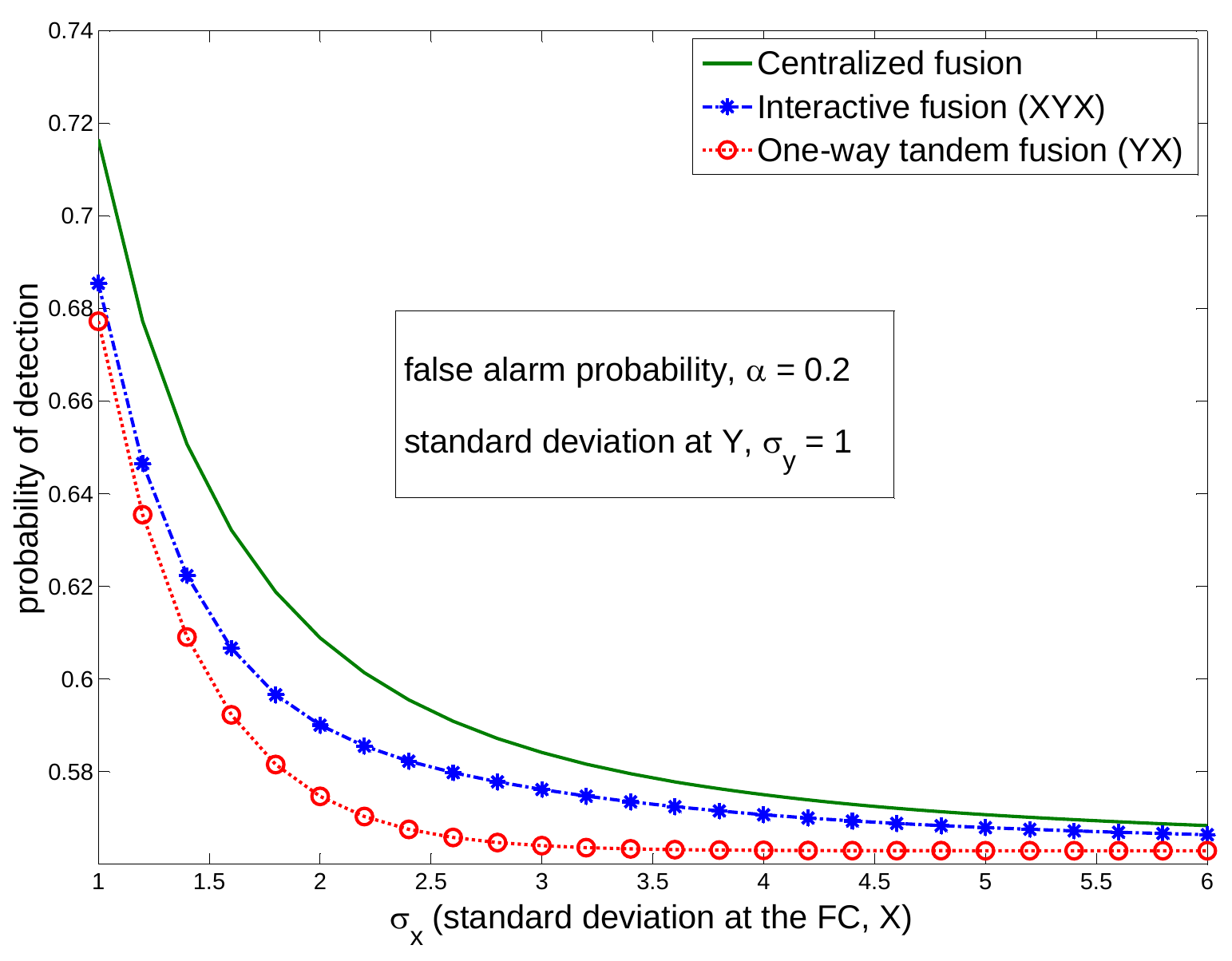}\\
  \caption{Performance of XYX and YX processes}\label{np-graph}
\end{figure}

 Consider the detection of a constant signal $s$ in white Gaussian noise with observations
\bea
\label{WGN}x=s+z_1,~~~~y=s+z_2,~~~~x,y\in\Real=\X=\Y,
\eea
where $z_1\sim N(0,\sigma_x^2),~~z_2\sim N(0,\sigma_y^2)$ and $z_1$ and $z_2$ are independent of each other, and the two hypotheses under test are
\bea
H_0:~s=0,~~~~H_1:~s=1.\nn
\eea
Fig. \ref{np-graph} shows the dependence of the probability of detection on $\sigma_x$ when $\sigma_y$ is fixed. The corresponding false alarm probability is $P_f=0.2$. The figure shows that the XYX process has strictly larger probability of detection compared with the YX process.

The curve corresponding to centralized fusion in Fig. \ref{np-graph} is obtained by repeating the same optimization procedure using (\ref{NP-problem}) and (\ref{NP-Lagrangian-0}), but with the probability of the centralized decision $w=\rho(x,y)$ given by $p_i(w=1)=\sum_{x,y}p(w=1|x,y)p_i(x,y)$. Here, the decision rule $p_{\txt{opt}}(w=1|x,y)=I_{R_{w=1}}(x,y)$, the constant false alarm probability constraint $\al=p_0(w=1)$, and the detection probability $P_d=p_1(w=1)$ can be easily written as
\be\ba
& R_{w=1}=\left\{(x,y): {x\over\sigma_x^2}+{y\over\sigma_y^2}>t=\ln\ld+{1\over 2\sigma_x^2}+{1\over 2\sigma_y^2}\right\},\\
& \al=\int_{-\infty}^\infty Q\left(\sigma_yt-{\sigma_y\over\sigma_x}{x\over\sigma_x}\right){e^{-{x^2\over 2\sigma_x^2}}\over\sqrt{2\pi\sigma_x^2}}dx,\\
& P_d=\int_{-\infty}^\infty Q\left(\sigma_yt-{\sigma_y\over\sigma_x}~{x\over\sigma_x}-{1\over\sigma_y}\right){e^{-{(x-1)^2\over 2\sigma_x^2}}\over\sqrt{2\pi\sigma_x^2}}dx,\nn
\ea\ee
where the threshold $t$ as a function of $\al$ is obtained by solving the constant false alarm probability constraint.

\section{The asymptotic Neyman-Pearson test}\label{section_KL}
Here, we will use scalar quantization since, as pointed out in \cite{LLG}, it is simpler and more efficient (in terms of a smaller processing delay) than vector quantization.

Consider $n$ observation samples~ $(x_1,y_1),...,(x_n,y_n)$, and suppose processing is carried out on a sample-by-sample basis. For the XYX process, the two sensors go through, for each $k=1,\cdots,n$, a decision process with $u_k=\gamma_k(x_k),~~v_k=\delta_k(y_k,u_k)$. The final decision at node X utilizes the entire observation sequence $x^n$ and the output sequence $v^n$ from node Y, i.e., $w=\rho(x^n,v^n)$. We have, therefore,
\bea
\label{np-test-eq1}p_i(w)=\sum_{x^n,v^n}p(w|x^n,v^n)p_i(x^n,v^n),
\eea
where $p(w|x^n,v^n)$ is determined by the final decision rule.

Meanwhile for the YX process, sensor $Y$ sends a decision sequence $v_k=\delta_k(y_k)$, $k=1,\cdots,n$, and X uses $v^n$ and its own observation $x^n$ to make the final decision $w=\rho(x^n,v^n)$. We again have the relation (\ref{np-test-eq1}).

As shown in \cite{akofor-chen-it-paper}, and based on \cite{cover-thomas}, the error exponent for the NP test is the KL distance
\bea
\label{kl-distance}D(p_0(x,v)\|p_1(x,v))=\sum_{x,v}p_0(x,v)\log{p_0(x,v)\over p_1(x,v)}.
\eea
This will be our objective function for the asymptotic performance of the NP test.

\subsection*{One-way tandem fusion (YX process)}\label{one-pass}
In the one-way tandem fusion network, as shown in Fig. \ref{one-and-two-way-diagram}(a), Y sends a decision $v=\delta(y)$ to X. The optimal decision $v$ is chosen so as to maximize the KL distance
\bea
&&K[x,v]=D\big(p_0(x,v)\|p_1(x,v)\big)
\eea
 at sensor X.

Since ~$p_i(x,v)=p_i(x)p_i(v),$ we have
\be
\label{KL-YX} K[x,v]=D\big(p_0(x)\|p_1(x)\big)+\sum_vp_0(v)~\log\big(p_0(v)/ p_1(v)\big)
\ee
where {~$p_i(v)=\sum_yp(v|y)p_i(y).$}

By application of Proposition \ref{region-proposition} in the optimization of (\ref{KL-YX}), we get the following result, the proof of which is given in \cite{akofor-chen-it-paper}.
\begin{thm}\label{XY_theorem}
The optimal decision rule at Y is $p_{\txt{opt}}(v|y)=I_{R_v}(y)$, where
\bea
\label{omit1}&&R_{v=1}=\bigg\{y:{p_1(y)\over p_0(y)}>\ld\bigg\},\\
\label{YX-threshold}&&\ld=\bigg(\log{\beta(1-\al)\over\al(1-\beta)}\bigg)\bigg/\bigg({\beta-\al\over \beta(1-\beta)}\bigg),
\eea
where $\al = P_0\left({p_1(y)/p_0(y)}>\ld\right)$ and $\beta = P_1\left({p_1(y)/p_0(y)}>\ld\right).$
\end{thm}

The maximum KL distance is given by
\be
\label{KYX0}K^{\txt{YX}}_{\max}=K[x]+\al^\ast\log{\al^\ast \over\beta^\ast}+(1-\al^\ast)\log{1-\al^\ast\over 1-\beta^\ast},
\ee
where $K[x]=D(p_0(x)\|p_1(x))$ and $\al^\ast$ and $\beta^\ast$ are the values of $\al$ and $\beta$ that maximize the KL distance.

\subsection*{Interactive fusion (XYX process)}\label{two-pass}
For the interactive fusion process in Fig. \ref{one-and-two-way-diagram}(b), X first sends a decision $u=\gamma(x)$ to Y. Then Y makes a decision $v=\delta(y,u)$ and sends it back to X. The optimal decisions $u$ and $v$ are chosen so as to maximize the KL distance $K[x,v]{\triangleq} K^{\txt{XYX}}$ in the final step at X. The KL distance can be written as
{\footnotesize
\be\ba
\label{KL-XYX}K^{\txt{XYX}}=D\big(p_0(x,v)\|p_1(x,v)\big)=D\big(p_0(x)\|p_1(x)\big)+\sum_xp_0(x)\sum_vp_0(v|x)~\log{p_0(v|x)\over p_1(v|x)},
\ea\ee
}
where $p_i(v|x)=\sum_{u}p(u|x)\sum_yp(v|y,u)p_i(y)$. Once more, by applying Proposition \ref{region-proposition} in the optimization of (\ref{KL-XYX}), we get the following result, the proof of which is given in \cite{akofor-chen-it-paper}.

\begin{thm}\label{XYX_theorem}
 For the XYX process, the optimal decision rule at sensor X is $p_{\txt{opt}}(u|x)=I_{R_u}(x)$, with decision region given by
\bea
\label{omit2}&R_{u=1}=\left\{x:\sum_uI_{R_u}(x)A_uB_u>0\right\},\\
\label{X-thresholds}&~~~~~~~~ A_u={\beta^{(2)}_u-\al^{(2)}_u\over \beta^{(2)}_u(1-\beta^{(2)}_u)},~~~~B_u={\beta^{(2)}_1-\beta^{(2)}_0\over \al^{(2)}_1-\al^{(2)}_0}-\ld^{(2)}_u,
\eea
and the optimal decision rule at sensor Y is $p_{\txt{opt}}(v|y,u)=I_{R_{v|u}}(y)$, with decision regions given by
\bea
\label{omit3}& R_{v=1|u}=\bigg\{y:{p_1(y)\over p_0(y)}>\ld^{(2)}_u\bigg\},\hspace{0.5cm}&\\
\label{Y-thresholds}&~~\ld^{(2)}_u=\bigg(\log{\beta^{(2)}_u(1-\al^{(2)}_u)\over\al^{(2)}_u(1-\beta^{(2)}_u)}\bigg)\bigg/\bigg({\beta^{(2)}_u-\al^{(2)}_u\over \beta^{(2)}_u(1-\beta^{(2)}_u)}\bigg),
\eea
where { $\al^{(2)}_u=P_0(R_{v=1|u})$}, and {$\beta^{(2)}_u=P_1(R_{v=1|u})$}.
\end{thm}

In terms of the decision regions, the KL distance (\ref{KL-XYX}) can be expressed as
\bea
&& K^{\txt{XYX}}=K[x]+\sum_{u,v}P_0(R_u)~P_0(R_{v|u})\log{P_0(R_{v|u})\over P_1(R_{v|u})}\nn\\
\label{KXYX0}&&~~~~=K[x]+\al^{(1)}f(\al_1^{(2)},\beta_1^{(2)})+(1-\al^{(1)})f(\al_0^{(2)},\beta_0^{(2)}),
\eea
where~ $ K[x]=D(p_0(x)\|p_1(x)),$~  $\al^{(1)}$ is a constant independent of the thresholds, and
\bea
\label{KL-function}f(\al,\beta)=\al \log{\al \over \beta }+(1-\al )\log{1-\al \over 1-\beta }.
\eea
Thus we have the following theorem.

\begin{prp}\label{XYX-trivial-region}~
 The YX and XYX processes achieve identical $K[x,v]$. That is,
 \bea
 K^{\txt{YX}}_{\max}= K^{\txt{XYX}}_{\max}.
 \eea
\end{prp}
\begin{proof}
The KL distances achieved by the two fusion systems, $K^{\txt{YX}}$ from (\ref{KYX0}) and $K^{\txt{XYX}}$ from (\ref{KXYX0}), are respectively
\bea
\label{KYX}K^{\txt{YX}}&=&K[x]+f(\alpha,\beta),\\
\label{KXYX}K^{\txt{XYX}}&=&K[x]+\al^{(1)}f(\al_1^{(2)},\beta_1^{(2)})+(1-\al^{(1)})f(\al_0^{(2)},\beta_0^{(2)}),
\eea
where the function $f(\al,\beta)$ is defined in (\ref{KL-function}).

Let $\alpha^*$ and $\beta^*$ be the optimal values that maximize $f(\al,\beta)$ in $K^{\txt{YX}}$. Comparing  (\ref{omit1})-(\ref{YX-threshold}) and  (\ref{omit3})-(\ref{Y-thresholds}), it is apparent that the same $\alpha^*$ and $\beta^*$ also maximize both $f(\alpha_1^{(2)},\beta_1^{(2)})$ and $f(\alpha_0^{(2)},\beta_0^{(2)})$ in $K^{\txt{XYX}}$. This is so since for each value of $u$, the threshold dependence of the LRT using $y$ is identical to that used in the YX process. Thus, the optimal decision on $v$ at Y for the XYX process simply ignores the input from $u$, leading to identical LRTs for both values of $u$.

\end{proof}

\begin{figure}[H]
\centering
\includegraphics[width=10cm,height=7cm]{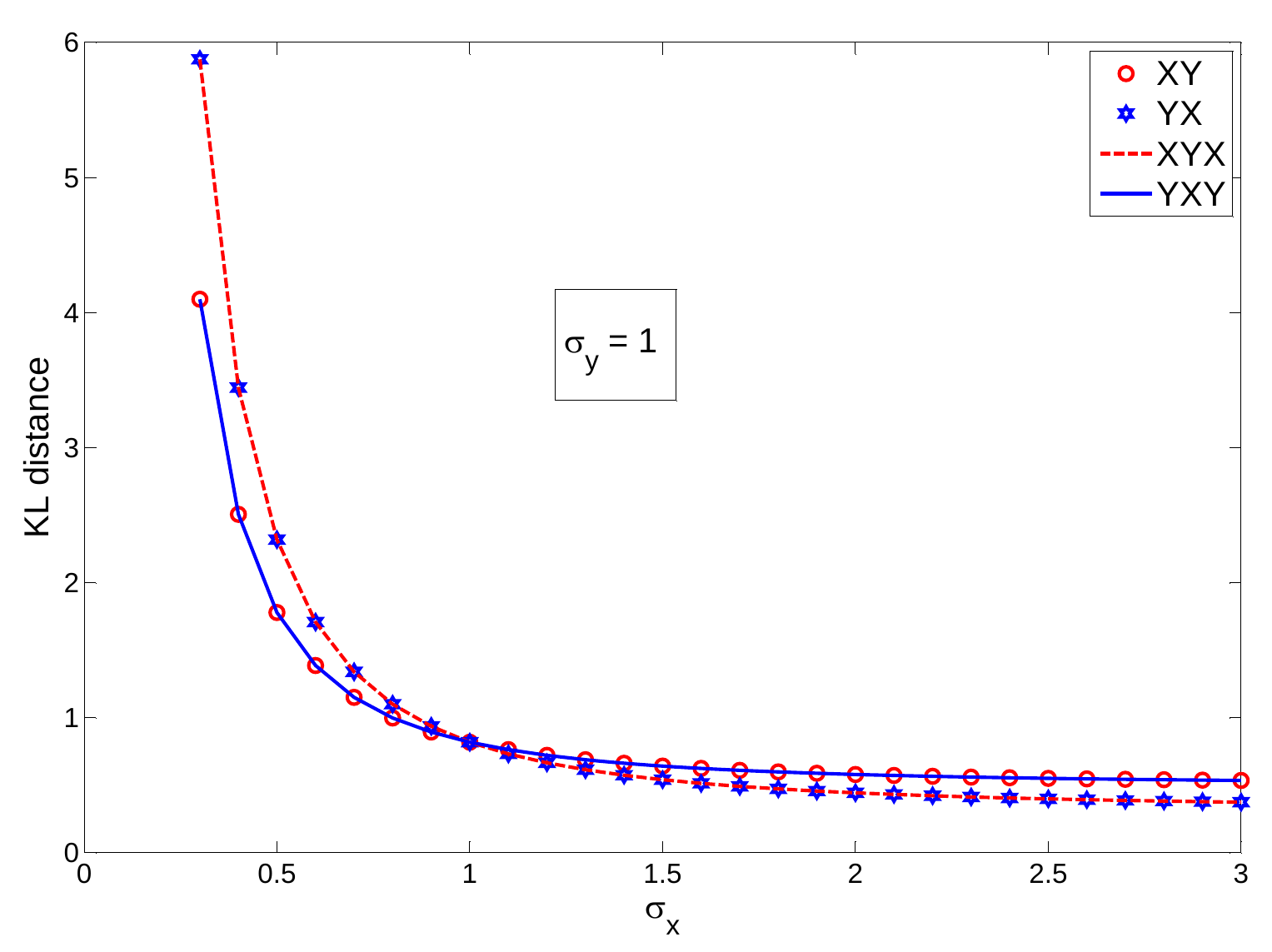}\\
\caption{Comparison of KL distances of one-way tandem fusion and interactive fusion with different communication directions.  For this plot, we fix $\sigma_y=1$ throughout while varying $\sigma_x$.}\label{kl-graph}
\end{figure}

Proposition \ref{XYX-trivial-region} holds for any probability distribution. The results for the constant signal in WGN under hypotheses (\ref{WGN}) are shown in Fig. \ref{kl-graph}, where the KL distances of YX and XYX processes coincide with each other. Also plotted are the KL distances of XY and YXY that also coincide with each other. An interesting observation from the plot is that the two sets of curves, each corresponding to making final decision at different nodes, intercept each other at the point when $\sigma_x=\sigma_y=1$. Thus for this example, it is always better to make the final decision at the sensor with better signal to noise ratio.

\section{Generalizations} \label{generalize}
In a two-sensor tandem network with a single round of interaction and $1$-bit sensor output, we have shown that interactive fusion may strictly improve the detection performance of fixed sample size NP test, but not the asymptotic performance of the large sample NP test. We now consider more realistic settings in which this result remains valid. These settings involve multiple round iterations, multiple sensors, and soft (i.e., multi-bit) sensor output.
\subsection*{Multiple-step memoryless interactive fusion (MIF)}\label{N-step process}
\begin{figure}
\centering
\scalebox{1.5}{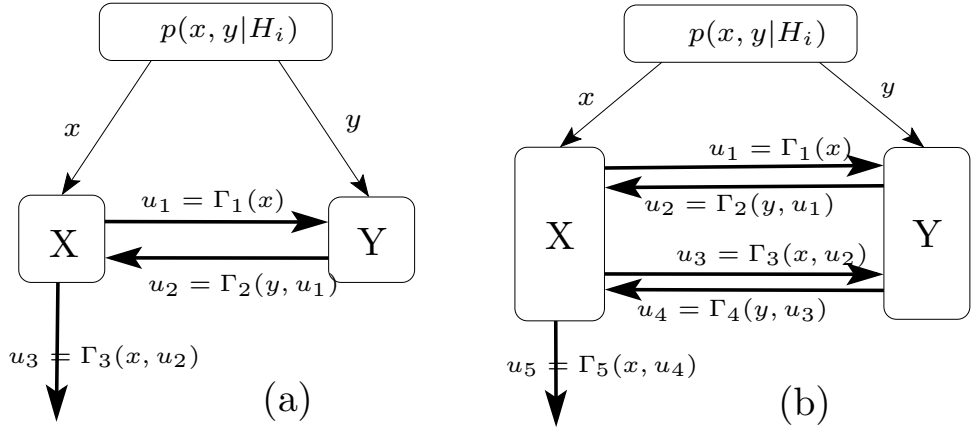}
\caption{Sample MIF processes: (a) $N=3$~ MIF (XYX process), and (b) $N=5$~ MIF (XYXYX process).}\label{fusion-diagrams-multiple-step}
\end{figure}

In multiple round interactive fusion, sensors exchange $1$-bit information repeatedly in $N>3$ steps. Without any restriction on memory, it is not difficult to see that interactive fusion may strictly outperform the one-way tandem fusion asymptotically. Indeed, for $N$ large enough, the performance of interactive fusion with memory will approach that of centralized detection.

However, there might be situations where the multiple round interactive fusion may proceed in a memoryless fashion, which we refer to as \emph{memoryless interactive fusion} (MIF). In this case, a sensor's decision at each step depends on its own observation and the latest decision (but not on earlier decisions) of the other sensor. For this memoryless processing model, we show that with respect to asymptotic detection performance, multiple-step interactive fusion still has no advantage over the one-way tandem fusion.

We begin with the expansion of the probability~ $p_i(u_N)=p(u_N|H_i)$~ of the final decision $u_N$. Denote any sequence $s_1,...,s_N$ by $s^N$. Let $u^N$ be the sequence of decisions in the MIF process XYXY$\cdots$YX involving two independent sensors X and Y, and let
\bea
\label{N-step-pdata}z^N\eqv(z_1,...,z_N)=(x,y,x,y,...,y,x)
\eea
be the corresponding sequence of observations used at processing, as shown in Fig. \ref{fusion-diagrams-multiple-step} for $N=3,5$. Here we assume $N$ is odd, thus, the decision process always starts with and ends at node $X$. Then because of the dependence structure $u_k=\Gamma_k(z_k,u_{k-1})$, $z_k=x$ when $k$ is odd, and $z_k=y$ when $k$ is even, we obtain
\be\ba
\label{odd-case1} &p_i(u_N)=\sum_{z^N,u^{N-1}}p_i(z^N)\prod_{k=1}^Np(u_k|z_k,u_{k-1})\\
&~~=\sum_{x,y,u^{N-1}}p_i(x,y)\prod_{r=1}^{(N-1)/2}\left[p(u_{2r-1}|x,u_{2r-2})p(u_{2r}|y,u_{2r-1})\right].
\ea\ee
Now using conditional independence, $p_i(x,y)=p_i(x)p_i(y)$, we get
\be
\label{odd-case2}   p_i(u_N|x)=\sum_{y,u^{N-1}}p_i(y)\prod_{r=1}^{(N-1)/2}\left[p(u_{2r-1}|x,u_{2r-2})p(u_{2r}|y,u_{2r-1})\right].
\ee
With this expansion of $p_i(u_N)$, the following lemma, based on Proposition \ref{region-proposition} and proved in \cite{akofor-chen-it-paper}, gives the peculiar nature of the resulting decision regions which are determined by an observation that is directly involved in the KL distance.

\begin{lmm}[Degenerate MIF decision regions]\label{N-step-lemma}
Let $u_N=\Gamma_N(x,u_{N-1})$ be the decision at the final step of a MIF process XYXY$\cdots$YX with independent observations $x$ and $y$. Let the objective function be given by the KL distance at the final step,
\bea
\label{N-step-KL-1} K[x,u_{N-1}]=\sum_{x,u_{N-1}}p_0(x,u_{N-1})\log{p_0(x,u_{N-1})\over p_1(x,u_{N-1})}.
\eea
Then all decision regions based on $x$, i.e., in the optimal decision rule
\be
p_{\txt{opt}}(u_{2r-1}|x,u_{2r-2})=I_{R_{u_{2r-1}|u_{2r-2}}}(x),\nn
\ee
with decisions ~$u_{2r-1}=\Gamma_{2r-1}(x,u_{2r-2}),~r=1,2,...,{N-1\over 2}$, have the following general form.
\bea
\label{trivial_regions} R_{u_{2r-1}=1|u_{2r-2}}=\left\{x:~\sum_{\al}I_{D_{\al}}(x) A_{\al,r,u_{2r-2}}>0\right\},
\eea
where $\{D_\al\}$ is a partition of the data space $\X$, and the coefficients $A_{\al,r,u_{2r-2}}$ are independent of $x$.
\end{lmm}

Notice that (\ref{omit2}) is a special case of (\ref{trivial_regions}). The following are some remarks about the degenerate decision regions (\ref{trivial_regions}):
\bit[leftmargin=0.5cm]
\item They depend on the distributions $p_0(x)$ and $p_1(x)$ only globally over $\X$, and not pointwise in $x$. Therefore given a single data point $x\in\X$, they cannot distinguish between $H_0$ and $H_1$.
\item They are determined by piecewise constant functions with discrete probability distributions, and hence cannot define independent continuous threshold parameters; i.e., they contain no independent thresholds.
\item They have piecewise constant probability; i.e., they have the same probability under both hypotheses.
\item Their only role is to reparametrize the thresholds of the other regions. Consequently, they cannot improve optimality of the KL distance (as the next lemma shows).
\eit
The following lemma, proved in \cite{akofor-chen-it-paper}, shows that the decision regions given by (\ref{trivial_regions}) are trivial in the sense that they do not participate in the decision process.
\begin{lmm}\label{covergence_lemma}
With respect to dependence on thresholds, the decision regions (\ref{trivial_regions}) of Lemma \ref{N-step-lemma} have piece-wise constant probabilities. Moreover, such probabilities play no role at convergence and therefore do not contribute to the overall decision process.
\end{lmm}

Therefore, careful analysis of the MIF process shows that whenever a sensor's data is explicitly summed over in the KL distance, the decision process becomes independent of that particular sensor's data. Since repetition of the decision process involving only one sensor's data cannot improve performance, it follows that MIF processing does not improve performance with respect to the KL distance.

\subsection*{Interactive fusion between the FC and multiple peripheral sensors}\label{multiple-sensors}
\begin{figure}
\centering
\scalebox{1.5}{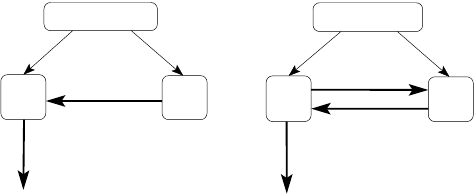}
\caption{(a) = one-way tandem fusion ($\vec{\txt{Y}}$X process), and (b) = interactive fusion (X$\vec{\txt{Y}}$X process)}\label{fusion-diagrams3-multiple-sensor}
\end{figure}

Consider our main setup in Fig. \ref{one-and-two-way-diagram} and maintain sensor X as the FC while replacing sensor $Y$ by K different sensors $\vec{\txt{Y}}=\{\txt{Y}_1,...,\txt{Y}_K\}$, with respective independent observations $\vec{y}=\{y_1,...,y_K\}$. The resulting system is shown in Fig. \ref{fusion-diagrams3-multiple-sensor}. For the $\vec{\txt{Y}}$X process, we have decisions $(\vec{v},w)\eqv(v_1,...,v_K,w)$ based on observations $(x,\vec{y})\eqv(x,y_1,...,y_K)$, where $\vec{v}=\vec{\delta}(\vec{y})=\big(\delta_1(y_1),...,\delta_K(y_K)\big)$ and $w=\rho(x,\vec{v})\eqv \rho(x,v_1,...,v_K)$. Similarly, for the X$\vec{\txt{Y}}$X process, the decisions $(u,\vec{v},w)\eqv(u,v_1,...,v_K,w)$ are based on observations $(x,\vec{y})\eqv(x,y_1,...,y_K)$, with $u=\gamma(x)$, $\vec{v}=\vec{\delta}(\vec{y},u)=\big(\delta_1(y_1,u),...,\delta_K(y_K,u)\big)$ and $w=\rho(x,\vec{v})\eqv \rho(x,v_1,...,v_K)$.

In the fixed sample size NP test with Lagrangian (\ref{NP-Lagrangian-0}), $p_i(w)$ is given by
{\footnotesize
\be\ba
p_i(w)=\sum_{x,{\vec{y}},{\vec{v}}}p(w|x,{\vec{v}})p({\vec{v}}|{\vec{y}})~p_i(x)p_i({\vec{y}})=\sum_{x,{\vec{y}},{\vec{v}}}p(w|x,{\vec{v}})~\prod_{k=1}^Kp(v_k|y_k)~p_i(x)~\prod_{k=1}^Kp_i(y_k)
\ea\ee
}
for the $\vec{\txt{Y}}$X process, and
{\small
\be\ba
&p_i(w)=\sum_{x,{\vec{v}},{\vec{y}},u}p(w|x,{\vec{v}})p({\vec{v}}|{\vec{y}},u)p(u|x)~p_i(x)p_i({\vec{y}})\\
&~~~~=\sum_{x,{\vec{v}},{\vec{y}},u}p(w|x,{\vec{v}})~\prod_{k=1}^Kp(v_k|y_k,u)~p(u|x)~p_i(x)~\prod_{k=1}^Kp_i(y_k)
\ea\ee
}
for the X$\vec{\txt{Y}}$X process.
It suffices to find the X$\vec{\txt{Y}}$X decision regions only since those for $\vec{\txt{Y}}$X can be deduced from them by simply deleting the first decision $u$. Using Proposition \ref{region-proposition} and following the same steps as in the proof of Theorem \ref{np-theorem}, we obtain the following. For the initial decision at X,~ $p_{\txt{opt}}(u|x)=I_{R_u}(x)$,~ with
\be\ba
R_{u=1}=\left\{x:~{\del^B S\over\del p(u=1|x)}>0\right\}=\left\{x:~{p_1(x)\over p_0(x)}Q(x)>\ld^{(1)}Q(x)\right\},
\ea\ee
where $\ld^{(1)}=\ld^{(0)}~{\prod_{k=1}^KP_0(R_{v_k=1|u=1})-\prod_{k=1}^KP_0(R_{v_k=1|u=0})\over \prod_{k=1}^KP_1(R_{v_k=1|u=1})-\prod_{k=1}^KP_1(R_{v_k=1|u=0})}$,\\ $Q(x)=-\sum_{\vec{v}}(-1)^{v_1+...+v_K}I_{R_{w=1|\vec{v}}}(x)$, while expressions for $\ld^{(0)}$ and the objective function $S$ are found in Appendix A of \cite{akofor-chen-it-paper}. For the decision at each $Y_k\in\{Y_1,...,Y_k\}$, we have ~$p_{\txt{opt}}(v_k|y_k,u)=I_{R_{v_k|u}}(y_k)$,~ with
\be\ba
 R_{v_k=1|u}=\left\{y_k:~{\del^B S\over\del p(v_k=1|y_k,u)}>0\right\}=\left\{y_k:~{p_1(y_k)\over p_0(y_k)}>\ld^{(2)}_{k,u}\right\},
\ea\ee
where
{\footnotesize
\be
\ld^{(2)}_{k,u}=\ld^{(0)}~{\sum_{{v\backslash v_k}}\left[P_0\left(R_{w=1|{v\backslash v_k},v_k=1}\cap R_u\right)-P_0\left(R_{w=1|{v\backslash v_k},v_k=0}\cap R_u\right)\right]\prod_{k'\neq k}P_0\left(R_{v_{k'}|u}\right) \over \sum_{{v\backslash v_k}}\left[P_1\left(R_{w=1|{v\backslash v_k},v_k=1}\cap R_u\right)-P_1\left(R_{w=1|{v\backslash v_k},v_k=0}\cap R_u\right)\right]\prod_{k'\neq k}P_1\left(R_{v_{k'}|u}\right) }.\nn
\ee
}
For the final decision at $X$,
{\small
\be\ba
\label{XYX-region-33} &R_{w=1|{\vec{v}}}=\left\{x:~{\del^B S\over\del p(w=1|x,{\vec{v}})}>0\right\}=\left\{x:~{p_1(x)\over p_0(x)}>\sum_u\ld_{{\vec{v}}u}^{(3)}~I_{R_u}(x)\right\},
\ea\ee
}
where { $\ld_{{\vec{v}}u}^{(3)}=\ld^{(0)}~{\prod_{k=1}^KP_0\left(R_{v_k|u}\right)\over \prod_{k=1}^K P_1\left(R_{v_k|u}\right)}$}.

\begin{figure}
\centering
\scalebox{1.5}{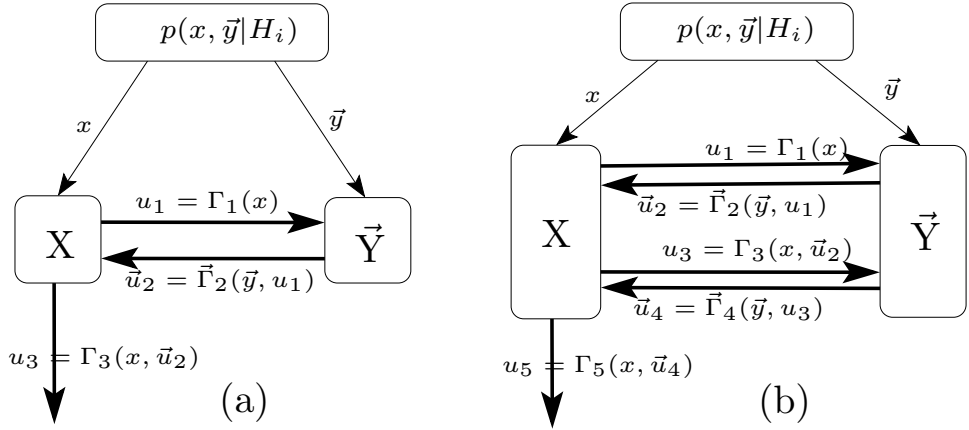}
\caption{Sample MIF processes with $K$ peripheral sensors: (a) $N=3$ multi-sensor MIF (X$\vec{\txt{Y}}$X process), and (b) $N=5$ milti-sensor MIF (X$\vec{\txt{Y}}$X$\vec{\txt{Y}}$X process).}\label{fusion-diagrams-multiple-sensor}
\label{fusion-diagrams-multiple-sensor2}
\end{figure}

Similarly, for the asymptotic NP test, the KL distance {\small $K^{\txt{X$\vec{\txt{Y}}$X}}\triangleq K[x,\vec{v}]=D\big(p_0(x,{\vec{v}})\|p_1(x,{\vec{v}})\big)$} can be expressed as
\bea
D\big(p_0(x)\|p_1(x)\big)+\sum_xp_0(x)\sum_{\vec{v}}p_0({\vec{v}}|x)~\log{p_0({\vec{v}}|x)\over p_1({\vec{v}}|x)},
\eea
where $p_i({\vec{v}}|x)=\sum_{u}p(u|x)\sum_yp({\vec{v}}|y,u)p_i(y)$. By the same steps as in the proof of Theorem \ref{XYX_theorem} for the X$\vec{\txt{Y}}$X process, the decision rule at sensor X is~ $p_{\txt{opt}}(u|x)=I_{R_u}(x)$,~ where
{\footnotesize
\be\ba
& R_{u=1}=\left\{x:{\del^B K[x,\vec{v}]\over\del p(u=1|x) }>0\right\}=\left\{x:\sum_uI_{R_u}(x)C_{u}>0\right\},\\
&C_{u}=\sum_{\vec{v}}\bigg(\sum_{u'}(-1)^{u'-1}P_0(R_{{\vec{v}}|u'})~\log{P_0(R_{{\vec{v}}|u})\over P_1(R_{{\vec{v}}|u})}-\sum_{u'}(-1)^{u'-1}P_1(R_{{\vec{v}}|u'}){P_0(R_{{\vec{v}}|u})\over P_1(R_{{\vec{v}}|u})}\bigg),
\ea\ee
}
and the pair of decision regions $R_{v=1|u}$ in the rule $p_{\txt{opt}}(v|y,u)=I_{R_{v|u}}(y)$ at sensor $Y$ has the following $K$ analogs corresponding to the sensors $\vec{\txt{Y}}$; for each $k=1,...,K,$~ we have $p_{\txt{opt}}(v_k|y_k,u)=I_{R_{v_k|u}}(y_k)$,~ with
{\small
\be\ba
& R_{v_k=1|u}=\bigg\{y_k:~{\del^B K[x,{\vec{v}}]\over\del p(v_k=1|y_k,u) }>0\bigg\}=\bigg\{y_k:{p_1(y_k)\over p_0(y_k)}>\ld^{(2)}_{ku}\bigg\},\hspace{0.5cm}&\\
&~~\ld^{(2)}_{ku}={\sum_{\vec{v}}(-1)^{v_k-1}\prod_{k'\neq k}P_0\left(R_{v_{k'}|u}\right)\log{P_0\left(R_{{\vec{v}}|u}\right)\over P_1\left(R_{{\vec{v}}|u}\right)}\over \sum_{\vec{v}}(-1)^{v_k-1}\prod_{k'\neq k}P_1\left(R_{v_{k'}|u}\right){P_0\left(R_{{\vec{v}}|u}\right)\over P_1\left(R_{{\vec{v}}|u}\right)}},
\ea\ee
}
where $P_i\left(R_{{\vec{v}}|u}\right) = \prod_{k=1}^KP_i\left(R_{v_k|u}\right)$.
The degenerate decision regions of Lemma \ref{N-step-lemma} maintain their form as well. Since the decision rules have the same critical features (including threshold structure), our conclusions hold for this more general setup as well. This includes the multiple-step MIF of Section \ref{N-step process} with $K$ peripheral sensors, shown in Fig. \ref{fusion-diagrams-multiple-sensor2} for $N=3,5$ steps.

\subsection*{Interactive fusion with soft sensor outputs}\label{multiple-bit}
We have shown in Section \ref{section_KL} that the $YX$ and $XYX$ processes have identical asymptotic detection performance when the output of each sensor is always binary. Now consider the other extreme case where the exchange of information is endowed with unlimited bandwidth. In that case, entire observations can be exchanged between sensors. Thus, both the YX and XYX processes again achieve exactly the same detection performance, namely, that of centralized detection. It remains to see if that is still the case for interactive fusion when soft information is exchanged, i.e., sensor outputs are of a multiple but finite number of bits.

Consider the case where $u$ and $v$ can take respectively $m$ and $l$ bits. Equivalently, we have $u\in \{0,1,...,2^m-1\}$ and $v\in \{0,1,...,2^l-1\}$. Improvement of performance of the fixed-sample NP test by interactive fusion is immediate by induction, since the single bit decisions are a particular case of the multiple bit decisions. Therefore we consider the situation for the asymptotic test.

By Proposition \ref{region-proposition}, the optimal decision rule at $X$ is~ $p_{\txt{opt}}(u|x)=I_{R_u}(x)$,~ with the decision regions given by
\be\ba
\label{XYX-region_4} &R_{u=k}=\bigcap_{k'\neq k}\left\{x:{\del K[x,v]\over\del p(u=k|x)}-{\del K[x,v]\over\del p(u=k'|x)}>0\right\},\\
&~~~~k=0,1,...,2^m-1,
\ea\ee
and the optimal rule at $Y$ is~ $p_{\txt{opt}}(v|y,u)=I_{R_{v|u}}(y)$,~ with the decision regions given by
\be\ba
\label{XYX-region_44}& R_{v=k|u}=\bigcap_{k'\neq k}\left\{y:{\del K[x,v]\over\del p(v=k|y,u)}-{\del K[x,v]\over\del p(v=k'|y,u)}>0\right\},\\
&~~~~k=0,1,...,2^l-1,
\ea\ee
where the objective function $K[x,v]$ is defined by (\ref{KL-XYX}). It is straightforward, with the help of equation (\ref{solution}), to verify that all of the critical features of our analysis remain unchanged. In particular, by the same procedure as in the proofs of Theorem \ref{XYX_theorem}  and Lemma \ref{N-step-lemma}, the decision regions $R_{u=k}$ in (\ref{XYX-region_4}) have the form
\be
R_{u=k}=\bigcap_{k'\neq k}\left\{x:\sum_{k''=0}^{2^m-1}I_{R_{u=k''}}(x)a_{k''k'}>0\right\},~~k=0,1,...,2^m-1,
\ee
which admits a piecewise constant probability. Hence multiple bit passing before the final decision does not alter our results.

\section*{Conclusion}
We have applied the decision theory developed in Chapter \ref{optimal-detection} to study two-sensor tandem fusion networks with conditionally independent observations. Based on the optimum decision structure in each case, we have shown that while interactive fusion improves performance of the fixed sample size NP test, it does not affect asymptotic performance as characterized by the error exponent of type II error.

Several extensions of the above result were considered. The lack of improvement in asymptotic detection performance of one-step interactive fusion was shown to extend to multiple-step memoryless interactive fusion. Furthermore, the result was shown to be valid in a more general setting where the FC simultaneously interacts with $K\geq 1$ independent sensors. Finally, the results we also shown to be true in the case of multi-bit sensor output.

\chapter{Optimal Fusion Architecture}\label{communication-direction}
\section{Introduction}
This section is based on \cite{akofor-chen-globalsip}, the references in which include \cite{zhu-akofor-chen-wcnc,song-et-al,song-et-al2,akofor-chen-icassp,akofor-chen-it-paper,Papastavrow&Athans:92AC}.

\begin{figure}[H]
\centering
\scalebox{1.5}{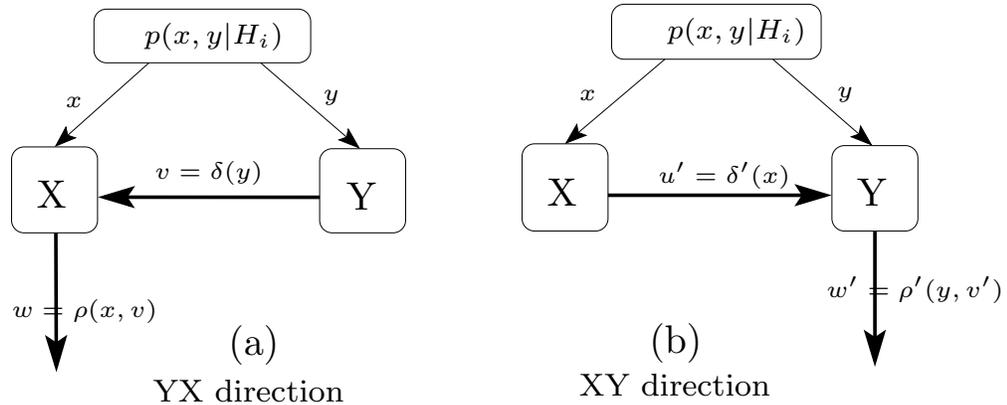}
\caption{Depiction of the communication directions}\label{comm-diagrams}
\end{figure}

Assume we wish to detect either a deterministic signal, or a Gaussian distributed random signal, in the presence of additive Gaussian noises using a two-sensor fusion system. That is, we have a distributed detection system with two sensors, one serving as a fusion center (FC) while the other as a peripheral sensor (PS) whose output is passed on to the FC for final decision making. As shown in Figure \ref{comm-diagrams}, a natural question about the preferred communication direction arises. What would be the optimal way of organizing the fusion system, i.e., which of the two sensors must serve as the FC for optimal detection performance?

We will show that for better detection performance at sufficiently low signal to noise ratio (SNR), the better sensor, i.e., the sensor with higher SNR, should serve as the fusion center. For the detection of a constant signal in additive Gaussian noises, it was also found in \cite{akofor-chen-icassp,zhu-akofor-chen-wcnc} that under the Neyman-Pearson and Bayesian criteria with conditionally independent observations, the sensor with lower SNR should serve as the FC. Note that these conclusions are false for general detection problems \cite{Papastavrow&Athans:92AC}.

In what follows, we first derive the optimal Bayesian test with dependent observations, which is valid for any continuous probability distribution that satisfies the assumptions of Proposition \ref{region-proposition}. Next, we consider the model with a random signal in additive Gaussian noises, and describe the topology of the resulting decision regions. Finally, we present computational results for low SNR.

\section{The optimal Bayesian test}\label{optimal-test}
Consider a system of two sensors X and Y, with two possible directions of communication as shown in Fig \ref{comm-diagrams}. In the YX direction, Fig. \ref{comm-diagrams} (a), sensor Y makes the first decision $v=\delta(y)$ with its observation $y$, and passes $v$ to sensor $X$. Sensor X, now equipped with $v$ and its own observation $x$, then makes the final decision $w=\rho(x,v)$. In the reverse direction XY, Fig. \ref{comm-diagrams} (b), sensor X makes the first decision $v'=\delta'(x)$, which is then used by sensor Y to make the final decision $w'=\rho'(y,v')$.

Without loss of generality, we will derive the test for the YX direction only. In addition, we consider only deterministic binary decisions since we are working under the assumptions of Proposition \ref{region-proposition}. The Bayesian cost function is given by
\bea
\label{bayes-cost}&& S[w]=\sum_{w,i}C_{wi}~p(w,H_i)=\sum_{w,i}C_{wi}\pi_i~p_i(w),
\eea
where $p_i(w)=\sum_{x,y,v}p(w|x,v)~p(v|y)~p_i(x,y)$, and $\pi_i=p(H_i)$.
We note that since we consider only deterministic decisions, after optimization $p_i(w)$ can be written as
\bea
p_i(w)=\sum_{x,y,v}I_{R_{w|v}}(x)~I_{R_v}(y)~p_i(x,y),
\eea
where $R_{w=1|v}$ are the decision regions at X and $R_{v=1}$ is the decision region at Y. We have the following result.

\begin{lmm}\label{region-lemma}
With simple binary hypotheses $H_i: (x,y)\sim p_i(x,y),~i=0,1,$ the optimal decision rule $p_{\txt{opt}}(v=1|y)=I_{R_{v=1}}(y),~p_{\txt{opt}}(w=1|x,v)=I_{R_{w=1|v}}(x),$ for the Bayesian decision problem with objective function (\ref{bayes-cost}) is given by the following decision regions. At Y,
{\small
\be\ba
\label{YX-region-at-Y}R_{v=1}=\left\{y:~{p_1(y)\over p_0(y)}{\sum_{x}[I_{R_{w=1|v=1}}(x)-I_{R_{w=1|v=0}}(x)]p_1(x|y)\over P_0(R_{w=1|v=1})-P_0(R_{w=1|v=0})}>\ld_0\right\},
\ea\ee
}and at X,
{\small
\be\ba
\label{YX-region-at-X} R_{w=1|v}=\left\{x:{p_1(x)\over p_0(x)}{\sum_{y}I_{R_v}(y)p_1(y|x)\over P_0(R_v)}> \ld_0\right\},
\ea\ee
}where $\ld_0={C_{10}-C_{00}\over C_{01}-C_{11}}{\pi_0\over\pi_1}$, and the summation $\sum_x,\sum_y$ over continuous random variables is by means of integration. For conditionally independent observations, the decision regions are given by simple likelihood ratio tests as follows:
{\small
\bea
&&\label{YX-region-at-Y-LRT}R_{v=1}=\left\{y:~{p_1(y)\over p_0(y)}>\ld^{(1)}\right\},\\
&&\label{YX-region-at-X-LRT} R_{w=1|v}=\left\{x:{p_1(x)\over p_0(x)}> \ld^{(2)}_v\right\},
\eea
}where $\ld^{(1)}=\ld_0{P_0(R_{w=1|v=1})-P_0(R_{w=1|v=0})\over P_1(R_{w=1|v=1})-P_1(R_{w=1|v=0})}$ and $\ld^{(2)}_v=\ld_0{P_0(R_v)\over P_1(R_v)}$.
\end{lmm}

\begin{proof}
By Proposition \ref{region-proposition}, the local decision regions are given by
{\small
\be\ba
&R_{v=1}=\left\{y:~{\del S[w]\over\del p(v=1|y)}<0\right\},\nn\\
&R_{w=1|v}=\left\{x:~{\del S[w]\over\del p(w=1|x,v)}<0\right\}.\nn
\ea\ee
}Due to the constraint $p(v=1|y)+p(v=0|y)=1$, we must use the differentiation rules
{\small
\be\ba
{\del p(v'|y')\over\del p(v|y)}=(-1)^{v'-v}\delta_{y'y},~~~~{\del p(w'|x',v')\over\del p(w|x,v)}=(-1)^{w'-w}\delta_{x'x}\delta_{v'v}.
\ea\ee
}Therefore, at Y
{\small
\be\ba
& {\del p_i(w)\over\del p(v=1|y)}=\sum_{x,v}(-1)^{v-1}p(w|x,v)~p_i(x,y)\\
&~~~~~~~~=\sum_{x,v}(-1)^{v-1}I_{R_{w|v}}(x)~p_i(x,y),\\
\label{YX-Y-derivatives}&~~\Ra~~\\
&{\del S[w]\over\del p(v=1|y)}= \sum_{w,i}C_{wi}\pi_i{\del p_i(w)\over\del p(v=1|y)}\\
&~~~~~~~~ =\sum_{w,i}C_{wi}\pi_i\sum_{x,v}(-1)^{v-1}I_{R_{w|v}}(x)~p_i(x,y).
\ea\ee
}Similarly, at X,
{\small
\be\ba
& {\del p_i(w)\over\del p(w=1|x,v)}=(-1)^{w-1}\sum_{y}p(v|y)~p_i(x,y)\\
&~~~~~~~~= (-1)^{w-1}\sum_{y}I_{R_v}(y)~p_i(x,y),\\
\label{YX-X-derivative}&~~\Ra~~\\
&{\del S[w]\over\del p(w=1|x,v)}=\sum_{w,i}C_{wi}\pi_i {\del p_i(w)\over\del p(w=1|x,v)}\\
&~~~~~~~~=\sum_{w,i}C_{wi}\pi_i(-1)^{w-1}\sum_{y}I_{R_v}(y)~p_i(x,y).
\ea\ee
}Straightforward simplification of (\ref{YX-Y-derivatives}) and (\ref{YX-X-derivative}) leads to (\ref{YX-region-at-Y}) and (\ref{YX-region-at-X}).

When the observations $x$ and $y$ are conditionally independent, i.e. $p_i(x,y)=p_i(x)p_i(y)$, then $p_i(x|y)=p_i(x)$ and $p_i(y|x)=p_i(y)$, in which case the decision regions (\ref{YX-region-at-Y}) and (\ref{YX-region-at-X}) reduce to (\ref{YX-region-at-Y-LRT}) and (\ref{YX-region-at-X-LRT}).

\end{proof}

The topologies of the decision regions (\ref{YX-region-at-Y}) and (\ref{YX-region-at-X}) are in general unknown for an arbitrary distribution, and they do not simplify to LRT's unless under very special circumstances. For the example with normal distributions that follows, we will be able to determine the topologies of the decision regions.

\section{Random signal in additive Gaussian noise}\label{example}
In this model, the sensor observations are given by
\bea
 x=s+z_1,~~~~y=s+z_2,
\eea
where
$z_1\sim N(0,\tau)$, $z_2\sim N(0,\ld)$, and the signal $s$ is determined by two hypotheses
\bea
H_0:~s=0,~~~~H_1:~s\sim N(\mu,\sigma_s^2).
\eea
Equivalently,
{\small
\be\ba
& H_0:~(x,y)\sim N\left(\left[
                    \begin{array}{c}
                      0 \\
                      0 \\
                    \end{array}
                  \right]
,\left[
   \begin{array}{cc}
     \tau & 0 \\
     0 & \ld \\
   \end{array}
 \right]
\right),\\
& H_1:~(x,y)\sim N\left(\left[
                    \begin{array}{c}
                      \mu \\
                      \mu \\
                    \end{array}
                  \right]
,\left[
   \begin{array}{cc}
     \sigma_s^2+\tau & \sigma_s^2 \\
     \sigma_s^2 & \sigma_s^2+\ld \\
   \end{array}
 \right]
\right).
\ea\ee
}

A very important parameter here is the signal variance $\sigma_s^2$. Notice that $\sigma_s^2$ determines the correlation coefficient of the bivariate normal distribution $p_1(x,y)=p(x,y|H_1)$. Thus, in the limit $\sigma_s^2\ra 0$, the observations become conditionally independent. Also, in the limit of large SNR, $\sigma_s\ra \infty$ with $\tau$ and $\ld$ finite, the YX and XY processes approach identical performance since the decision rule (while still not determined by simple LRT's) becomes independent of $\tau$ and $\ld$.

The above random signal model was considered in \cite{song-et-al} and \cite{song-et-al2}, where the decision of the first sensor Y was simply assumed to be based on a LRT, i.e.,
{\small
\be\ba
\label{model-Y-region-song}&R_{v=1}=\left\{y:~{p_1(y)\over p_0(y)}>\ld_0\right\},~~~~\ld_0={C_{10}-C_{00}\over C_{01}-C_{11}}{\pi_0\over\pi_1}.
\ea\ee
}The optimal decision rule, as derived in Theorem III.1 below, looks drastically different from the simple LRT. For the decision region at X however, our method (due to Lemma \ref{region-lemma}) gives the decision regions (\ref{model-X-region1}) and (\ref{model-X-region2}) which are of the same structure as those of \cite{song-et-al} and \cite{song-et-al2}. Nevertheless, the difference in the decision region at Y alone may lead to remarkably different conclusions about the preferred communication direction. This is because the analysis done in \cite{song-et-al} is no longer valid with our decision rule given by Theorem \ref{region-theorem-for-example}.

For simplicity, we assume that we have real samples $x,y\in\Real$. The proof of the following theorem uses an idea due to \cite{song-et-al} to determine the topologies of the decision regions.
\begin{thm}[]\label{region-theorem-for-example}
The decision regions for this model are the following. At Y, there are thresholds $T_v^\pm\in\Real,~T_v^-<T_v^+,$ such that
{\footnotesize
\be\ba
\label{model-Y-region}R_{v=1}=\left\{y:{\!p_1(y)\over p_0(y)}{Q({T_1^--\mu_1(y)\over \sigma_1})-Q({T_1^+-\mu_1(y)\over \sigma_1})-\left[Q({T_0^--\mu_1(y)\over \sigma_1})\!-\!Q({T_0^+-\mu_1(y)\over \sigma_1})\right]\over Q({T_1^-\over \sqrt{\tau}})\!-\!Q({T_1^+\over \sqrt{\tau}})\!-\!\left[Q({T_0^-\over \sqrt{\tau}})-Q({T_0^+\over \sqrt{\tau}})\right]}\!>\!\ld_0\right\},
\ea\ee
}where~ {\small $\mu_1(y)={y+\mu\ld/\sigma_s^2\over 1+\ld/\sigma_s^2}$}, {\small $\sigma^2_1=\tau+{\ld\over 1+\ld/\sigma_s^2}$}, and $\ld_0$ is as defined in Lemma \ref{region-lemma}.

At X, there are thresholds $t^\pm\in\Real,~t^-<t^+,$ such that
{\small
\bea
\label{model-X-region1}R_{w=1|v=1}=\left\{x:{p_1(x)\over p_0(x)}{1-[Q({t^--\mu_2(x)\over\sigma_2})-Q({t^+-\mu_2(x)\over\sigma_2})]\over 1-[Q({t^-\over\sqrt{\ld}})-Q({t^+\over\sqrt{\ld}})]}> \ld_0\right\},\\
\label{model-X-region2}R_{w=1|v=0}=\left\{x:{p_1(x)\over p_0(x)}{Q({t^--\mu_2(x)\over\sigma_2})-Q({t^+-\mu_2(x)\over\sigma_2})\over Q({t^-\over\sqrt{\ld}})-Q({t^+\over\sqrt{\ld}})}> \ld_0\right\},~~~~~~~~
\eea
}where {\small $\mu_2(x)={x+\mu\tau/\sigma_s^2\over 1+\tau/\sigma_s^2}$} and {\small $\sigma_2^2=\ld+{\tau\over 1+\tau/\sigma_s^2}$}.
\end{thm}
\begin{proof}
The general form of the decision regions follows from Lemma \ref{region-lemma}. From (\ref{YX-region-at-Y}) and (\ref{YX-region-at-X}), and following the proof of Lemma B.1, \cite{song-et-al}, let us define the following functions, which are the logarithms of the left hand sides of the defining inequalities of the decision regions.
{\small
\be\ba
&f(y)=\log{p_1(y)\over p_0(y)}+\log{\sum_{x}[I_{R_{w=1|v=1}}(x)-I_{R_{w=1|v=0}}(x)]p_1(x|y)\over P_0(R_{w=1|v=1})-P_0(R_{w=1|v=0})},\\
& g_v(x)=\log{p_1(x)\over p_0(x)}+\log{\sum_{y}I_{R_v}(y)~p_1(y|x)\over P_0(R_v)},\nn
\ea\ee
}where $p_0(y)=n(y|0,\ld)$, $p_1(y)=n(y|\mu,\ld)$, $p_0(x)=n(x|0,\tau)$, $p_1(x)=n(x|\mu,\tau)$, $p_1(x|y)=n(x|\mu_1(y),\sigma_1^2)$, and $p_1(y|x)=n(y|\mu_2(x),\sigma_2^2)$ are normal pdfs.

Observe that $f(y)$ and $g_v(x)$ are convex, as can be seen by verifying that $f''(y)\geq 0$ and $g''_v(x)\geq 0$. That is,
{\small
\be\ba
& f'(y)={y\over\ld}-{y-\mu\over \sigma_s^2+\ld}+{\sum_{x}\left({x-\mu'_1(y)\over\sigma_1^2}\right)[I_{R_{w=1|v=1}}(x)-I_{R_{w=1|v=0}}(x)]p_1(x|y)\over \sum_{x}[I_{R_{w=1|v=1}}(x)-I_{R_{w=1|v=0}}(x)]p_1(x|y)}\nn\\
&~~~~={y\over\ld}-{y-\mu\over \sigma_s^2+\ld}+\left\langle {x-\mu'_1(y)\over\sigma_1^2}\right\rangle_y,\nn\\
&~~\Ra~~\nn\\
&f''(y)={\sigma_s^2\over\ld(\sigma_s^2+\ld)}+\left\langle \left({x-\mu'_1(y)\over\sigma_1^2}-\left\langle {x-\mu'_1(y)\over\sigma_1^2}\right\rangle_y\right)^2\right\rangle_y\geq 0,\nn
\ea\ee
}where $\langle~\rangle_y$ denotes the self-evident conditional expectation involved, and
similarly,
{\small
\be\ba
& g'_v(x)={y\over\tau}-{y-\mu\over \sigma_s^2+\tau}+{\sum_y\left({y-\mu'_2(x)\over\sigma_2^2}\right)I_{R_v}(y)p_1(y|x)\over \sum_yI_{R_v}(y)p_1(y|x)}\nn\\
&~~~~={y\over\tau}-{y-\mu\over \sigma_s^2+\tau}+\left\langle {y-\mu'_2(x)\over\sigma_2^2}\right\rangle_x,\nn\\
&~~\Ra~~\nn\\
&g''_v(x)={\sigma_s^2\over\tau(\sigma_s^2+\tau)}+\left\langle \left({y-\mu'_2(x)\over\sigma_2^2}-\left\langle {y-\mu'_2(x)\over\sigma_2^2}\right\rangle_x\right)^2\right\rangle_x\geq 0.\nn
\ea\ee
}Therefore, the decision regions take the form
\bea
\label{form-of-regions} R_{v=1}=[t^-,t^+]^c,~~~~R_{w=1|v}=[T^-_v,T^+_v]^c,~~v=0,1,
\eea
where $t^\pm,~T^\pm_v$ are thresholds depending on $\ld_0$ and the distribution parameters $\{\mu,\tau,\ld,\sigma_s^2\}$, and $A^c$ denotes complement of the set $A$. Substituting (\ref{form-of-regions}) in (\ref{YX-region-at-Y}) and (\ref{YX-region-at-X}), we obtain the decision regions in terms of Q-functions as stated in the theorem.
\end{proof}

It is not difficult to see that as $\sigma_s^2\ra 0$, the decision regions (\ref{model-Y-region}), (\ref{model-X-region1}), and (\ref{model-X-region2}) are determined by simple LRT's as expected. Based on the derived topologies of the decision regions, we will now estimate and compare detection performance for the two directions, YX and XY.

\section{Computational results for low signal to noise ratio}\label{results}
Having determined the global nature of the decision regions $R_{v=1}=[t^-,t^+]^c$, $R_{w=1|v}=[T^-_v,T^+_v]^c$, we now compute the threshold values that minimize the cost function $S[w]$. The optimal value of the cost function is given by

{\small
\be\ba
& S_{\txt{opt}}[w]=\sum_{w,i}C_{wi}\pi_i\sum_{x,y,v}I_{R_{w|v}}(x)~I_{R_v}(y)~p_i(x,y)\nn\\
&~~=\pi_0\sum_{w}C_{w0}\sum_{x,y,v}I_{R_{w|v}}(x)~I_{R_v}(y)~p_0(x,y)\nn\\
&~~~~~~~~+\pi_1\sum_{w}C_{w1}\sum_{x,y,v}I_{R_{w|v}}(x)~I_{R_v}(y)~p_1(x,y)\nn\\
&~~=\pi_0\sum_{w}C_{w0}\sum_{v}p_0(R_{w|v})~p_0(R_v)+\pi_1\sum_{w}C_{w1}\sum_{x,y,v}I_{R_{w|v}}(x)~I_{R_v}(y)~p_1(x,y)\nn\\
&~~=\pi_0C_{00}+\pi_0(C_{10}-C_{00})\sum_{v}p_0(R_{w=1|v})~p_0(R_v)+\pi_1C_{01}\nn\\
&~~~~~~~~-\pi_1(C_{01}-C_{11})\sum_{x,y,v}I_{R_{w=1|v}}(x)~I_{R_v}(y)~p_1(x,y)\nn\\
&~~=\min_{T}~S(T),~~~~T=(t^-,t^+,T^-_0,T^-_1,T^+_0,T^+_1).
\ea\ee
}Consider the special but important case of minimizing the error probability (corresponding to $C_{00}=C_{11}=0,~C_{10}=C_{01}=1$). We have
{\small
\be\ba
&P_e[w]=\pi_0\sum_{v}p_0(R_{w=1|v}\times R_v)+\pi_1\sum_{v}p_1(R_{w=0|v}\times R_v)\nn\\
&~~=\pi_0\sum_{v}p_0(R_{w=1|v}\times R_v)+\pi_1-\pi_1\sum_{v}p_1(R_{w=1|v}\times R_v)]\nn\\
&~~=\pi_0[p_0(R_{w=1|v=1}\times R_{v=1})-p_0(R_{w=1|v=0}\times R_{v=1})\nn\\
&~~~~~~+p_0(R_{w=1|v=0})]+\pi_1-\pi_1[p_1(R_{w=1|v=1}\times R_{v=1})\nn\\
&~~~~~~-p_1(R_{w=1|v=0}\times R_{v=1})+p_1(R_{w=1|v=0})].
\ea\ee
}The inequalities ~ $t^-<t^+$, ~$T^-_0<T^+_0$,~ and ~$T^-_1<T^+_1$~ are constraints on the optimization problem.

\begin{figure}[H]
\begin{center}
   \includegraphics[width=10cm,height=7cm]{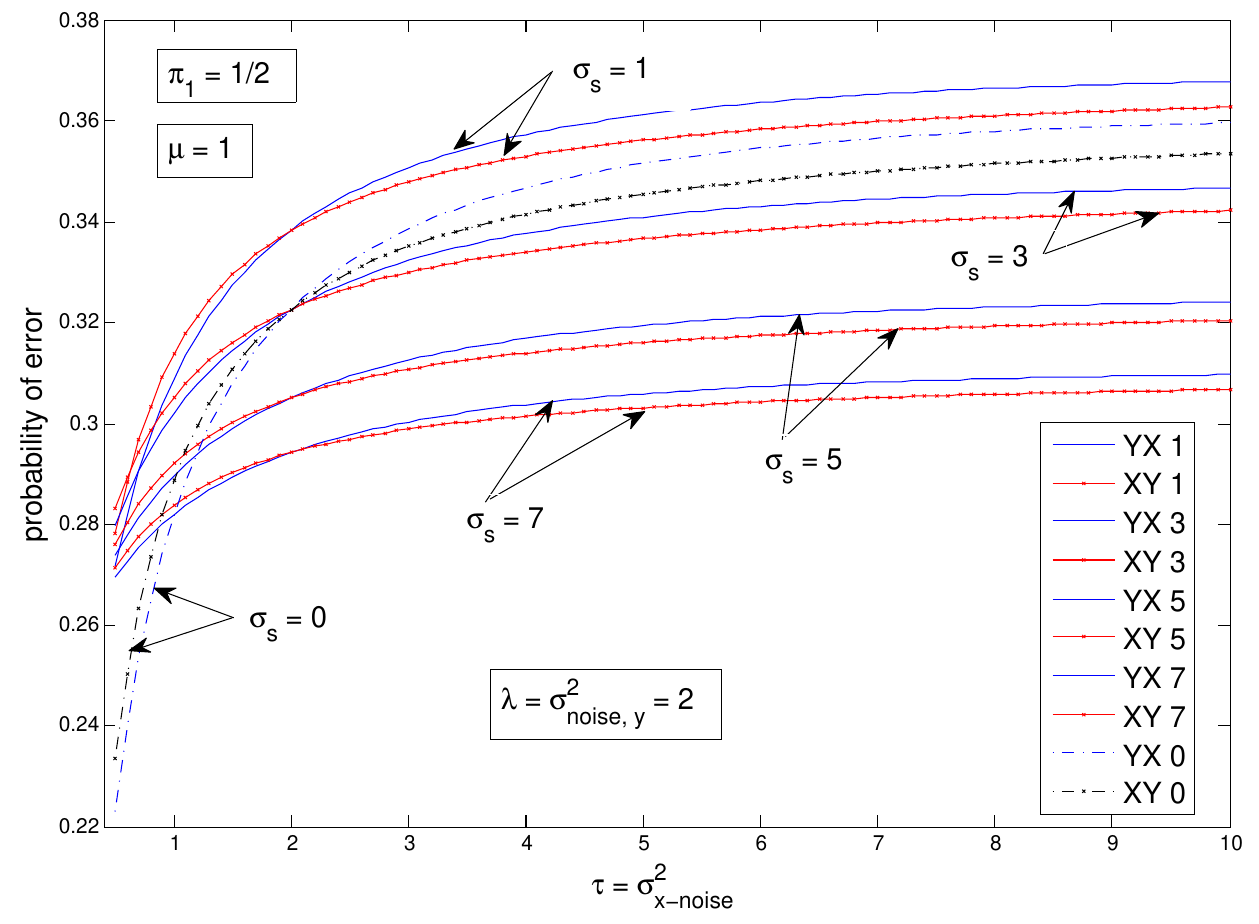}\\
  \caption{Performance of YX and XY directions}\label{low-sigma-s}
\end{center}
\end{figure}

Results for signal variance $\sigma_s^2$ near $0$ are given in Fig. \ref{low-sigma-s}. In this figure, the prior probability $\pi_1=1/2$, but the same behavior is observed for various values of $\pi_1\neq 1/2$. For various values of $\sigma_s=1,3,5,7$, the behavior is identical to that for $\sigma_s=0$ where the observations are conditionally independent. For each value of $\sigma_s$, the figure clearly shows that when $\tau<\ld$ (i.e., when X is the better sensor), YX performs better as it has smaller probability of error, meanwhile XY performs better when $\tau>\ld$ (i.e., when Y is the better sensor). Hence the better sensor is the preferred FC for $0\leq\sigma_s\leq 7$. There is no apparent reason why this result should not extend to large $\sigma_s$, especially as our decision rules are valid for all $\sigma_s$. However, the large $\sigma_s$ regime requires a more efficient code for numerically computing the optimal threshold values than what is available. Therefore we have deferred to future work on the large correlation regime.

\section*{Conclusion}\label{conclusion}
Based on the results of Chapter \ref{optimal-detection}, have derived the general form of the optimal Bayesian test for a two sensor tandem fusion network with dependent observations. Application of this test to a random signal in additive Gaussian noise shows that for small correlation strength, measured by the signal to noise ratio, the sensor with the cleaner data is still the preferred fusion center. This is in agreement with the case of independent observations.


\chapter{Detection over Acyclic Graphs}\label{acyclic-graph-detection}
\section{Introduction}
This section is based on ongoing work on distributed detection over acyclic directed graphs. It is a continuation of the discussion of sensor network rules initiated in Section \ref{sensor-network-rules}. In the Bayesian framework, and under mild assumptions on the dependence structure of the local sensor decisions, we obtain decision rules for arbitrary directed graphs. We also briefly study large sample asymptotic analysis, and derive associated Chernoff and Kullback information measures. These information measures are then used to define online sensor comparison with respect to asymptotic optimality in the network.

It is certainly the case that detection of a phenomenon by a network of distributed sensors is analytically more complex than detection of the same phenomenon by a centralized sensor network. This complexity arises for a number of reasons which include the following.
\bit
\item[(i)] The decision rules for different sensors are coupled in such a way that exact analysis may be impossible, especially when the observations of the sensors are conditionally dependent.

\item[(ii)] For a given sensor $X$ in the network, the number of decision thresholds for $X$ grows exponentially with the number of sensors that transmit their decisions to $X$. If $I_X$ is the number of sensors transmitting to $X$, then $X$ requires at least $2^{I_X}$ thresholds for binary processing.

\item[(iii)] When sensor $X$ is allowed to transmit more than one type of message (i.e., different sensors receive different messages from $X$), the number of decision channels connecting parents to offsprings of $X$ increases. If $O_X$ is the number of distinct messages $X$ is transmitting, this leads to a total of at least $2^{O_X}$ more computational steps in binary processing. Moreover, $X$ now requires at least $O_X\times 2^{I_X}$ thresholds for binary processing. An even more interesting consequence of the ability of sensors to transmit multiple messages is that decision processing over a closed path in a directed graph becomes nontrivial.
\eit

\begin{figure}[H]
\centering
\scalebox{1}{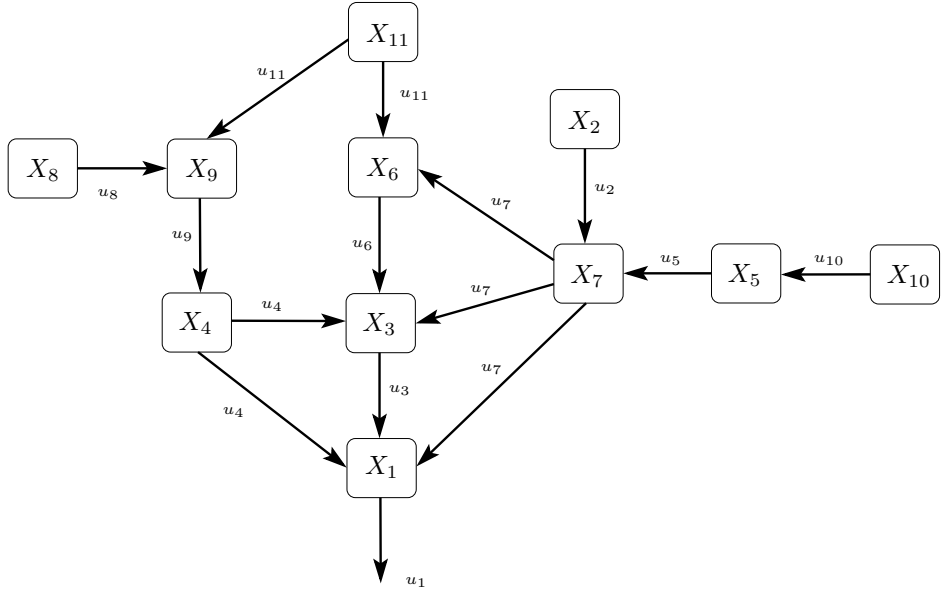} 
\caption{An 11-node acyclic directed graph}\label{acyclic-graph}
\end{figure}

If we relax (i) and (iii) by considering only conditionally independent observations, and also require that each sensor transmits only one type of message, then the decision rules even for the most elaborate graphical networks closely resemble those of simple networks. Under these assumptions, our main objective is to determine the optimal decision threshold structure in any sensor network for which desired communication and fusion patterns already exist in the form of directed graphs. See Fig \ref{acyclic-graph} for example.

Another objective will be to define online sensor comparison based on large sample asymptotic analysis. Let us refer to this type of comparison as AAR (\emph{asymptotic accuracy rate}, or \emph{Chernoff information}) comparison. AAR comparison can serve as an alternative to the usual ROC (receiver operating curve) sensor comparison in the large sample regime. AAR comparison works for M-ary decisions, and as we will see, it is robust with respect to Bayesian cost structure and prior probability.

We can also use AAR comparison to compare graphical network patterns in the following way. Suppose we are given $K$ sensors $X^n=(X_1,...,X_K)$. Let $\G(X^n,L)$ be the collection of all network patterns $G$ with at most $L$ direct links between any two sensors from $X^n$. For any given network pattern $G\in\G(X^n,L)$, define an online fusion center for $G$ to be the sensor $X_G$ in $G$ that has the largest AAR. Then it is natural to define the optimal network pattern to be the network pattern $G\in\G(X^n,L)$ such that the AAR of $X_G$ is larger than that of $X_{G'}$ for all $G'\in \G(X^n,L)$.

Distributed detection over directed graphs with at most one path between any two nodes was studied in \cite{TPKbI,TPKbII}, where some optimal control techniques were developed and optimal communication architectures were discussed. We consider distributed detection over arbitrary acyclic directed graph networks. Meanwhile the methods of \cite{TPKbI,TPKbII} penalized error at every root node in a tree, for simplicity, we consider only the costs at a single fusion center. Because we are interested mainly in the fused decision, all graphs are assumed to be connected. Bayesian detection over graph networks (feedback/memory included) was discussed in Section 4.5 of \cite{Varshney:book}, where decision rules were derived under binary hypothesis testing. We obtain decision rules under M-ary hypothesis testing but, for computational purposes, we likewise restrict to binary decisions. Because our main focus is on the graphical structure, we also assume the sensors take only one set of observations so that the decision process is \emph{static}, i.e., memory and feedback are not included in our discussion, although multiple processing steps can occur if we allow for cyclic communication paths in the network.

For an illustration of our notation, consider the 11-node directed  graph of Fig \ref{acyclic-graph}. Every arrow denotes a communication channel and the direction in which information must flow. Each node $X_j$ represents a sensor whose observation we denote by $x_j\in \X_j$, where the alphabet $\X_j$ is for simplicity taken to be the vector space $\Real^{n_j}$ for a positive integer $n_j$. The decision of sensor $X_j$ is denoted by $u_j=\gamma_j(x_j,\td{u}_j)$, where $\gamma_j$ is an integer-valued function and $\td{u}_j$ is the set of decisions of all parents of $X_j$. The decision $u_j$ is passed on to every offspring of $X_j$. For example, in Fig \ref{acyclic-graph}, sensor $X_7$'s decision $u_7=\gamma_7(x_7,\td{u}_7)$, where $\td{u}_7=\{u_2,u_5\}$, is transmitted to the sensors $X_1$, $X_3$, and $X_6$. By convention, if there are $n$ nodes in a graph, we consider $X_1$ to be the fusion center, after which the labeling and placement of all other nodes $X_2,X_3,....,X_n$ may come in any order.

In the work of \cite{Papastavrow&Athans:92AC} and related literature, e.g., Sections 4.4 and 4.5 of \cite{tsitsiklis-0}, two sensors are compared in terms of their stand-alone receiver operating characteristic (ROC) curves. ROC curve comparison is useful if the sensor quality is determined by more than one parameter, including variance and prior probability. It is also useful when each sensor has a separate local objective and the overall network objective is designed in such a way that it depends on these local objectives. A downside of ROC curve comparison is that it only works for binary decisions, since in that case the detection probability and false alarm probability are related by a single threshold parameter. In our analysis, we assume for simplicity that (i) the overall network objective is independent of any local objectives, (ii) every sensor uses the same prior probability, (iii) a sensor's quality is determined solely by the variance of its observation.

In what follows we first describe graphical networks and provide the threshold structure of decision rules for acyclic graph networks, including effects of communication over nonideal channels. This is followed by a short discussion on large sample analysis, from which the obtained Chernoff and Kullback information measures are then used to define online comparison among individual sensors and comparison of network patterns.

\emph{Notation:} Let us recall the notation introduced in Section \ref{optimal-detection-intro}. Sensors are labeled using upper case letters $X$, $Y$, $Z$, and so on. Random variables, as well as their values, are denoted by lower case letters $x$, $y$, $z$, $u$, $v$, $w$, etc. Also, we do not distinguish between summation and integration symbols, i.e., if $u$ is discrete and $x$ is continuous, we write $\sum_{u,x}=\sum_u\sum_x$, where $\sum_u$ denotes summation over $u$, and $\sum_x$ denotes integration over $x$. Similarly, if $u$, $v$ are discrete and $x$, $y$ are continuous, we write $\delta_{(u,x)(v,y)}=\delta_{uv}\delta_{xy}$, where $\delta_{uv}$ is the Kronecker delta, while $\delta_{xy}$ is the Dirac delta.

\section{Graphical networks}\label{graphical}
Recall that optimal Bayesian decision rules for acyclic graphs were derived in Section \ref{sensor-network-rules}, and presented in Theorem \ref{acyclic-prop}. The following observations about an acyclic graph network are what made that result possible.
\begin{enumerate}
\item For any given pair of sensors $X_i,X_j$ in a network, multiple directed paths between $X_i$ and $X_j$ do not lead to any intersections between different decision regions as long as these paths have the same sense, i.e., they do not form a loop.
\item Processing in a closed path can be nontrivial (i.e., distributed, or decentralized) only if a sensor on that path is allowed to send multiple messages.
\end{enumerate}
Therefore the simple network decision rules we have seen in the previous chapters, as well as their experimental implementation, can be readily extended to arbitrary directed graphs provided we make one more simplifying assumption about the sensor network, besides conditional independence of observations. The assumption is that every sensor passes the same message to its offsprings. This eliminates networks containing closed processing paths.

Note, however, that a sensor can of course be allowed (if necessary) to send multiple messages in a directed acyclic graph, and that the single-message decision rules we will obtain can easily be extended to multiple-message decision rules for any acyclic directed graph.

\subsection*{Acyclic directed graphs}\label{graphical-acyclic}
Consider a directed graph $G=(V,\rho)$, where $V=\{X_1,...,X_n\}$ are vertices,
\bea
V\times V=\{(X_i,X_j):i=1,...,n,~j=1,...,n\}\nn
\eea
is the set of placeholders (i.e., possibilities) for directed arrows connecting the vertices [thus $(X_i,X_j)$ represents the possibility of an arrow directed from $X_i$ to $X_j$], and the map
\bea
\rho:V\times V\ra\{0,1\},~(X_i,X_j)\mapsto a_{ij}=\rho(X_i,X_j)\nn
\eea
indicates the presence or absence of an arrow, i.e., $a_{ij}=1$ says an arrow pointing from $X_i$ to $X_j$ exists, while $a_{ij}=0$ means such an arrow does not exist. For computational purposes, the links in the graph are more conveniently written in matrix form: the matrix of $G$ is given by
\bea
M_G=\sum_{i=1}^n\sum_{j=1}^n\rho(X_i,X_j)e_{ij}=\sum_{i=1}^n\sum_{j=1}^na_{ij}e_{ij},\nn
\eea
where $e_{ij}$ is the matrix with $(i,j)$th entry $1$, and $0$ for all other entries. For example, the matrix of the graph in Fig \ref{acyclic-graph} is
{\footnotesize
\bea
\left[
  \begin{array}{ccccccccccc}
     \cdot & \cdot & \cdot  & \cdot  & \cdot  & \cdot  & \cdot  & \cdot & \cdot  & \cdot & \cdot \\
    \cdot & \cdot & \cdot  & \cdot  & \cdot  & \cdot  & 1 & \cdot & \cdot  & \cdot & \cdot \\
    1 & \cdot & \cdot  &  \cdot & \cdot  & \cdot  & \cdot  & \cdot &  \cdot & \cdot & \cdot \\
    1 & \cdot & 1 &  \cdot &  \cdot & \cdot  & \cdot  & \cdot &  \cdot & \cdot &  \cdot\\
     \cdot & \cdot &  \cdot &  \cdot &  \cdot & \cdot  & 1 & \cdot &  \cdot & \cdot & \cdot \\
    \cdot  & \cdot & 1 &  \cdot &  \cdot &  \cdot &  \cdot & \cdot &  \cdot & \cdot & \cdot \\
    1 & \cdot & 1 &  \cdot & \cdot  & 1 & \cdot  & \cdot & \cdot  & \cdot & \cdot \\
     \cdot & \cdot & \cdot  & \cdot  &  \cdot & \cdot  &  \cdot & \cdot & 1 & \cdot & \cdot \\
     \cdot & \cdot &  \cdot & 1 &  \cdot &  \cdot & \cdot  & \cdot &  \cdot & \cdot & \cdot \\
     \cdot & \cdot &  \cdot &  \cdot & 1 &  \cdot &  \cdot & \cdot &  \cdot & \cdot &  \cdot\\
     \cdot & \cdot & \cdot  &  \cdot & \cdot  & 1 &  \cdot &  \cdot& 1 & \cdot & \cdot \\
  \end{array}
\right]\nn
\eea
}
Notice that (i) the number of 1-entries is the number of arrows in the graph, (ii) the $1$'s in the $j$th column correspond to the parents of $X_j$, and (iii) the $1$'s in the $i$th row correspond to offsprings of $X_i$. The total number of thresholds is
\bea
2^3+2^0+2^3+2^1+2^1+2^2+2^2+2^0+2^2+2^0+2^0=36.\nn
\eea
These observations are crucial for developing a general optimization code for the graphical networks.

Denoting the observation and decision of $X_j$ by $x_j$ and $u_j$ respectively, the dependence structure of the decisions is given by
\bea
u_j=u_j(x_j,\td{u}_j),~~~~\td{u}_j=\{u_i:i=1,...,n,~a_{ij}=1\},\nn
\eea
where $\td{u}_j$ consists of the decisions of the parents of $X_j$ (i.e, all nodes $\td{X}_j=\{X_i:i=1,...,n,~a_{ij}=1\}$ bearing arrows into $X_j$) in the graph $G$.

If a node $X_k$ has $I_k$ parents (i.e., in-degree) then its number of thresholds is $2^{I_k}$. Since we have assumed that each node passes the same message to all of its offsprings, the total number of thresholds is
\bea
\sum_{k=1}^n2^{I_k}.\nn
\eea

As before, let $X_1$ be the fusion center. Also let $\vec{x}=(x_1,...,x_n)$ and $\vec{u}=(u_1,...,u_n)$, and consider the risk function
\be\ba
& S=\sum_{u_1,h}C_{u_1h}p(u_1,h)=\sum_{\vec{x},\vec{u},h}C_{u_1h}~\prod_{i=1}^np(u_i|x_i,\td{u}_i)~p_h(\vec{x})\pi_h\nn\\
&~~~~\sr{(a)}{=}\sum_{\vec{x},\vec{u},h}C_{u_1h}~\prod_{i=1}^np(u_i|x_i,\td{u}_i)~\prod_{i=1}^np_h(x_i)\pi_h,\nn
\ea\ee
where step (a) holds for conditionally independent observations.

\begin{thm}\label{acyclic-theorem}
The binary decision rule  for the network $\{X_1,...,X_n\}$ viewed as an acyclic directed graph are as follows. We have~ $p_{\txt{opt}}(u_k|x_k,\td{u}_k)=I_{R_{u_k|\td{u}_k}}(x_k)$,~ with decision regions
\be\ba
&R_{u_{k}=1|\td{u}_{k}}=\left\{x_{k}:{\del^BS\over\del p_{\txt{opt}}(u_{k}=1|x_{k},\td{u}_{k})}<0\right\}=\left\{x_{k}:{p_1(x_{k})\over p_0(x_{k})}>\ld^{(k)}_{\td{u}_{k}}\right\},\nn\\
\ea\ee
where ~~$\ld_{\td{u}_1}^{(1)}=\ld~{\sum_{\vec{u}\backslash(u_1,\td{u}_1)}\prod_{i\neq 1}p_0(R_{u_{i}|\td{u}_{i}})\over\sum_{\vec{u}\backslash(u_1,\td{u}_1)}\prod_{i\neq 1}p_1(R_{u_{i}|\td{u}_{i}})}$,~~ $\ld={(C_{10}-C_{00})(1-\pi)\over(C_{01}-C_{11})\pi}$,~ and
{\small
\bea
\ld_{\td{u}_k}^{(k)}=\ld~{\sum\limits_{\vec{u}\backslash\{u_1,\td{u}_{k}\}}(-1)^{u_{k}-1}p_0(R_{u_1=1|\td{u}_1})\prod\limits_{i\not\in\{1,k\}}p_0(R_{u_{i}|\td{u}_{i}})\over \sum\limits_{\vec{u}\backslash\{u_1,\td{u}_{k}\}}(-1)^{u_{k}-1}p_1(R_{u_1=1|\td{u}_1})\prod\limits_{i\not\in\{1,k\}}p_1(R_{u_{i}|\td{u}_{i}})},~~~~\txt{for}~~k\in \{2,3,...,n\}.\nn
\eea
}

The optimal value of the risk function is
{\footnotesize
\be\ba
\label{acyclic-graph-riskeq}& S_{\txt{opt}}=\sum_{\vec{x},\vec{u},h}C_{u_1h}~\prod_{i=1}^nI_{R_{u_{i}|\td{u}_{i}}}(x_{i})~\prod_{i=1}^np_h(x_{i})~\pi_h=\sum_{\vec{u},h}C_{u_1h}~\prod_{i=1}^np_h(R_{u_{i}|\td{u}_{i}})~\pi_h\\
&~~~~\sr{(a)}{=}\pi\left[1-\sum_{\vec{u}\backslash u_1}p_1(R_{u_1=1|\td{u}_1})\prod_{i\neq 1}p_1(R_{u_{i}|\td{u}_{i}})\right]+(1-\pi)\sum_{\vec{u}\backslash u_1}p_0(R_{u_1=1|\td{u}_1})\prod_{i\neq 1}p_0(R_{u_{i}|\td{u}_{i}}),
\ea\ee
}
where step (a) holds for 0-1 cost.
\end{thm}
\begin{proof}
This is the result of Theorem \ref{acyclic-prop} simplified for binary decisions.
\end{proof}

\begin{figure}[H]
\centering
\scalebox{1}{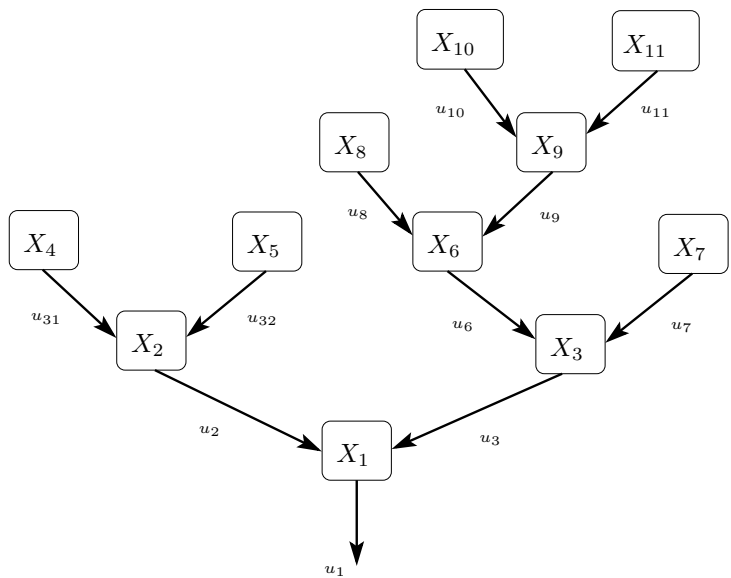}
\caption{An 11-node binary tree network}\label{binary-tree}
\end{figure}

\begin{figure}[H]
\centering
\scalebox{1}{\includegraphics[width=10cm,height=7cm]{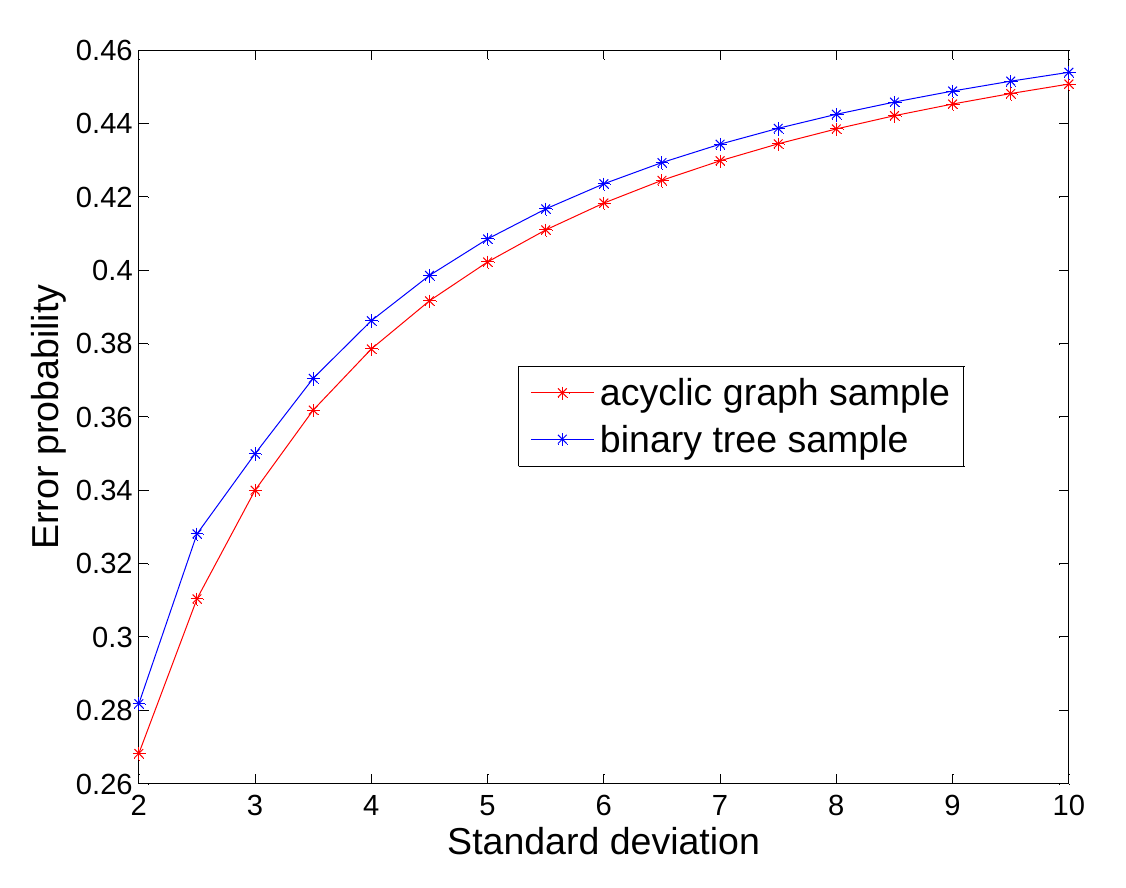}}\\
\caption{Relative performance of the sample acyclic graph of Fig \ref{acyclic-graph} and the sample binary tree of Fig \ref{binary-tree}. The graphs each contain 11 sensors but the acyclic graph has greater connectivity and thus performs better.}\label{sample_graphs}
\end{figure}

For a numerical outcome of Theorem \ref{acyclic-theorem}, consider the following example.
\begin{example*}\label{gaussian-example}
Let the observations of the sensors $(X_1,...,X_n)$ be given by
\bea
\label{ge-eq1}x_k=s+b_k,~~~~k=1,...,n,\nn
\eea
where $b_k\sim N(0,\sigma_k^2)$ and the hypotheses are
\bea
\label{ge-eq2}H_0:s=0,~~~~H_1:s=1.\nn
\eea
\end{example*}
For the above example, Fig \ref{sample_graphs} displays the relative performance of the acyclic graph in Fig \ref{acyclic-graph} and the binary tree in Fig \ref{binary-tree}. For this numerical computation, we have assumed the sensors are identical, i.e., $\sigma_k=\sigma_1$ for all $k$.

We also have the following result, which has been obtained in certain fusion settings by \cite{chen-willett,chen-chen-varshney}.
\begin{thm}[Nonideal channel effects]\label{nonideal-channel}
Let $X_1,X_2,\cdots,X_n$ be a sensor network in the form of an acyclic directed graph. Then the Bayesian decision rules for the network have the same form as in Theorem \ref{acyclic-theorem}, except that everywhere, we replace $p_h(R_{u_k|\td{u}_k})$ with $\sum_{\td{u}_k'}p_h(R_{u_k|\td{u}_k'})p(\td{u}_k'|\td{u}_k)$, where $\left[p(\td{u}_k'|\td{u}_k)\right]$ is a $2^{I_k}\times 2^{I_k}$ multi-channel transition matrix, with $I_k$ the number of parents of $X_k$. In particular, the objective function (\ref{acyclic-graph-riskeq}) becomes
\be\ba
\label{acyclic-graph-riskeq2}& S_{\txt{opt}}=\sum_{\vec{u},\vec{\td{u}}',h}\pi_hC_{u_1h}~\prod_{k=1}^np_h(R_{u_k|\td{u}_k'})~p(\td{u}_k'|\td{u}_k).
\ea\ee

\end{thm}
\begin{proof}
If $X_j$ is transmitting its decision $u_j$ to $X_i$ through a channel
\bea
g_{ji}:u_j\mapsto u'_j=g_{ji}(u_j),\nn
\eea
where $g_{ji}$ is a random function independent of the observations,
then $u'_j=g_{ij}(u_j)\in\td{u}'_i$. The objective function can be expanded as
\be\ba
&S=\sum_{u_1,h}C_{u_1h}p(u_1,h)=\sum_{\vec{u},\vec{\td{u}}',\vec{x},h}C_{u_1h}p(\vec{u},\vec{x},\vec{\td{u}}',h)\nn\\
&~~=\sum_{\vec{u},\vec{\td{u}}',\vec{x},h}\pi_hC_{u_1h}~\prod_{k=1}^np(u_k|x_k,\td{u}_k')~p(\td{u}_k'|\td{u}_k)~p_h(\vec{x}),\nn
\ea\ee
where $\vec{u}=(u_1,...,u_n)$ are the decisions of the sensors, meanwhile $\vec{\td{u}}'=(\td{u}_1',...,\td{u}_n')$ are the messages received by the sensors, i.e., $\td{u}'_k$ consists of the messages channeled to $X_k$ by its parents. The decisions have the dependence structure $u_k=u_k(x_k,\td{u}_k')$. Thus a straightforward application of the decision procedure developed in Section \ref{sensor-network-rules} yields the desired result, and in particular, (\ref{acyclic-graph-riskeq2}) holds.
\end{proof}

\subsection*{On directed graphs with cycles}\label{graphical-cyclic}
\begin{figure}[H]
\centering
\scalebox{0.7}{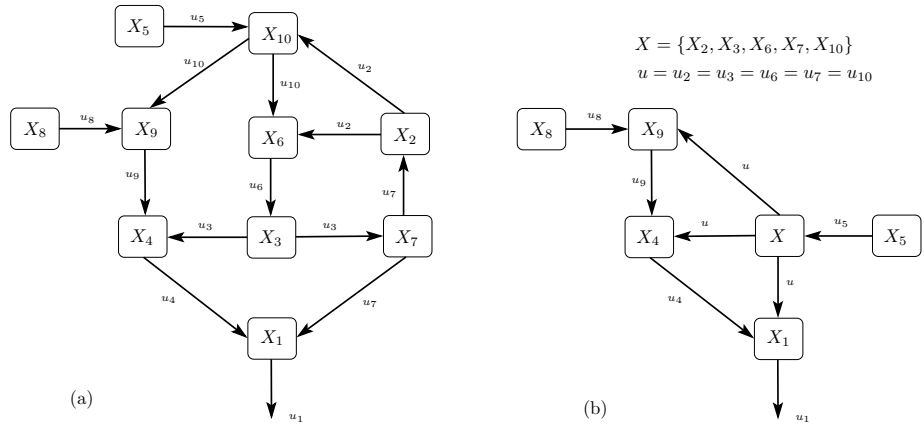}
\caption{A 10-node directed graph containing cycles. The graphs (a) and (b) are equivalent.}\label{cyclic-graph}
\end{figure}

Let us again assume that every sensor passes the same message to its offsprings. Then without any further constraint on the \emph{dependence structure} of the decisions (\emph{as determined by the arrows in the graph}), a closed path in a graph is trivial in the sense that it is equivalent to a centralized sub-collection of sensors. The situation may be described more precisely as follows.

\begin{lmm}\label{reduction-lemma}
In a graphical sensor network, suppose a closed path consists of $r$ sensors $X_{i_1},X_{i_2},...,X_{i_r}$, with respective observations $x_{i_1},x_{i_2},...,x_{i_r}$. Suppose further that
\bit
\item the dependence structure of the decisions of the sensors is determined only by the directed arrows in the graph, and
\item every sensor in the network passes the same message to its offsprings.
\eit
Then the subnetwork $Y=\{X_{i_1},X_{i_2},...,X_{i_r}\}$ behaves like a centralized sensor system with all observations $y=\{x_{i_1},x_{i_2},...,x_{i_r}\}$ available at the same location.
\end{lmm}
\begin{proof}
If no constraint (apart from that imposed by the arrows) is placed on the dependence struture of the decisions, then in particular, there is no \emph{timing constraint}. Thus information will continue to flow in the closed path until the decision of very sensor has achieved maximum performance, which is \emph{predicted} to be the performance of a centralized network formed by the $r$ sensors.
\end{proof}

The above discussion shows that directed graphs containing cycles are equivalent to acyclic directed graphs, unless there are extra communication constraints to render the cyclic processing nontrivial.

\section{Large sample asymptotic analysis}\label{large-sample}
Fix $k\in \{1,...,n\}$. Let $X_k$ make a $T$-observations sample, {\small $x_k^T=(x_{k,1},...,x_{k,t},...,x_{k,T})$}, and make a corresponding sequence of decisions
\be
u_k^T=u_k^T(x_k^T,\td{u}_k^{T-1})=(u_{k,1},...,u_{k,t},...,u_{k,T}).\nn
\ee
The decisions are made in a \emph{sequential} manner, i.e., a decision is made after each sample. Thus for each $t\in \{1,...,T\}$,
\bea
u_{k,t}=u_{k,t}(x_k^t,\td{u}_k^{t-1})\in\{0,1\}^{r_k}\simeq \{1,2,...,2^{r_k}\}
\eea
is an $r_k$-bit random variable. Here, $r_k$ is sensor $X_k$'s transmission rate (in bits per sample).

Let $M_k$ be the size of a linear indexing set for the value set of the decision sequence $u_k^T$, i.e., as a random variable, $u_k^T$ can take on $M_k$ possible values simply labeled as
$\{u_k^T(m_k):m_k=1,...,M_k\}=\{u_k^T(1),...,u_k^T(M_k)\}$. Then the number of bits per sample $r_k$ can be written as
\bea
r_k={1\over T}\log_2 M_k~~~~\txt{or}~~~~M_k=2^{r_kT}.\nn
\eea
Thus for each $m_k\in \{1,...,M_k\}$, we have
\be\ba
&u_k^T(m_k)\in \left(\{0,1\}^{r_k}\right)^T=\{0,1\}^{Tr_k}\simeq \{1,2,...,2^{r_k}\}^T\nn\\
&~~~\simeq\{1,2,\cdots,2^{Tr_k}\}=\{1,2,\cdots,M_k\},\nn\\
&~\Ra~~u_k^T:(x_k^T,\td{u}_k^{T-1})\sr{\simeq}{\longmapsto} m_k\in\{1,2,\cdots,M_k\},\nn
\ea\ee
where all powers of sets are Cartesian, and $\simeq$ denotes equivalence with respect to cardinality (i.e., $A\simeq B$ means ``the sets $A$ and $B$ have the same number of elements''). Furthermore, we may wish to impose a rate constraint
\bea
{1\over T}\log_2M_k=r_k\leq R_k,
\eea
where $R_k$ is a fixed bit rate representing the maximum number of bits per sample that $X_k$ can transmit to its offsprings.

For simplicity, let the bit rate of the fusion center $X_1$ be set at $r_1=1$. Then the error probabilities (based on the final decision $u_{1,T}$) at the fusion center are
\be\ba
&\al_T=p(u_{1,T}=1|H_0)=p_0(u_{1,T}=1),~~~~\txt{(False alarm)},\nn\\
&\beta_T=p(u_{1,T}=0|H_1)=p_1(u_{1,T}=0),~~~~\txt{(Missed detection)}.\nn
\ea\ee
With a constraint $\al_T\leq \vep_T$ on the false alarm, we define the error exponent for missed detection as
\be\ba
b(\vep,R_1,...,R_n)=\liminf_{T\ra\infty}~-{1\over T}\log\left(\inf_{A_T}~\beta_T\right),
\ea\ee
where
\be\ba
&A_T=\bigg\{\left\{p\big(u_k^T|x_k^T,\td{u}_k^{T-1}\big)\right\}_k:\{r_k\leq R_k\}_k,~\al_T\leq \vep_T\bigg\}\nn\\
&~~~~\simeq \bigg\{\left\{p\big(m_k|x_k^T,\td{m}_k\big)\right\}_k:\{r_k\leq R_k\}_k,~\al_T\leq \vep_T\bigg\}\nn
\ea\ee
is the set of possible decision strategies. That is, we have
\bea
\beta_T\approx e^{-b(\vep,R_1,...,R_n)T}~~~~\txt{for large}~~T.\nn
\eea

In the Bayesian formulation, the risk function at $X_1$ has the expansion
\be\ba
\label{sequential-bayes-objective}&S_T=\sum_{u_{1,T},h}C_{u_{1,T}h}~p(u_{1,T},h)=\sum_{u_{1,T},h}\pi_hC_{u_{1,T}h}~p(u_{1,T}|h)\\
&~~=\sum_{u_{1,T},h}\pi_hC_{u_{1,T}h}~p_h(u_{1,T})=\sum_{\vec{u}^T,\vec{x}^T,h}\pi_hC_{u_{1,T}h}~p_h(\vec{u}^T,\vec{x}^T)\\
&~~=\sum_{\vec{u}^T,\vec{x}^T,h}\pi_hC_{u_{1,T}h}\prod_{k=1}^np\left(u_k^T|x_k^T,\td{u}_k^{T-1}\right)~p_h\left(\vec{x}^T\right)\\
&~~=\sum_{\vec{m},\vec{x}^T,h}\pi_hC_{u_{1,T}h}\prod_{k=1}^np\left(m_k|x_k^T,\td{m}_k\right)~p_h\left(\vec{x}^T\right)\\
&~~=\sum_{\vec{m},h}~\sum_{\left\{x_k^T\in R_{m_k|\td{m}_k}\right\}_k}\pi_hC_{u_{1,T}h}~p_h\left(\vec{x}^T\right),
\ea\ee
where $\vec{u}^T=(u_1^T,\cdots,u_n^T)$, $\vec{x}^T=(x_1^T,\cdots,x_n^T)$, and $\vec{m}=(m_1,...,m_n)$. The main challenge is to design the decision regions $\{R_{m_k|\td{m}_k}:k=1,...,n\}$ in an optimal way. For large $T$, the law of large numbers says these regions can be approximated by sets of sequences $x_k^T$ whose \emph{empirical distributions}
\bea
p(a|x_k^T)={|\{t:~x_{k,t}=a\}|\over T}={\txt{cardinality}\{t:~x_{k,t}=a\}\over T},~~~~a\in\X_k,\nn
\eea
are close (in a certain sense) to the marginal distributions for $p_h\left(\vec{x}^T\right)$. Using this reasoning, it is possible to derive bounds on the error exponent $b(\vep,R_1,...,R_n)$.

We can use a likelihood ratio quantizer to provide an upper bound for $S_T$ as follows. Let

{\small
\be\ba
R^{(LR)}_{m_k|\td{m}_k}=\left\{x_k^T:\ld^{(k)}_{m_k|\td{m}_k}<l_h(x_k^T)<\ld^{(k)}_{m_k+1|\td{m}_k}\right\},\nn
\ea\ee
}where~ $l_h(x_k^T)={\max_{h'}p_{h'}(x_k^T)\over p_h(x_k^T)}$, and the thresholds $\ld$ depend on $T$. Then for any $\vep_k>0$, we can choose $T$ large enough so that
\be\ba
&S_T\leq \sum_{u_1,h}\pi_hC_{u_{1,T}h}\prod_{k=1}^ne^{-T~\left[L^{(k)}_h-\vep_k-r_k\right]},\nn\\
&L^{(k)}_h=\lim_{T\ra\infty}\max_{p(m_k)}~-{1\over T}\sum_{m_k} p(m_k)~\log~p_h\left(\ld^{(k)}_{m_k|\td{m}_k}<l_h(x_k^T)<\ld^{(k)}_{m_k+1|\td{m}_k}\right).\nn
\ea\ee
 Thus, if
\be
r_k<L^{(k)}_h-\vep_k,\nn
\ee
then $S_T\ra 0$ as $T\ra \infty$. The quantity $L^{(k)}_h$ is an example of an information measure at $X_k$, since it is a function of the \emph{likelihood ratio statistic} $l_h(x_k^T)$ and determines \emph{estimation accuracy} to some extent. Concrete examples of information measures are the following.

\section{Chernoff information and Kullback-Leibler distance}\label{CK-info-section}
The discussion here is related to the discussion in Section 11.8 of \cite{cover-thomas}. The Chernoff information is derived as follows (Note here that neither the hypothesis $h$ nor the decision $u_{1,T}$ needs to be binary). With the usual optimization variables $p(u_k|x_k^T,\td{u}_k^T)$ in mind, we have
\be\ba
& \min S_T=\min\sum_{u_{1,T},h}C_{u_{1,T}h}~p(u_{1,T},h)\nn\\
&~=\min\sum_{u_{1,T},h,x_1^T,\td{u}_1^T}C_{u_{1,T}h}~p(u_{1,T}|x_1^T,\td{u}_1^T)p_h(x_1^T,\td{u}_1^T)\pi_h\nn\\
&~\sr{(s1)}{=}\sum_{u_{1,T},h,x_1^T,\td{u}_1^T}C_{u_{1,T}h}~I_{R_{u_{1,T}|\td{u}_1^T}}(x_1^T)p_h(x_1^T,\td{u}_1^T)\pi_h\nn\\
&~\sr{(s2)}{=}\sum_{u_{1,T},h,x_1^T,\td{u}_1^T}C_{u_{1,T}h}~I_{R_{u_{1,T}}}(x_1^T,\td{u}_1^T)p_h(x_1^T,\td{u}_1^T)\pi_h\nn\\
&~\sr{(s3)}{=}\sum_{x_1^T,\td{u}_1^T}\min_{u_{1,T}}\sum_h\pi_hC_{u_{1,T}h}~p_h(x_1^T,\td{u}_1^T)\nn\\
&~\sr{(s4)}{=}\sum_{x_1^T,\td{u}_1^T}\min_{\ld}~ e^{\sum_{u_{1,T}}\ld_{u_{1,T}}\log\sum\limits_h\pi_hC_{u_{1,T}h}~p_h(x_1^T,\td{u}_1^T)}\nn\\
&~=\min_\ld\sum_{x_1^T,\td{u}_1^T}e^{\sum_{u_{1,T}}\ld_{u_{1,T}}(x_1^T,\td{u}_1^T)\log\sum_h\pi_hC_{u_{1,T}h}~p_h(x_1^T,\td{u}_1^T)},\nn
\ea\ee
where $\sum_{u_{1,T}}\ld_{u_{1,T}}=1$. Step (s1) is due to the optimal decision rule at $X_1$. Step (s2) is simply step (s1) along with the assumption that the optimal rule at $\td{X}_1$ is already given, so that both $x_1^T$ and $\td{u}_1^T$ are treated observational data by $X_1$. Thus, we have introduced a \emph{conditionally optimal} rule at $X_1$, with decision regions $R_{u_{1,T}}$. Step (s3) is merely a reinterpretation of the (conditionally) optimal rule at $X_1$. Step (s4) is due to a familiar mathematical identity $\min(|a_1|,|a_2|,...)=\min\limits_{\ld:\sum \ld_i=1}~\prod_i|a_i|^{\ld_i}$.

We define the \emph{Chernoff information} for testing $M$ hypotheses at $X_1$ to be
\be\ba
&\C(p_0,p_1,\cdots,p_{M-1}) = \lim_{T\ra\infty}~-{1\over T}\log~\min(S_T)\nn\\
&~\sr{(s)}{=}-\min_\ld\!\left[\log\sum_{x_1,\td{u}_1}e^{\sum_{u_1}\ld_{u_1}(x_1,\td{u}_1)\log\left(\max\limits_h~\!c_{u_1,h}~\!p_h(x_1,\td{u}_1)\right)}\right],\nn
\ea\ee
where $\sum_{u_1}\ld_{u_1}=1$,
\be\ba
c_{u_1,h}=\lim_{T\ra\infty}[\pi_hC_{u_{1,T}h}]^{1/T}=\lim_{T\ra\infty}[C_{u_{1,T}h}]^{1/T}~\in~\{0,1\},\nn
\ea\ee
and step (s) assumes the independence $p_h(x_1^T,\td{u}_1^T)=\prod_{t=1}^Tp_h(x_{1,t},\td{u}_{1,t})$, as well as uses the well known identity
\bea
\lim_{T\ra\infty}\left(\sum_i|a_i|^T\right)^{1/T}=\max_i~|a_i|.\nn
\eea
Note that if $X_k$, $k\neq 1$, is viewed as an independent fusion center with risk
\be
S_T^{(k)}=\sum_{u_{k,T},h}C_{u_{k,T}h}^{(k)}~p(u_{k,T},h),\nn
\ee
then the above procedure can be repeated to obtain the Chernoff information at $X_k$ as
\be\ba
&\C^{(k)}(p_0,p_1,\cdots,p_{M-1}) = \lim_{T\ra\infty}~-{1\over T}\log~\min(S_T^{(k)})\nn\\
&~=-\min_\ld\left[\log\sum_{x_k,\td{u}_k}e^{\sum_{u_k}\ld_{u_k}(x_k,\td{u}_k)\log\left(\max\limits_hc_{u_k,h}^{(k)}~\!\!p_h(x_k,\td{u}_k)\right)}\right].\nn
\ea\ee

When $h$ and $u_1$ are both binary, and we consider $0$-$1$ costs, then $c_{u_1,h}=1-\delta_{u_1,h}$, and we obtain the usual expression for Chernoff information,

{\small
\be\ba
&\C(p_0,p_1)=-\min_{0\leq\ld\leq 1}\left[\log\sum_{x_1,\td{u}_1}p_0(x_1,\td{u}_1)e^{\ld(x_1,\td{u}_1)\log{p_1(x_1,\td{u}_1)\over p_0(x_1,\td{u}_1)}}\right]\nn\\
&~\sr{(s)}{\leq} \sum_{x_1,\td{u}_1}p_0(x_1,\td{u}_1)\log{p_0(x_1,\td{u}_1)\over p_1(x_1,\td{u}_1)}=D(p_0|p_1),\nn
\ea\ee
}as given in Section 11.9 of \cite{cover-thomas}, where step (s) is due to Jensen's inequality. Note that because the situation is symmetric, we also have $\C(p_0,p_1)\leq D(p_1|p_0)$. Thus, we have the \emph{Kullback-Leibler (KL) distance} bound
\be
\label{KL-bound}\C(p_0,p_1)\leq \min\big(D(p_0|p_1),D(p_1|p_0)\big).
\ee
Recalling that $D(p_0|p_1)$ is the best possible error exponent for the Neyman-Pearson test (Chernoff-Stein Lemma, Section 11.8 of \cite{cover-thomas}), the above inequality shows that sacrificing an arbitrarily small false alarm detection performance can improve (asymptotic) missed detection performance.

The KL bound (\ref{KL-bound}) can be obtained for more general $h$ and $u_1$ (again with the help of Jensen's inequality) as follows. Let $h$, $u_1$ have the same alphabet of size $M$, and let $c_{u_1,h}=1-\delta_{u_1,h}$. Then
\be\ba
&\C(p_0,p_1,\cdots,p_{M-1})&=&~-\min_\ld\!\left[\log\sum_{x_1,\td{u}_1}\!e^{\sum_{u_1}\ld_{u_1}(x_1,\td{u}_1)\log\left(\max\limits_h~\!c_{u_1,h}~\!p_h(x_1,\td{u}_1)\right)}\right]\nn\\
&~~~~&\leq &~ -\min_\ld\left[\log\sum_{x_1,\td{u}_1}\max\limits_h~e^{\sum_{u_1}\ld_{u_1}(x_1,\td{u}_1)\log\left(c_{u_1,h}~p_h(x_1,\td{u}_1)\right)}\right]\nn\\
&~~~~&\leq&~ (M-1)\sum_{x_1,\td{u}_1}\min\limits_{h\neq 0}~p_0(x_1,\td{u}_1)\log\left({p_0(x_1,\td{u}_1)\over p_h(x_1,\td{u}_1)}\right).\nn
\ea\ee
Since the above inequality holds if $0$ on the right hand side is replaced by any value of $h$, we get
\be
\label{KL-bound-general}\C(p_0,p_1,\cdots,p_{M-1})\leq (M-1)\min_h\sum_{x_1,\td{u}_1}\min\limits_{h'\neq h}~p_h(x_1,\td{u}_1)\log\left({p_h(x_1,\td{u}_1)\over p_{h'}(x_1,\td{u}_1)}\right).\nn
\ee

\subsection*{Online sensor comparison and asymptotically optimal network patterns}\label{sensor-comparison}
Since Chernoff information does not depend on the prior, we may use it to describe (asymptotically) optimal network patterns. Here, ``\emph{optimal}'' will mean ``\emph{asymptotically optimal}''. Note that by definition, every sensor is a \emph{local fusion center} (LFC). On the other other hand, a sensor may or may not be a \emph{global fusion center} (GFC), which is defined as follows.

Consider a set of sensors $X^n=\{X_1,...,X_n\}$. We may rank sensors according to quality such that $X_k$ is better than $X_{k'}$ if
\be
\C^{(X_k)}(p)\geq \C^{(X_{k'})}(p),\nn
\ee
where $\C^{(X_k)}(p)$ denotes $\C^{(X_k)}(p_0,p_1,...,p_{M-1})$, i.e., the Chernoff information of $X_k$. Consider a \emph{network pattern} $G=(V_G,A_G)$ over $X^n$, where $V_G=X^n$ is the set of vertices and $A_G=\{\rho_{kk'}:k,k'=1,...,n\}$ is the set of arrows: there is an arrow $X_k\ra X_{k'}$ if and only if $\rho_{kk'}=1$, and $\rho_{kk'}=0$ otherwise. Let $\C^{(X)}_G(p)$ denote the chernoff information of a sensor $X$ in $G$.

$X$ is a \emph{global fusion center} in $G$, written $G\leq X$, if
\be
\C^{(X)}_G(p)=\mathop{\max}\limits_{X'\in G}~\C^{(X')}_G(p).\nn
\ee
That is, a GFC is any sensor with maximal Chernoff information.

Let $\G$ be a set of network patterns over $X^n$. For example, we may consider $\G$ to be the set of all network patterns over $X^n$ with at most $L$ edges, i.e.,
\be
\G=\G(X^n,L)=\{G=(X^n,A_G):|A_G|\leq L\}.\nn
\ee
.  Let $\G_X=\{G\in\G:~G\leq X\}$ be the set of network patterns in each of which $X$ is a GFC. Then we have the following problems:
\begin{enumerate}
\item Find an optimal network pattern in $\G$ with $X$ as a GFC.
\item Find an optimal fusion center with respect to $\G$.
\end{enumerate}
For problem 1, $G_X$ is an optimal pattern with $X$ as a GFC if
\be\ba
\C_{G_X}^{(X)}(p)=\max_{G\in\G_X}~\C_G^{(X)}(p)=\max_{G\in\G:~G\leq X}~\C_G^{(X)}(p).\nn
\ea\ee
For problem 2, $X$ is an \emph{optimal fusion center} (with respect to $\G$) if
\be\ba
\C_{G_X}^{(X)}(p)=\max_{Y\in X^n}~\C_{G_Y}^{(Y)}(p)=\max_{Y\in X^n}~\max_{G\in\G_Y}~\C_G^{(Y)}(p).\nn
\ea\ee

We note that the optimality defined here is robust with respect to the Bayesian cost structure and prior probability, since the Chernoff information depends neither on cost structure nor on prior probability.

\subsection*{Conclusion}
We have studied distributed detection over sensor networks in the form of acyclic directed graphs. It was found that the decision rules for such networks are not more complicated than those for simple networks, provided we assume that each sensor sends the same message to all sensors receiving from it. This is still true regardless of whether sensor observations are conditionally independent or not. Information measures associated with large sample analysis of error probability were used to define sensor comparison and to define asymptotic optimality of network patterns.


\chapter{Conclusion}\label{conclusion}
\section{Main results and application}
Based on familiar notions of optimization and statistics (Chapters \ref{optimization},\ref{info-theory}), we have developed a detection network optimization technique (Chapter \ref{optimal-detection}) that can be applied in a variety of distributed detection systems. The obtained decision procedure provides necessary and sufficient conditions for, i.e., a complete characterization of, optimality in any decision optimization problem for which the underlying decision objective function is differentiable, monotonic, and convex in decision probabilities. This defines the scope of applicability of the result.

Our decision optimization procedure was applied in the following three distributed detection settings.
\begin{enumerate}
\item \emph{Interactive distributed detection (Chapter \ref{interactive-detection})}: Under the Neyman-Pearson framework, we studied effects of a single round of interaction through exchange of 1-bit decisions between two sensors, both in the fixed sample case and in the large sample case. We observed that without any communication rate constraints, interaction improves performance of the fixed sample test but not its asymptotic performance. These results were generalized to cases involving multiple rounds of memoryless interaction, multiple sensors in parallel, and exchange of multibit decisions.
\item \emph{Optimal fusion architecture  (Chapter \ref{communication-direction})}: In the Bayesian framework, we derived the optimal decision rule for the detection of a deterministic or Gaussian signal in Gaussian noise by a two-sensor tandem fusion network. We found that for low SNR, the sensor with higher SNR should serve as the fusion center.
\item \emph{Detection by acyclic graph networks (Chapter \ref{acyclic-graph-detection})}: We showed that in a sensor network in the form of an acyclic directed graph, if each sensor transmits the same message to all of its off-springs in the network, then the optimal decision rules for such a network are similar to those of simple tandem and parallel networks. This is still true when the sensors communicate through non-ideal channels. In a brief study of large sample asymptotic analysis we derived Chernoff and Kullback information measures and used these measures to define a scheme for comparison among sensors and among sensor network patterns. This type of comparison scheme can be used, in particular, to determine (asymptotically) optimal sensor distributions within a given class of sensor network patterns.
\end{enumerate}

\section{Future research}
Here we present a number of problems indicating possible directions for future research. These problems come from the major parts of this thesis, and they include natural extensions of our optimal signal detection procedure as well as additional problems arising from the three main applications we have studied.

Based directly on our decision procedure of Section \ref{opt-hyp-testing}, the following are a number of possible research directions.
\begin{enumerate}
\item \emph{Sequential detection}: Using our decision optimization method, we would like to describe both centralized and distributed sequential detection from scratch. We expect our method to provide a relatively simple description of the sequential detection problem, especially in the distributed setting.

\item \emph{Deeper study of randomization of decision rules and dependence of detection performance on randomization parameters}: In particular, when randomization does make a difference, we would like to be able to select the best possible (i.e., an optimal) randomized decision rule in an automatic way using our decision procedure.
\item \emph{Optimization of non-differentiable convex decision functions}: For simplicity, we assumed differentiability of the decision function in our analysis. However, we expect that the same decision process should still apply, with minimal adjustments, when the objective function is non-differentiable. In particular, the (partial) derivative of the decision function should be replaced by its subdifferential.
\item \emph{Extension to harmonic decision functions}: Given that some key properties of convex functions which made our analysis possible are possessed by all subharmonic functions, we expect that our optimization procedure for monotonic convex decision functions can be readily extended to a similar optimization procedure for monotonic subharmonic decision functions.
\item \emph{Relation to the optimization of submodular set functions}: We would like to explore connections between our optimization procedure and the optimization of submodular set functions used for the determination of optimal sensor placement within a given network pattern.
\end{enumerate}

Each of our three main applications also gave rise to a research problem as follows.
\begin{enumerate}[resume]
\item \emph{Interactive distributed detection}: What is the cost incurred by additional rounds of interaction? The main point here is that although interaction can strictly improve performance of the fixed sample test, the additional communication steps involved can be costly. A natural way to account for this would be to modify the original objective function, say by including an additional term in it.

\item \emph{Optimal Fusion architecture}: What is the optimal communication direction for large SNR? What is needed here is simply an efficient computational algorithm, since the optimal decision rule is already available.

\item \emph{Acyclic graph detection}: What is the optimal sensor distribution that achieves uniform reliability (asymptotically) in a sensor network? Here, uniform reliability refers to the situation where the value of the local Chernoff information is the same at each sensor.
\end{enumerate}

\section{Comments on sequential detection}
At first sight, dynamic programming (DP) methods appear to be the ideal choice for sequential detection problems. This is mainly because DP itself is sequential in nature. The DP approach is an \emph{inductive} approach which, for more practical reasons, typically dictates that optimization should be done incrementally after each observation sample. Unfortunately, however, DP is suboptimal in general. Moreover, despite its built in sequential nature, DP does not necessarily provide the most convenient description of the sequential detection problem.

It is theoretically more convenient to consider a \emph{deductive} approach in which we assume that an exhaustive collection of \emph{sequential detection plans or strategies}, each consisting of a (possibly infinite) sequence of observation samples along with a corresponding randomly generated sequence of decisions, are already available. In such an approach, it is necessary to \emph{explicitly} specify additional constraints ensuring that at any given step, processing continues to the next step if and only if the decision outputs of all previous steps each failed to meet the stopping criterion. (Note that such stopping constraints are \emph{implicit} in an inductive approach such as DP). One then proceeds to select the sequential detection plan that optimizes the underlying objective function of the sequential test.

Note however that the inductive approach is preferred over the deductive approach when knowledge of available sequential detection strategies is severely limited. When knowledge of available sequential detection strategies is unlimited, then with respect to optimality, the deductive approach is always preferred over the inductive approach.

Based on the above discussion, we conclude that our optimization technique developed in Chapter \ref{optimal-detection} can provide a more convenient description of the sequential detection problem, as compared with a dynamic programming approach. To apply our method in sequential detection, we need to first specify a decision function, a natural choice of which is some measure of stopping time as a function of the probabilities of the decision sequences. Next we specify stopping constraints, which should greatly reduce the number of nontrivial optimization variables (i.e., probabilities of the decision sequences). Finally, we optimize the decision function by selecting an optimal decision strategy (i.e., a decision sequence or strategy whose probability optimizes the decision function). If the decision function is a monotonic convex function of probabilities of the decision sequences, then either Proposition \ref{region-proposition} or Corollary \ref{region-corollary} (depending on whether we are maximizing the decision function or minimizing it) will provide a complete characterization of the solution of the optimization problem.


\addcontentsline{toc}{part}{Bibliography}

\hrulefill
\begin{center}
\bit
\item[] \hspace{0cm}{\Large About the author}
\vspace{0.5cm}
\item[] \textbf{Name:}~~ Earnest Akofor
\vspace{0.5cm}

\item[] \textbf{Degrees awarded:}~
\bit
\item[] Ph.D. Physics, Syracuse University, 2010
\vspace{0.2cm}

\item[] Diploma in Mathematical Sciences, AIMS, 2004
\vspace{0.2cm}

\item[] Diploma in High Energy Physics, ICTP, 2003
\vspace{0.2cm}

\item[] B.Sc. Physics, University of Buea, 2001
\eit
\vspace{0.5cm}

\item[] \textbf{Professional experience:}~
\bit
\item[] Peer Review, IJDSN (2015) and IEEE Transactions (2015-2016)
\vspace{0.2cm}

\item[] Research Assistant, Syracuse University, 2011-2016
\vspace{0.2cm}

\item[] Adjunct Instructor, ITT Technical Institute, 2011
\vspace{0.2cm}

\item[] Teaching \& Research Assistant, Syracuse University, 2004-2009
\eit
\eit
\end{center}

\end{document}